\documentclass[10pt-default]{article}
\usepackage{amsmath}
\usepackage{amssymb}
\usepackage{amsfonts}
\usepackage[dvips]{graphicx}

\input{tcilatex}

\begin{document}

{\huge \ }

{\huge New Strings for Old Veneziano }

{\huge Amplitudes IV. Connections }

{\huge With Spin Chains and Other}

{\Huge Stochastic Systems}

$\ \ \ \ \ \ \ \ \ \ \ \ \ \ \ \ \ \ \ \ \ \ \ \ \ \ \ \ \ \ \ \ \ $

Arkady Kholodenko

\textit{375 H.L.Hunter Laboratories, Clemson University,}

\textit{Clemson, \ SC} 29634-0973, U.S.A.

E-mail: string@clemson.edu

\bigskip

{\large Abstract: }In a series of recently published papers we reanalyzed
the existing treatments of the Veneziano and Veneziano-like amplitudes and
the models associated with these amplitudes. In this work we demonstrate
that the already obtained \ new partition function for these amplitudes can
be exactly mapped into that for the Polychronakos-Frahm (P-F) spin chain
model which, in turn, is obtainable from the Richardon-Gaudin (R-G) XXX
model. Reshetikhin and Varchenko demonstrated that such a model is
obtainable as a leading approximation in their WKB-type analysis of
solutions of \ the Knizhnik-Zamolodchikov (K-Z) equations. The linear
independence of solutions of these equations is controlled by determinants
(discovered by Varchenko) whose explicit form up to a constant coincides
with the Veneziano (or Veneziano-like) amplitudes. In the simplest case,
when K-Z equations are reducible to the Gauss hypergeometric equation, the
determinantal conditions coincide with those which were discovered by Kummer
in 19-th century. Kummer's results admit physical interpretation crucial for
providing needed justification associating determinantal formula(s) with
Veneziano-like amplitudes. General results are illustrated by many examples.
These include but are not limited \ to only high energy physics since all
high energy physics scattering processes can be looked upon from much
broader stochastic theory of random fragmentation and coagulation processes
\ recently undergoing active development in view of its applications in
disciplines ranging from ordering in spin glasses and population genetics to
computer science, linguistics and economics, etc. In this theory \ Veneziano
amplitudes play a central (universal) role since they are the
Poisson-Dirichlet-type distributions for these processes (analogous to the
more familiar Maxwell distribution for gases).


\ 

\ 

\pagebreak

\bigskip

\textbf{Contents}

\textbf{1}.\textbf{Introduction }

\textbf{2}.\textbf{Combinatorics of Veneziano amplitudes and spin chains:}

\ \textbf{\ qualitative considerations}

\textbf{3}.\textbf{Connection with the Polychronakos-Frahm (}P-F\textbf{)
spin }

\ \ \ \textbf{chain model}

\textbf{4}.\textbf{Connections with WZNW model and XXX s=1/2 Heisenberg }

\ \ \ \textbf{antiferromagnetic spin chain}

\ \ \ 4.1 \ General remarks

\ \ \ 4.2 Method of generating functions andq-deformed harmonic oscillator

\ \ \ 4.3 The limit $q\rightarrow 1^{\pm }$ and emergence of the
Stiltjes-Wiegert polynomials

\ \ \ 4.4 ASEP, q-deformed harmonic oscillator and spin chains

\ \ \ 4.5 Crossover between the XXZ and XXX spin chains: connections

\ \ \ \ \ \ \ \ with the KPZ and EW equations and the lattice Liouville model

\ \ \ 4.6 ASEP, vicious random walkers and string models

\textbf{5}. \textbf{Gaudin model as linkage between the WZNW model and K-Z }

\ \ \ \ \textbf{equations. Recovery of the Veneziano-like amplitudes}

\ \ \ \ 5.1 General remarks

\ \ \ \ 5.2 Gaudin magnets, K-Z equation and P-F spin chain

\ \ \ \ 5.3 The Shapovalov form

\ \ \ \ 5.4 Mathematics and physics of the Bethe ansatz equations for XXX

\ \ \ \ \ \ \ \ Gaudin model according to works by Richardson. Connections

\ \ \ \ \ \ \ \ with the Veneziano model

\ \ \ \ \ 5.5 Emergence of the Veneziano-like amplitudes as consistency
condition

\ \ \ \ \ \ \ \ for N=1 solutions of K-Z equations. Recovery of the pion-pion

\ \ \ \ \ \ \ \ scattering amplitude

\ \textbf{6.} \textbf{Discussion. Unimaginable \ ubiquity of the
Veneziano-type}

\ \ \ \ \ \textbf{amplitudes in Nature}

\ \ \ \ \ 6.1 \ General remarks

\ \ \ \ \ 6.2. Random fragmentation and coagulation

\ \ \ \ \ \ \ \ \ \ \ \ processes and the Dirichlet distribution

\ \ \ \ \ 6.3 \ The Ewens sampling formula and Veneziano amplitudes

\ \ \ \ \ 6.4 \ Stochastic models for second order chemical reaction

\ \ \ \ \ \ \ \ \ \ \ kinetics involving Veneziano-like amplitudes

\ \ \ \ \ 6.4.1Quantum mechanics, hypergeometric functions and the

\ \ \ \ \ \ \ \ \ \ \ \ \ Poisson-Dirichlet distribution

\ \ \ \ \ 6.4.2 Hypergeometric functions, Kummer series expansions and

\ \ \ \ \ \ \ \ \ \ \ \ \ Veneziano-like amplitudes

\ \ \ \ \textbf{A. Basics of ASEP}

\ \ \ \ \ \ \ \ \ A.1 Equations of motion and spin chains

\ \ \ \ \ \ \ \ \ A.2 Dynamics of ASEP and operator algebra

\ \ \ \ \ \ \ \ \ A.3 Steady -state and q-algebra for the deformed harmonic
oscillator

\ \ \ \ \ \textbf{B. Linear independence of solutions of K-Z equation}

\ \ \ \ \ \textbf{C. Connections between the gamma and Dirichlet }

\ \ \ \ \ \ \ \ \ \ \ \ \textbf{distributions}

\ \ \ \ \ \textbf{D. Some facts from combinatorics of the symmetric group}

\ \ \ \ \ \ 

\ \ \ \ \ \ \ \ \ \ \ 

\section{Introduction}

\bigskip

Since time when quantum mechanics (QM) was born (in 1925-1926) two seemingly
opposite approaches for description of atomic and subatomic physics were
proposed respectively by Heisenberg and Schr\"{o}dinger. Heisenberg's
approach is aimed at providing an affirmative answer to the following
question: Is combinatorics of spectra (of obsevables) \ provides sufficient
\ information about microscopic system so that dynamics of such a system can
be described in terms of known macroscopic concepts? Schrodinger's approach
is exactly opposite and is aimed at providing an affirmative answer to the
following question: Using some plausible mathematical arguments is it
possible to find an equation which under some prescribed restrictions will
reproduce the spectra of observables? Although it is widely believed that
both approaches are equivalent, already Dirac in his lectures on quantum
field theory [1] noticed (without much elaboration) that Schrodinger's
description of QM contains a lot of "dead wood" which can be safely disposed
altogether. According to Dirac \ "Heisenberg's picture of QM is good because
Heisenberg's equations of motion make sense".

To our knowledge, Dirac's comments were completely ignored, perhaps, because
he had not provided enough evidence making Heisenberg's description of QM
superior to that of Schrodinger's. In recent papers [2,3] we found examples
supporting Dirac's claims. From the point of view of combinatorics, there is
not much difference in description of QM, quantum field theory and string
theory. Therefore, in this paper we choose the Heisenberg's point of view on
string theory using results of our recent works in which we re analyzed the
existing treatments connecting Veneziano (and Veneziano-like) amplitudes
with the respective string-theoretic models. As result, we were able to find
new tachyon-free models reproducing Veneziano (and Veneziano-like)
amplitudes. In this work the result of our \ papers [4-6] which will be
called as Part I, Part II and Part III respectively, are \ developed further
to bring them in correspondence with those proposed by other authors.
Without any changes in the already developed formalism, we were able to
connect our results with an impressive number of string-theoretic models,
including the most recent ones. Nevertheless, below we argue that, although
physically plausible, the established connections (in the way they are
typically treated in physics literature) are mathematically ill founded. To
correct this deficiency, in Section 5 we use some works by mathematicians.
Of particular importance for us are the works by Reshetikhin and Varchenko
[7] \ and by Varchenko summarized in Varchenko's \ MIT lecture notes [8].
These works enabled us to relate Veneziano (and Veneziano-like) amplitudes
(e.g. those describing $\pi \pi $ scattering) to Knizhnik-Zamolodchikov
(K-Z) equations and, hence, to WZNW models. This is achieved by employing
known connections between the WZNW models and spin chains. In the present
case, between the K-Z equations and the XXX-type Richardson-Gaudin magnetic
chains as described in Section 5. Sections 2-4 contain mathematically less
sophisticated results aimed at providing needed physical \ motivations and
background. For this purpose in section 2 we replaced mathematically
sophisticated derivation of the \ Veneziano partition function by
considerably simpler combinatorial derivation of such function. As a by
product of this effort we were able to uncover the connections with spin
chains already at this stage of our investigation. To strengthen this
connection, in Section 3 we demonstrate that \ the obtained Veneziano
partition function \ coincides with the Polychronakos-Frahm (P-F) partition
function for the ferromagnetic spin chain model. Although such \ a spin
chain was extensively studied in literature, we discuss different paths in
Section 4 aimed at establishing links between the P-F spin chain and variety
of string-theoretic models, including the most recent ones. This is achieved
by mapping \ combinatorial and analytical properties \ of the P-F spin
chains into analogous properties of spin chains used for description of the
stochastic process known as asymptotic simple exclusion process (ASEP). To
make our presentation self-contained, we provide in Appendix A basic
information on ASEP sufficient for understanding the results discussed in
the main text. In addition, in the main text we provide some information on
Kardar-Parisi-Zhang (KPZ) and Edwards-Wilkinson (EW) equations which are
just different well defined macroscopic limits of the microscopic ASEP
equations. We do this with purpose of \ reproducing variety of
string-theoretic models, including the most recent ones. Such a success have
not deterred us from looking at other, more rigorous (mathematically)
approaches. These are discussed in Sections 5 and, in part, in Section 6.
These sections are interrelated and contain the most important results of
this paper. While the content of Section 5 was already briefly discussed,
the content of Section 6 provides the strongest independent support to the
results and conclusions of Section 5. At the same time, this section can be
read independently of the rest of the paper since it contains some important
facts from \ the theory of random fragmentation and coagulation processes
[9-11] which is currently in the process of rapid development because of its
wide applications ranging from theory of spin glasses and population
genetics to computer science, linguistics and economics, etc. In high energy
physics this theory was developed for some time by Mekjian, e.g. see [12]
and references therein. \ Since our Section 6 is not a review, our treatment
of topics discussed in it is markedly different from that developed in
Mekjian's papers and is subordinated to the content of Section 5.
Specifically, the main result of Section 5 is the deteminantal formula,
equation (5.47), which up to a constant coincides with the Veneziano (or
Veneziano-like) amplitude. A special case of this formula produces known
pion-pion scattering amplitude. In Section 6 we argue that: 1. Veneziano
amplitudes play the central role in the theory of random fragmentation and
coagulation processes where they are known as the Poisson-Diriclet (P-D)
probability distributions. 2. \ The discrete spectra of all exactly solvable
quantum mechanical (QM) problems can be rederived in terms of some P-D
stochastic processes. This is so because all exactly solvable QM problems
involve some kind of orthogonal polynomials-all derivable from the Gauss
hypergeometric function which admits an interpretation in terms of the P-D
process. 3. Since in the simplest case the K-Z equations are reducible to
the hypergeometric equations, the processes they describe are also of P-D
type. 4. In the case of \ Gauss hyprgeometric equation, the determinantal
formula (5.47) is reduced to that obtained by Kummer in 19th century. To
facilitate understanding \ and appreciation of these facts and to
demonstrate utility of the obtained results\ beyond the scope of high energy
physics, in Section 6 we discuss some applications of the developed
formalism to genetics and chemical kinetics.

\section{Combinatorics of Veneziano amplitudes and spin chains: qualitative
considerations}

In Part I, we noticed that the Veneziano condition for the 4-particle
amplitude given by 
\begin{equation}
\alpha (s)+\alpha (t)+\alpha (u)=-1,  \tag{2.1}
\end{equation}%
where $\alpha (s)$, $\alpha (t),\alpha (u)$ $\in \mathbf{Z}$, can be
rewritten in more mathematically suggestive form. To this purpose, following
[13], we need to consider additional homogenous equation of the type 
\begin{equation}
\alpha (s)m+\alpha (t)n+\alpha (u)l+k\cdot 1=0  \tag{2.2}
\end{equation}%
with $m,n,l,k$ being some integers. By adding this equation to (2.1) we
obtain, 
\begin{equation}
\alpha (s)\tilde{m}+\alpha (t)\tilde{n}+\alpha (u)\tilde{l}=\tilde{k} 
\tag{2.3a}
\end{equation}%
or, equivalently, as%
\begin{equation}
n_{1}+n_{2}+n_{3}=\hat{N},  \tag{2.3b}
\end{equation}%
where all entries \textit{by design} are nonnegative integers. For the
multiparticle case this equation should be replaced by%
\begin{equation}
n_{0}+\cdot \cdot \cdot +n_{k}=N  \tag{2.4}
\end{equation}%
so that \textsl{combinatorially the task lies in finding all nonnegative
integer combinations} \textsl{of} $n_{0},...,n_{k}$ \textsl{producing}
(2.4). It should be noted that such a task makes sense as long as $N$ is
assigned. But the actual value of $N$ is \textit{not} \textit{fixed} and,
hence, can be chosen quite arbitrarily. Equation (2.1) is a simple statement
about the energy -momentum conservation. Although the numerical entries in
this equation can be changed as we just explained, the actual physical
values can be \ subsequently re obtained by the appropriate coordinate
shift. Such a procedure should be applied to the amplitudes of conformal
field theories (CFT) \ with some caution \ since the periodic ( or
antiperiodic, etc.) boundary conditions cause energy and momenta to become a 
\textit{quasi -energy} and a \textit{quasi momenta} (as \ it is known from
solid state physics).

The arbitrariness of selecting $N$ reflects a kind of gauge freedom. As in
other gauge theories, we may try to fix the gauge by using some physical
considerations. These include, for example, an observation made in Part I
that the 4 particle amplitude is zero if any two entries into (2.1) are the
same. This \ fact prompts us to arrange the entries in (2.3b) in accordance
with their magnitude, i.e. $n_{1}\geq n_{2}\geq n_{3}.$ More generally, we
can write: $n_{0}\geq n_{1}\geq \cdot \cdot \cdot \geq n_{k}\geq 1\footnote{%
The last inequality: $n_{k}\geq 1,$ is chosen only for the sake of
comparison with the existing literature conventions, e.g. see Ref.[15\textbf{%
]}.}$.

In Section 6 we demonstrate that if the entries in this sequence of
inequalities are treated as random nonnegative numbers subject to the
constraint (2.4), these constrains are necessary and sufficient for recovery
of the probability density for such set of random numbers. This density \ is
known in mathematics as Dirichlet distribution\footnote{%
For reasons explained in Section 6 it is also called the Poisson-Dirichlet
distribution.} [9-11,14]. Without normalization, integrals over this
distribution coincide with Veneziano amplitudes.

Provided that (2.4) holds, we shall call such a sequence a \textit{partition 
}and shall denote it as\textit{\ }$\mathit{n\equiv }(n_{0},...,n_{k})$. If $%
n $ is partition of $N$, then we shall write $n\vdash N$. It is well known
[15,16] that there is one- to -one correspondence between the Young diagrams
and partitions. We would like to use this fact in order to design a
partition function capable of reproducing the Veneziano (and Veneziano-like)
amplitudes. Clearly, such a partition function should also make physical
sense. Hence, we would like to provide some qualitative arguments aimed at
convincing our readers that such a partition function does exist and is
physically sensible.

We begin with observation that there is one- to- one correspondence between
the Young tableaux and directed random walks\footnote{%
Furthermore, it is possible to map bijectively such type of random walk back
into Young diagram with only two rows, e.g. read [17], page 5. This allows
us to make a connection with spin chains at once. In this work we are not
going to use this route to spin chains in view of the simplicity of \
alternative approaches discussed in this section.}. It is useful to recall
details of this correspondence now. To this purpose we need to consider a
square lattice and to place on it the Young diagram associated with some
particular partition.

Let us choose some $\tilde{n}\times \tilde{m}$ rectangle\footnote{%
Parameters $\tilde{n}$ and $\tilde{m}$ will be specified shortly below.} so
that the Young diagram occupies the left part of this rectangle. We choose
the upper left vertex of the rectangle as the origin of the $xy$ coordinate
system whose $y$ axis (South direction) is directed downwards and $x$ axis
is directed Eastwards. Then, the South-East boundary of the Young diagram
can be interpreted as directed (that is without self intersections) random
walk which begins at $(0,-\tilde{m})$ and ends at $(\tilde{n},0).$
Evidently, such a walk completely determines the diagram. The walk can be
described by a sequence of 0's and 1's.\ Say, $0$ for the $x-$ step move and
1 for the $y-$ step move. The totality $\mathcal{N}$ of Young diagrams which
can be placed into such a rectangle is in one-to-one correspondence with the
number of arrangements of 0's and 1's whose total number is $\tilde{m}+%
\tilde{n}$. Recalling the Fermi statistics, the number $\mathcal{N}$ can be
easily calculated and is given by $\mathcal{N}=(m+n)!/m!n!\footnote{%
We have suppressed the tildas for $n$ and $m$ in this expression since these
parameters are going to be redefined below anyway.}$. It can be represented
in two equivalent ways%
\begin{eqnarray}
(m+n)!/m!n! &=&\frac{(n+1)(n+2)\cdot \cdot \cdot (n+m)}{m!}\equiv \left( 
\begin{array}{c}
n+m \\ 
m%
\end{array}%
\right)  \notag \\
&=&\frac{(m+1)(m+2)\cdot \cdot \cdot (n+m)}{n!}\equiv \left( 
\begin{array}{c}
m+n \\ 
n%
\end{array}%
\right) .  \TCItag{2.5}
\end{eqnarray}

Let now $p(N;k,m)$ be the number of partitions of $N$ into $\leq k$ \
nonnegative parts, each not larger than $m$. Consider the generating
function of the following type 
\begin{equation}
\mathcal{F}(k,m\mid q)=\dsum\limits_{N=0}^{S}p(N;k,m)q^{N}  \tag{2.6}
\end{equation}%
where the upper limit $S$\ will be determined shortly below. It is shown in
Refs.[15,16] that $\mathcal{F}(k,m\mid q)=\left[ 
\begin{array}{c}
k+m \\ 
m%
\end{array}%
\right] _{q}\equiv \left[ 
\begin{array}{c}
k+m \\ 
k%
\end{array}%
\right] _{q}$ where, for instance,$\left[ 
\begin{array}{c}
k+m \\ 
m%
\end{array}%
\right] _{q=1}=\left( 
\begin{array}{c}
k+m \\ 
m%
\end{array}%
\right) \footnote{%
On page 15 of the book by Stanley [16], one can find that the number of
solutions $N(n,k)$ in \textit{positive} integers to $y_{1}+...+y_{k}=n+k$ is
given by $\left( 
\begin{array}{c}
n+k-1 \\ 
k-1%
\end{array}%
\right) $ while the number of solutions in \textit{nonnegative} integers to $%
x_{1}+...+x_{k}=n$ is $\left( 
\begin{array}{c}
n+k \\ 
k%
\end{array}%
\right) .$ Careful reading of Page 15 indicates however that the last number
refers to solution in nonnegative integers of the equation $%
x_{0}+...+x_{k}=n $. This fact was used essentially in (1.21) of Part I.}.$
From this result it should be clear that the expression $\left[ 
\begin{array}{c}
k+m \\ 
m%
\end{array}%
\right] _{q}$ is the $q-$analog of the binomial coefficient $\left( 
\begin{array}{c}
k+m \\ 
m%
\end{array}%
\right) .$ In literature [15,16] this $q-$ analog is known as the \textit{%
Gaussian} coefficient. Explicitly, it is defined as%
\begin{equation}
\left[ 
\begin{array}{c}
a \\ 
b%
\end{array}%
\right] _{q}=\frac{(q^{a}-1)(q^{a-1}-1)\cdot \cdot \cdot (q^{a-b+1}-1)}{%
(q^{b}-1)(q^{b-1}-1)\cdot \cdot \cdot (q-1)}  \tag{2.7}
\end{equation}%
for some nonegative integers $a$ and $b$. From this definition we anticipate
that the sum defining generating function $\mathcal{F}(k,m\mid q)$ in (2.6)
should have only \textit{finite} number of terms. Equation (2.7) allows easy
determination of the upper limit $S$ in the sum (2.6). It is given by $km$.
This is just the area of the $k\times m$ rectangle. In view of the
definition of $p(N;k,m)$, the number $m=N-k$. Using this fact (2.6) can be
rewritten as: $\mathcal{F}(N,k\mid q)=\left[ 
\begin{array}{c}
N \\ 
k%
\end{array}%
\right] _{q}.$This expression happens to be the Poincare$^{\prime }$
polynomial for the Grassmannian $Gr(m,k)$ of the complex vector space 
\textbf{C}$^{N}$of dimension $N$ as can be seen from page 292 of the book by
Bott and Tu, [18]\footnote{%
To make a comparison it is sufficient to replace parameters $t^{2}$ and $n$
in \ Bott and Tu book by $q$ and $N.$}. From this (topological) point of
view the numerical coefficients, i.e. $p(N;k,m),$ in the $q$ expansion of
(2.6) should be interpreted as Betti numbers of this Grassmannian. They can
be determined recursively using the following property of the Gaussian
coefficients [\textbf{4}], page 26,%
\begin{equation}
\left[ 
\begin{array}{c}
n+1 \\ 
k+1%
\end{array}%
\right] _{q}=\left[ 
\begin{array}{c}
n \\ 
k+1%
\end{array}%
\right] _{q}+q^{n-k}\left[ 
\begin{array}{c}
n \\ 
k%
\end{array}%
\right] _{q}  \tag{2.8}
\end{equation}%
and taking into account that $\left[ 
\begin{array}{c}
n \\ 
0%
\end{array}%
\right] _{q}=1.$ We refer our readers to Part II for mathematical proof that 
$\mathcal{F}(N,k\mid q)$ is indeed the Poincare$^{\prime }$ polynomial for
the complex Grassmannian. With this fact proven, we notice that, due to
relation $m=N-k,$ it is sometimes more convenient \ for us to use the
parameters $m$ and $k$ rather than $N$ and $k$. With such a replacement we
obtain: 
\begin{eqnarray}
\mathcal{F}(k,m &\mid &q)=\left[ 
\begin{array}{c}
k+m \\ 
k%
\end{array}%
\right] _{q}=\frac{(q^{k+m}-1)(q^{k+m-1}-1)\cdot \cdot \cdot (q^{m+1}-1)}{%
(q^{k}-1)(q^{k-1}-1)\cdot \cdot \cdot (q-1)}  \notag \\
&=&\dprod\limits_{i=1}^{k}\frac{1-q^{m+i}}{1-q^{i}}.  \TCItag{2.9}
\end{eqnarray}%
This result is of central importance. In our work, Part II, considerably
more sophisticated mathematical apparatus was used to obtain it (e.g. see
equation (6.10) of this reference and arguments leading to it).\ 

In the limit : $q\rightarrow 1$ (2.9) reduces to $\mathcal{N}$ as required.
To make connections with results known in physics literature we need to re
scale $q^{\prime }s$ in (2.9), e.g. let $q=t^{\frac{1}{i}}.$ Substitution of
such an expression back into (2.9) and taking the limit $t\rightarrow 1$
again produces $\mathcal{N}$ in view of (2.5). This time, however, we can
accomplish more. By noticing that in (2.4) the actual value of $N$
deliberately is not yet fixed and taking into account that $m=N-k$ we can
fix $N$ by fixing $m$. Specifically, we would like to choose $m=1\cdot
2\cdot 3\cdot \cdot \cdot k$ and with such a choice we would like\ to
consider a particular term in the product (2.9), e.g.%
\begin{equation}
S(i)=\frac{1-t^{1+\frac{m}{i}}}{1-t}.  \tag{2.10}
\end{equation}%
In view of our "gauge fixing" the ratio $m/i$ is a positive integer by
design. This means that we are having a geometric progression. Indeed, if we
rescale $t$ again : $t\rightarrow t^{2},$ we then obtain:%
\begin{equation}
S(i)=1+t^{2}+\cdot \cdot \cdot +t^{2\hat{m}}  \tag{2.11}
\end{equation}%
with $\hat{m}=\frac{m}{i}.$ Written in such a form the \ above sum is just
the Poincare$^{\prime }$ polynomial for the complex projective space \textbf{%
CP}$^{\hat{m}}.$ This can be seen by comparing pages 177 and 269 of the book
by Bott and Tu [18]. Hence, at least for some $m$'s, \textit{the Poincare}$%
^{\prime }$\textit{\ polynomial for the Grassmannian in just the product of} 
\textit{the Poincare}$^{\prime }$\textit{\ polynomials for the complex
projective spaces of known dimensionalities}. For $m$ just chosen, in the
limit $t\rightarrow 1,$ we reobtain back the number $\mathcal{N}$ as
required. This physically motivating process of gauge fixing just described
can be replaced by more rigorous mathematical arguments. The recursion
relation (2.8) introduced earlier indicates that this is possible. The \
mathematical details leading to factorization which we just described can be
found, for instance, in the Ch-3 of lecture notes by Schwartz [19]. The
relevant physics emerges by noticing that the partition function $Z(J)$ for
the particle with spin $J$ is given by [20] 
\begin{eqnarray}
Z(J) &=&tr(e^{-\beta H(\sigma )})=e^{cJ}+e^{c(J-1)}+\cdot \cdot \cdot
+e^{-cJ}  \notag \\
&=&e^{cJ}(1+e^{-c}+e^{-2c}+\cdot \cdot \cdot +e^{-2cJ}),  \TCItag{2.12}
\end{eqnarray}%
where $c$ is known constant. Evidently, up to a constant, $Z(J)\simeq S(i).$
Since\ mathematically the result (2.12) is the Weyl character formula, this
fact brings the classical group theory into our discussion. More
importantly, because the partition function for the particle with spin $J$
can be written in the language of N=2 supersymmetric quantum mechanical model%
\footnote{%
We hope that no confusion is made about the meaning of N in the present case.%
}, as demonstrated by Stone [20] and others [21], the connection between the
supersymmetry and the classical group theory is evident. \ It was developed
in Part III.

In view of arguments presented above, the Poincare$^{\prime }$ polynomial
for the Grassmannian can be interpreted as a partition function for some
kind of a spin chain made of \ apparently independent spins of various
magnitudes\footnote{%
In such a context it can be vaguely considered as a variation on the theme
of the Polyakov rigid string (Grassmann $\sigma $ model, Ref.[22], pages
283-287), except that now it is \textit{exactly solvable} in the qualitative
context \ just described and, below, in mathematically rigorous context.}.
These qualitative arguments we would like to make more mathematically and
physically rigorous. The first step towards this goal is made in the next
section.

\section{Connection with the Polychronakos-Frahm spin chain model}

\bigskip

The Polychronakos-Frahm (P-F) spin chain model \ was originally proposed by
Polychronakos and described in detail in [23]. Frahm [24] \ motivated by the
results of Polychronakos made additional progress in elucidating the
spectrum and thermodynamic properties of this model so that it had become
known as the P-F model. Subsequently, many other researchers have
contributed to our understanding of this exactly integrable spin chain
model. Since this paper is not a review, we shall quote only works on P-F
model which are of immediate relevance.

Following [23], we begin with some description of the P-F model. Let $\sigma
_{i}^{a}$ ($a=1,2,...,n^{2}-1)$ be $SU(n)$ spin operator of i-th particle
and let the operator$\ \sigma _{ij}$ be responsible for a spin exchange
between particles $i$ and $j,$ i.e.%
\begin{equation}
\sigma _{ij}=\frac{1}{n}+\tsum\limits_{a}\sigma _{i}^{a}\sigma _{j}^{a}. 
\tag{3.1}
\end{equation}%
In terms of these definitions, the Calogero-type model Hamiltonian can be
written as [25,26] 
\begin{equation}
\mathcal{H}=\frac{1}{2}\tsum\limits_{i}(p_{i}^{2}+\omega
^{2}x_{i}^{2})+\tsum\limits_{i<j}\frac{l(l-\sigma _{ij})}{\left(
x_{i}-x_{j}\right) ^{2}},  \tag{3.2}
\end{equation}%
where $l$ is some parameter. The P-F model is obtained from the above model
in the limit $l\rightarrow \pm \infty $ . Upon proper rescaling of $\mathcal{%
H}$ in (3.2), in this limit one obtains%
\begin{equation}
\mathcal{H}_{\mathcal{P-F}}=-sign(l)\tsum\limits_{i<j}\frac{\sigma _{ij}}{%
\left( x_{i}-x_{j}\right) ^{2}},  \tag{3.3}
\end{equation}%
where the coordinate $x_{i}$ minimizes the potential for the rescaled
Calogero model\footnote{%
The Calogero model is obtainable from the Hamiltonian (3.2) if one replaces
the spin exchange operator $\sigma _{ij}$ by 1. Since we are interested in
the large $l$ limit, one can replace the factor $l(l-1)$ by $l^{2}$ in the
interaction term.}, that is%
\begin{equation}
\omega ^{2}x_{i}^{{}}=\tsum\limits_{i<j}\frac{2}{\left( x_{i}-x_{j}\right)
^{3}}.  \tag{3.4}
\end{equation}%
It should be noted that $\mathcal{H}_{\mathcal{P-F}}$ is well defined
without such a minimization, that is for arbitrary real parameters $x_{i}$.
This fact will be further explained in Section 5.\ In the large $l$ limit
the spectrum of $\mathcal{H}$ is decomposable as 
\begin{equation}
E=E_{\mathcal{C}}+lE_{\mathcal{P-F}}  \tag{3.5}
\end{equation}%
where $E_{C}$ is the spectrum of the spinless Calogero model while $E_{%
\mathcal{P-F}}$ is the spectrum of the P-F model. In view of such a
decomposition, the partition function for the Hamiltonian $\mathcal{H}$ at
temperature $T$ can be written as a product: Z$_{\mathcal{H}}(T)=$Z$_{%
\mathcal{C}}(T)$Z$_{\mathcal{P-F}}(T/l)$. From here, one formally obtains
the result: 
\begin{equation}
Z_{\mathcal{P-F}}(T)=\lim_{l\rightarrow \infty }\frac{Z_{\mathcal{H}}(lT)}{%
Z_{\mathcal{C}}(T)}.  \tag{3.6}
\end{equation}%
It implies that the spectrum of the P-F spin chain can be obtained if both
the total and \ the Calogero partition functions can be calculated. In [23]
Polychronakos argued that $Z_{\mathcal{C}}(T)$ is essentially a partition
function of $\mathit{N}$ noninteracting harmonic oscillators. Thus, we
obtain 
\begin{equation}
Z_{\mathcal{C}}(N;T)=\tprod\limits_{i=1}^{N}\frac{1}{1-q^{i}},\text{ }q=\exp
(-\beta ),\beta =\left( k_{B}T\right) ^{-1}.  \tag{3.7}
\end{equation}%
Furthermore, the partition function $Z_{\mathcal{H}}(T)$ according to
Polychronakos can be obtained using $Z_{\mathcal{C}}(N;T)$ as follows.
Consider the grand partition function of the type%
\begin{equation}
\Xi =\tsum\limits_{N=0}^{\infty }Z_{n}(N;T)y^{N}\equiv \left(
\tsum\limits_{L=0}^{\infty }Z_{\mathcal{C}}(L;T)y^{L}\right) ^{n}  \tag{3.8}
\end{equation}%
where $n$ is the number of flavors\footnote{%
That is $n$ the same number as $n$ in $SU(n).$}. \ Using this definition we
obtain%
\begin{equation}
Z_{n}(N;T)=\sum_{\Sigma _{i}k_{i}=N}\prod\limits_{i=1}^{n}Z_{\mathcal{C}%
}(k_{i};T).  \tag{3.9}
\end{equation}%
Next, Polychronakos identifies $Z_{n}(N;T)$ with Z$_{\mathcal{H}}(T)$. Then,
with help of (3.6) the partition function $Z_{\mathcal{P-F}}(T)$ is obtained
straightforwardly as 
\begin{equation}
Z_{\mathcal{P-F}}(N;T)=\sum_{\Sigma _{i}k_{i}=N}\frac{\tprod%
\limits_{i=1}^{N}(1-q^{i})}{\prod\limits_{i=1}^{n}\prod%
\limits_{r=1}^{k_{i}^{{}}}(1-q^{r})}.  \tag{3.10}
\end{equation}%
Consider this result for a special case: $n=2$. It is convenient to evaluate
the ratio first before calculating the sum. Thus, we obtain:%
\begin{equation}
\frac{\tprod\limits_{i=1}^{N}(1-q^{i})}{\prod\limits_{i=1}^{2}\prod%
\limits_{r=1}^{k_{i}^{{}}}(1-q^{r})}=\frac{(1-q)\cdot \cdot \cdot (1-q^{N})}{%
(1-q)\cdot \cdot \cdot (1-q^{k})(1-q)\cdot \cdot \cdot (1-q^{N-k})}\equiv 
\mathcal{F}(N,k\mid q).  \tag{3.11}
\end{equation}%
where the Poincare$^{\prime }$ polynomial $\mathcal{F}(N,k\mid q)$ for the
Grassmanian of the complex vector space \textbf{C}$^{N}$ of dimension $N$ \
was obtained in the previous section. Indeed (3.11) can be trivially brought
into the same form as given in our equation (2.9) using the relation $m+k=N$%
. To bring (2.9) in correspondence with equation (4.1) of Polychronakos
[23], we use the second equality (2.9) \ in which we make a substitution: $%
m=N-k.$ After this replacement, (3.10) acquires the following form%
\begin{equation}
Z_{\mathcal{P-F}}^{f}(N;T)=\sum\limits_{k=0}^{N}\prod\limits_{i=0}^{k}\frac{%
1-q^{N-i+1}}{1-q^{i}}  \tag{3.12}
\end{equation}%
coinciding with equation (4.1) by Polychronakos. This equation corresponds
to the ferromagnetic version of the P-F spin chain model. To obtain the
antiferromagnetic version of the model requires us only to replace $q$ by $%
q^{-1}$ in (3.12) and to multiply the whole r.h.s. by some known power of $q$%
. Since this factor will not affect thermodynamics, following Frahm [24], we
shall ignore it. As result, we obtain 
\begin{equation}
Z_{\mathcal{P-F}}^{af}(N;T)=\sum\limits_{k=0}^{N}q^{\left( N/2-k\right)
^{2}}\prod\limits_{i=0}^{k}\frac{1-q^{N-i+1}}{1-q^{i}},  \tag{3.13}
\end{equation}%
in accord with Frahm's equation (21). \ This result is analyzed further in
the next section.

\section{Connections with WZNW model and XXX \ s=1/2 \ Heisenberg
antiferromagnetic \ spin chain}

\subsection{General remarks}

\bigskip To establish these connections we follow work by Hikami [27]. For
this purpose, we introduce the notation%
\begin{equation}
\left( q\right) _{n}=\prod\limits_{i=1}^{n}(1-q^{i})  \tag{4.1}
\end{equation}%
allowing us to rewrite (3.13) in the equivalent form%
\begin{equation}
Z_{\mathcal{P-F}}^{af}(N;T)=\sum\limits_{k=0}^{N}q^{\left( N/2-k\right)
^{2}}\prod\limits_{i=0}^{k}\frac{1-q^{N-i+1}}{1-q^{i}}=\sum%
\limits_{k=0}^{N}q^{\left( N/2-k\right) ^{2}}\frac{\left( q\right) _{N}}{%
\left( q\right) _{k}\left( q\right) _{N-k}}.  \tag{4.2}
\end{equation}%
Consider now the limiting case ($N\rightarrow \infty )$ of the obtained
expression. \ For this purpose we need to take into account that [\textbf{%
andrews1}]%
\begin{equation}
\lim_{N\rightarrow \infty }\left[ 
\begin{array}{c}
N \\ 
k%
\end{array}%
\right] _{q}=\frac{1}{\left( q\right) _{k}}.  \tag{4.3}
\end{equation}%
To use this asymptotic result in (4.2) it is convenient to consider
separately the cases of $N$ being even and odd. For instance, if $N$ is
even, we can write: $N=2m.$ In such a case we can introduce new summation
variables: $l=k-m$ and/or $l=m-k.$ Then, in the limit $N\rightarrow \infty $
(that is m$\rightarrow \infty )$ we obtain asymptotically%
\begin{equation}
Z_{\mathcal{P-F}}^{af}(\infty ;T)=\frac{1}{\left( q\right) _{\infty }}%
\sum\limits_{i=-\infty }^{\infty }q^{i^{2}}.  \tag{4.4a}
\end{equation}%
in accord with [27]. Analogously, if $N=2m+1$, we obtain instead 
\begin{equation}
Z_{\mathcal{P-F}}^{af}(\infty ;T)=\frac{1}{\left( q\right) _{\infty }}%
\sum\limits_{i=-\infty }^{\infty }q^{\left( i+\frac{1}{2}\right) ^{2}}. 
\tag{4.4b}
\end{equation}%
According to Melzer [28] and Kedem, McCoy and Melzer [29], the obtained
partition functions coincide with the Virasoro characters for SU$_{1}$(2)
WZNW \ model describing the conformal limit of the XXX (s=1/2)
antiferromagnetic spin chain [30]. Even though equations (4.4a) and (4.4b)
provide the final result, they do not reveal their physical content. This
task was accomplished in part in the same papers where connection with the
excitation spectrum of the XXX antiferromagnetic chain \ was made. To avoid
repetitions, \ below we arrive at these results using different arguments.
By doing so many new and unexpected connections \ with other \ stochastic
models will be uncovered.

\subsection{Method of generating functions and q-deformed harmonic oscillator%
}

We begin with definitions. In view of (2.9),(3.12) and (4.2), we would like
to introduce the Galois number $G_{N}$ via 
\begin{equation}
G_{N}=\sum\limits_{k=0}^{N}\left[ 
\begin{array}{c}
N \\ 
k%
\end{array}%
\right] _{q}.  \tag{4.5}
\end{equation}%
This number can be calculated recursively as it was shown by Goldman and
Rota [31] with the result%
\begin{equation}
G_{N+1}=2G_{N}+\left( q^{N}-1\right) G_{N-1}.  \tag{4.6}
\end{equation}%
Alternative proof is given by Kac and Cheung [32]. To calculate $G_{N}$ we
have to take into account that $G_{0}=1$ and $G_{1}=2.$ These results can be
used as a reference when one attempts to calculate the related Rogers-Szego\
(R-S) polynomial $H_{N}(t)$ defined as [33]%
\begin{equation}
H_{N}(t;q):=\sum\limits_{k=0}^{N}\left[ 
\begin{array}{c}
N \\ 
k%
\end{array}%
\right] _{q}t^{k}  \tag{4.7}
\end{equation}%
so that $H_{N}(1)=G_{N}\footnote{%
For brevity, unless needed explicitly, we shall suppress the argument $q$ in 
$H_{N}(t;q).$}.$ Using [32] once again, we find that $H_{N}(t)$ obeys the
following recursion relation%
\begin{equation}
H_{N+1}(t)=(1+t)H_{N}(t)+\left( q^{N}-1\right) tH_{N-1}(t)  \tag{4.8}
\end{equation}%
which for $t=1$ coincides with (4.6) as required. The above recursion
relation is supplemented with initial conditions. These are : $H_{0}=1$ and $%
H_{1}=1+t$.

At this point we would like to remind our readers that for $t=1$ according
to (3.12) we obtain: $Z_{\mathcal{P-F}}^{f}(N;T)=G_{N}$. Hence, by
calculating $H_{N}(t)$ we shall obtain the partition function for the P-F
chain.

To proceed with such calculations, we follow [34]. In particular, consider
first an auxiliary recursion relation for the Hermite polynomials:%
\begin{equation}
H_{n+1}(x)=2xH_{n}(x)-2nH_{n-1}(x)  \tag{4.9a}
\end{equation}%
supplemented by the differential relation 
\begin{equation}
\frac{d}{dx}H_{n}(x)=2nH_{n-1}(x)  \tag{4.9b}
\end{equation}%
which, in view of (4.9a), can be conveniently rewritten as 
\begin{equation}
H_{n+1}(x)=(2x-\frac{d}{dx})H_{n}(x).  \tag{4.9c}
\end{equation}%
This observation prompts us to introduce the raising operator $R=2x-\dfrac{d%
}{dx}$ so that we obtain: 
\begin{equation}
R^{n}H_{0}(x)=H_{n}(x).  \tag{4.10}
\end{equation}%
The lowering operator can be now easily obtained again using (4.9). We get 
\begin{equation}
\frac{1}{2}\frac{d}{dx}H_{n}(x)\equiv LH_{n}(x)=nH_{n-1}(x)  \tag{4.11}
\end{equation}%
so that $[L,R]=1$ as required. \ Based on this,the number operator $N$ can
be obtained as $N=RL$ so that $NH_{n}(x)=nH_{n}(x)$ or, explicitly, using
provided definitions, we obtain 
\begin{equation}
(\frac{d^{2}}{dx^{2}}-2x\frac{d}{dx}+2n)H_{n}(x)=0.  \tag{4.12}
\end{equation}%
Evidently, we can write: $R\mid n>=\mid n+1>,$ $L\mid n>=\mid n-1>$ and , $%
<m\mid n>=n!\delta _{mn}$ as usual.

We would like now to transfer all these results to our main object of
interest- the recursion relation (4.8). To this purpose, we introduce the
difference operator $\Delta $ via 
\begin{equation}
\Delta H_{N}(t):=H_{N}(t)-H_{N}(qt).  \tag{4.13}
\end{equation}%
Using definition (4.7) we obtain now 
\begin{equation}
\Delta H_{N}(t)=(1-q^{N})tH_{N-1}(t)  \tag{4.14}
\end{equation}%
where we took into account that 
\begin{equation}
\left[ 
\begin{array}{c}
N \\ 
k%
\end{array}%
\right] _{q}=\left[ 
\begin{array}{c}
N \\ 
N-k%
\end{array}%
\right] _{q}.  \tag{4.16}
\end{equation}%
Using this result in (4.8) we obtain at once 
\begin{equation}
H_{N+1}(t)=[(1+t)-\Delta ]H_{N}(t).  \tag{4.17}
\end{equation}%
This, again, can be looked upon as a definition of a raising operator so
that we can formally rewrite (4.17) as 
\begin{equation}
\mathcal{R}H_{N}(t)=H_{N+1}(t).  \tag{4.18}
\end{equation}%
The lowering operator can be defined now as 
\begin{equation}
\mathcal{L}:=\frac{1}{x}\Delta  \tag{4.19}
\end{equation}%
so that 
\begin{equation}
\mathcal{L}H_{N}(t)=(1-q^{N})H_{N-1}(t).  \tag{4.20}
\end{equation}%
The action of the number operator $\mathcal{N}=\mathcal{RL}$ is \ now
straightforward, i.e.%
\begin{equation}
\mathcal{N}H_{N}(t)=(1-q^{N})H_{N}(t).  \tag{4.21}
\end{equation}%
Following Kac and Cheung [32] we introduce the $q-$derivative via 
\begin{equation}
D_{q}f(x):=\frac{f(qx)-f(x)}{x(q-1)}.  \tag{4.22}
\end{equation}%
By combining this result with (4.13) we obtain, 
\begin{equation}
D_{q}f(x)=\frac{\Delta f(x)}{x(1-q)}.  \tag{4.23}
\end{equation}%
This allows us to rewrite the raising and lowering operators in terms of $q-$%
derivatives. Specifically, we obtain:%
\begin{equation}
\mathcal{\tilde{R}}\text{ :}\mathcal{=}(1+t)-(1-q)tD_{q}  \tag{4.24}
\end{equation}%
and 
\begin{equation}
\mathcal{\tilde{L}}:=D_{q}.  \tag{4.25}
\end{equation}%
While for the raising operator rewritten in such a way equation (4.18) still
holds, for the lowering operator $\mathcal{\tilde{L}}$ we now obtain:%
\begin{equation}
\mathcal{\tilde{L}}H_{N}(t)=\frac{1-q^{N}}{1-q}H_{N-1}(t)\equiv \lbrack
N]H_{N-1}(t).  \tag{4.26}
\end{equation}%
The number operator $\mathcal{N}_{q}$ is acting in this case as 
\begin{equation}
\mathcal{N}_{q}H_{N}(t)=[N]H_{N}(t).  \tag{4.27}
\end{equation}%
We would like to connect these results with those available in literature on 
$q-$deformed harmonic oscillator. Following Chaichan et al [35], we notice
that the undeformed oscillator algebra is given in terms of the following
commutation relations 
\begin{equation}
aa^{+}-a^{+}a=1  \tag{4.28a}
\end{equation}%
\begin{equation}
\lbrack N,a]=-a  \tag{4.28b}
\end{equation}%
and%
\begin{equation}
\lbrack N,a^{+}]=a^{+}.  \tag{4.28c}
\end{equation}%
In these relations it is not assumed a priori that $N=a^{+}a$ and,
therefore, this algebra is formally different from the traditionally used $%
[a,a^{+}]=1$ for the harmonic oscillator. This observation allows us to
introduce the central element $Z=N-a^{+}a$ which is zero for the standard
oscillator algebra. The deformed oscillator algebra can be obtained now
using equations (4.28) in which one should replace (4.28a) by
[floreanni\&vinet] 
\begin{equation}
aa^{+}-qa^{+}a=1.  \tag{4.28d}
\end{equation}%
Consider now the combination $K:=$ $\mathcal{LR}$-$q\mathcal{N}$ acting on $%
H_{N}$ using previously introduced definitions. A simple calculation
produces an operator identity: $\mathcal{LR}-q\mathcal{N=}1$ so that we can
formally make a provisional identification : $\mathcal{L\rightarrow }a$ and $%
\mathcal{R\rightarrow }a^{+}.$ To proceed, we need to demonstrate that with
such an identification equations (4.28 b,c) hold as well. For this to
happen, we should \ properly normalize our wave function in accord with
known procedure for the harmonic oscillator where we have to use $\mid n>=%
\dfrac{1}{\sqrt{n!}}\left( a^{+}\right) ^{n}\mid 0>$. In the present case,
we have to use $\mid N>=\dfrac{1}{\sqrt{[N]!}}\left( \mathcal{R}\right)
^{n}\mid 0>$ as the basis wavefunction while making an identification: $\mid
0>=$ $H_{0}(t).$ The eigenvalue equation (4.27), when written explicitly,
acquires the following form:%
\begin{equation}
\lbrack tD_{q}^{2}-\frac{1+t}{1-q}D_{q}+\frac{[N]}{1-q}]H_{N}(t)=0. 
\tag{4.29}
\end{equation}

\subsection{The limit $q\rightarrow 1^{\pm }$ and emergence of the
Stieltjes-Wigert polynomials}

\bigskip Obtained results need further refinements for the following
reasons. Although the recursion relations (4.8), (4.9) look similar, in the
limit $q\rightarrow 1^{\pm }$ (4.8) is not transformed into (4.9).
Accordingly, (4.29) is not converted into equation for the Hermite
polynomials known for harmonic oscillator. \ Fortunately, the situation can
be repaired in view of recent paper by Karabulut [37] who spotted and
corrected some error in the influential earlier paper by Macfarlane [39].
Following [38] we define the translation operator $T(s)$ as $T(s):=e^{s\frac{%
\partial }{\partial x}}$. Using this definition, the creation $a^{\dag }$
and annihilation $a$ operators are defined as follows%
\begin{equation}
a^{\dag }=\frac{1}{\sqrt{1-q}}[q^{x+\frac{1}{4}}-T^{\frac{-1}{2}}(s)]T^{%
\frac{-1}{2}}(s)  \tag{4.30a}
\end{equation}%
where $T^{\frac{1}{2}}(s)=e^{\frac{s}{2}\frac{\partial }{\partial x}}$ and,
accordingly, 
\begin{equation}
a=\frac{1}{\sqrt{1-q}}T^{\frac{1}{2}}(s)[q^{x+\frac{1}{4}}-T^{\frac{1}{2}%
}(s)].  \tag{4.30b}
\end{equation}%
Under such conditions, the inner product is defined in the standard way,
that is%
\begin{equation}
(f,g)=\int\limits_{-\infty }^{\infty }f^{\ast }(x)g(x)dx  \tag{4.31}
\end{equation}%
so that $(q^{x})^{\dag }=q^{x}$ and $(\partial /\partial x)^{\dag
}=-(\partial /\partial x)$ thus making the operator $a^{\dag }$ to be a
conjugate of $a$ in a usual way.The creation-annihilation operators just
defined satisfy commutation relation (4.28 d). At the same time, the
combination $a^{\dag }a\ $\ while acting on the wave functions $\Psi _{n%
\text{ }}$\ (to be defined below)\ \ produces equation similar to (4.27),
that is 
\begin{equation}
a^{\dag }a\Psi _{n\text{ }}=[n]\Psi _{n\text{ }}\equiv \lambda _{n}\Psi _{n%
\text{ }}.  \tag{4.32}
\end{equation}%
\ Furthermore, it can be shown that \ \ \ 
\begin{equation}
a\Psi _{n\text{ }}=\sqrt{\lambda _{n}}\Psi _{n-1\text{ }}\text{ \ \ and \ }%
a^{\dag }\Psi _{n\text{ }}=\sqrt{\lambda _{n+1}}\Psi _{n+1\text{ }} 
\tag{4.33}
\end{equation}%
\ in accord with previously obtained results. Next, we would like to obtain
the wave function $\Psi _{n\text{ }}$ explicitly. To this purpose we start
with the ground state $a\Psi _{0\text{ }}=0$ and use (4.30b) to get (for
s=1/2)\footnote{%
The rationale for choosing s=1/2 is explained in the same reference.} the
following result 
\begin{equation}
\Psi _{0\text{ }}(x+\frac{1}{2})=q^{\frac{1}{4}+x}\Psi _{0\text{ }}(x). 
\tag{4.34}
\end{equation}%
\ \ Let $w(x)$ be some yet unknown function. Then, it is appropriate to look
for solution of (4.34) in the form%
\begin{equation}
\Psi _{0\text{ }}(x)=const\text{ }\cdot w(x)q^{x^{2}},  \tag{4.35a}
\end{equation}%
provided that the function $w(x)$ is periodic: \ $w(x)=w(x+1/2).$ The
normalized ground state function acquires then the following look%
\begin{equation}
\Psi _{0\text{ }}(x)=\alpha _{w}w(x)q^{x^{2}},  \tag{4.35b}
\end{equation}%
where the constant $\alpha _{w}$ is given by 
\begin{equation}
\alpha _{w}=\left( \int\limits_{-\infty }^{\infty }dx\left\vert
q^{x^{2}}w(x)\right\vert ^{2}\right) ^{-\frac{1}{2}}.  \tag{4.35c}
\end{equation}%
Using this result,\ \ $\Psi _{n\text{ }}$ can be constructed in a standard
way through use of the raising operators. There is, however, a faster way to
obtain the desired result.\ \ To this purpose, in view of (4.35b), suppose
that $\Psi _{n\text{ }}(x)$ can be decomposed as follows 
\begin{equation}
\Psi _{n\text{ }}(x)=\frac{\alpha _{w}w(x)}{\sqrt{(q,q)_{n}}}%
\sum\limits_{k=0}^{\infty }C_{k}^{n}(q)(-1)^{k}q^{\left( n-k\right)
/2}q^{(x-k)^{2}}  \tag{4.36}
\end{equation}%
\ \ \ where\ $(q,q)_{n}=(1-q)(1-q^{2})\cdot \cdot \cdot (1-q^{n})$ and $%
C_{k}^{n}(q)$ is to be determined as follows. By applying the operator \ $%
a^{\dag }/$\ $\sqrt{\lambda _{n+1}}$ to (4.36)\ and taking into account that
\ $T^{-\frac{1}{2}}$\ $w(x)=w(x-1/2)=w(x)$ (in \ view of the periodicity of $%
w(x)$) we end up with the recursion relation for $C_{k}^{n}(q)$:%
\begin{equation}
C_{k}^{n+1}(q)=q^{k}C_{k}^{n}(q)+C_{k-1}^{n}(q).  \tag{4.37}
\end{equation}%
\ This relation should be compared with that given by (2.8). Andrews [33],
page 35, demonstrated that (2.8) and (4.37) are equivalent. Hence, we obtain:%
\begin{equation}
C_{k}^{n}(q)=\left[ 
\begin{array}{c}
n \\ 
k%
\end{array}%
\right] _{q}.  \tag{4.38}
\end{equation}%
\ This implies that, indeed, up to a constant, the obtained wavefunction
should be related to the Rogers-Szego polynomial. This relation is
nontrivial however. We would like to discuss it in some detail now.

Following [37,39], let $q=e^{-c^{2}},$ where $c$ is some nonegative number.
Introduce the distributed Gaussian polynomials via%
\begin{equation}
\Phi _{n}(x)=\sum\limits_{k=0}^{\infty
}C_{k}^{n}(q)(-1)^{k}q^{-k/2}q^{(x-k)^{2}}.  \tag{4.39}
\end{equation}%
These polynomials satisfy the following orthogonality relation:%
\begin{equation}
\int\limits_{-\infty }^{\infty }\Phi _{n}(x)\Phi _{m}(x)dx=\left\Vert \Phi
_{n}(x)\right\Vert ^{2}\delta _{mn}  \tag{4.40}
\end{equation}%
with\footnote{%
Notice that $\alpha =\left( \int\limits_{-\infty }^{\infty
}dxq^{2x^{2}}\right) ^{-\frac{1}{2}}=\left( \frac{\pi }{2c^{2}}\right) ^{-%
\frac{1}{4}}$} 
\begin{equation}
\left\Vert \Phi _{n}(x)\right\Vert =\left( \frac{\pi }{2c^{2}}\right) ^{%
\frac{1}{4}}q^{-\frac{n}{2}}\sqrt{(q,q)_{n}}.  \tag{4.41}
\end{equation}%
This result calls for change in normalization of $\Phi _{n}(x),$ i.e$.,\phi
_{n}(x)=\frac{\Phi _{n}(x)}{\left\Vert \Phi _{n}(x)\right\Vert }.$ Under
such conditions $\phi _{n}(x)$ coincides with $\Psi _{n\text{ }}(x),$
provided that $w(x)=1.$ Introduce new variable: $u=q^{-2x},$ and consider a
shift: $\Phi _{n}(x)\rightarrow \Phi _{n}(x-s).$ Using (4.39), we can write 
\begin{equation}
\Phi _{n}(x-s)=u^{s}\exp \{-\left( \ln u\right) ^{2}/(-4\ln q)\}P_{n}(u;s) 
\tag{4.42a}
\end{equation}%
where 
\begin{equation}
P_{n}(u;s)=\sum\limits_{k=0}^{\infty
}C_{k}^{n}(q)(-1)^{k}q^{-k/2}q^{(s+k)^{2}}u^{k}.  \tag{4.42b}
\end{equation}%
The orthogonality relation (4.40) is converted then into 
\begin{equation}
\int\limits_{0}^{\infty }duu^{2s-1}\exp \{-\left( \ln u\right) ^{2}/(-4\ln
q)\}P_{n}(u;s)P_{m}(u;s)=\delta _{mn}.  \tag{4.43}
\end{equation}%
In view of (4.43) consider now a special case: $s=1/2$. Then, the weight
function is known as $\QTR{sl}{lognormal}$ distribution and polynomials $%
P_{n}(u;1/2)$ \ (up to a constant ) are known as Stieltjes-Wigert (S-W)
polynomials. Their physical relevance will be discussed below in Subsection
4.6. In the meantime, we introduce the Fourier transform of $f(x)$ in the
usual way as 
\begin{equation}
\int\limits_{-\infty }^{\infty }dx\exp (2\pi i\theta x)f(x)=f(\theta ) 
\tag{4.44}
\end{equation}%
Then, the Parseval relation implies: 
\begin{equation}
\int\limits_{-\infty }^{\infty }\Phi _{n}(x)\Phi
_{m}(x)dx=\int\limits_{-\infty }^{\infty }\Phi _{n}(\theta )\Phi _{m}(\theta
)dx=\left\Vert \Phi _{n}(x)\right\Vert ^{2}\delta _{mn},  \tag{4.45a}
\end{equation}%
causing%
\begin{equation}
\Phi _{n}(\theta )=\left( \frac{\pi }{c^{2}}\right) ^{\frac{1}{4}}\exp
(-\left( \pi /c\right) \theta ^{2})\sum\limits_{k=0}^{\infty
}C_{k}^{n}(q)(-q^{-\frac{1}{2}}e^{2\pi i\theta })^{k}.  \tag{4.45b}
\end{equation}%
Comparison between these results and (4.7) produces%
\begin{equation}
\int\limits_{-\infty }^{\infty }H_{n}(-q^{-\frac{1}{2}}e^{-2\pi i\theta
};q)H_{m}(-q^{-\frac{1}{2}}e^{-2\pi i\theta };q)\exp (-2\left( \pi /c\right)
\theta ^{2})=\left( \frac{c}{2\pi }\right) ^{\frac{1}{2}}q^{-n}(q,q)_{n}%
\delta _{mn}  \tag{4.46a}
\end{equation}%
which can be alternatively rewritten as 
\begin{equation}
\int\limits_{0}^{1}H_{n}(-q^{-\frac{1}{2}}e^{-2\pi i\theta })H_{m}(-q^{-%
\frac{1}{2}}e^{-2\pi i\theta })\vartheta _{3}(2\pi \theta ;q)d\theta
=q^{-n}(q,q)_{n}\delta _{mn}  \tag{4.46b}
\end{equation}%
with $\vartheta _{3}(\theta ,q)=\sum\limits_{n=-\infty }^{\infty
}q^{n^{2}/2}e^{in\theta }$ . That is $\vartheta _{3\text{ \ }}$is one of the
Jacobi's theta functions. In order to use the obtained results, it is useful
to compare them against those, known in literature already, e.g. see [40].
Equation (4.46a) is in agreement with (5) of [40] if we make
identifications: $\kappa =\pi $ and $c=\sqrt{2}\kappa ,$ where $\kappa $ is
the parameter introduced in this reference. With help of such an
identification we can proceed with comparison. For this purpose, following
40] we introduce yet another generating function 
\begin{equation}
S_{n}(t;q):=\sum\limits_{k=0}^{n}\left[ 
\begin{array}{c}
n \\ 
k%
\end{array}%
\right] _{q}q^{k^{2}}t^{k}  \tag{4.47}
\end{equation}%
so that the S-W polynomials can be written now as [41], page 197, 
\begin{equation}
\mathcal{\tilde{S}}_{n}(t;q)=(-1)^{n}q^{\frac{n}{2}}(\sqrt{(q,q)_{n}}%
)^{-1}P_{n}(t;\frac{1}{2})\equiv (-1)^{n}q^{\frac{2n+1}{{}}}(\sqrt{(q,q)_{n}}%
)^{-1}S_{n}(-q^{\frac{1}{2}}t;q),  \tag{4.48}
\end{equation}%
provided that $0<q<1.$ Comparison between generating functions (4.7) and
(4.47) allows us to write as well%
\begin{equation}
S_{n}(t;q^{-1})=H_{n}(tq^{-n};q),\text{ or, equivalently, }%
H_{n}(t;q^{-1})=S_{n}(q^{-n}t;q)  \tag{4.49}
\end{equation}%
Using this result we can rewrite the recursion relation (4.8) for $%
H_{n}(t;q) $ in terms of the recursion relation for $S_{n}(t;q)$ if needed
and then to repeat all the arguments with creation and annihilation
operators, etc. For the sake of space, we leave this option as an exercise
for our readers. Instead, to finish our discussion we would like to show how
the obtained polynomials reduce to the usual Hermite polynomials in the
limit $q\rightarrow 1^{-}.$ For this purpose we would like to demonstrate
that the recursion relation (4.8) is actually the recursion relation for the
continuous $q-$Hermite polynomials [42,43]. This means that we have to
demonstrate that under some conditions (to be specified) the recursion (4.8)
is equivalent to 
\begin{equation}
2xH_{n}(x\mid q)=H_{n+1}(x\mid q)+(1-q^{n})H_{n-1}(x\mid q).  \tag{4.50}
\end{equation}%
known for q-Hermite polynomials. To demonstrate the equivalence we assume
that $x=\cos \theta $ in (4.50) and then, let $z=e^{i\theta }.$ Furthermore,
we assume that 
\begin{equation}
H_{n}(x\mid q)=z^{n}H_{n}(z^{-2};q),  \tag{4.51}
\end{equation}%
allowing us to obtain,%
\begin{equation}
(z+z^{-1})z^{n}H_{n}=z^{n+1}H_{n+1}+(1-q^{n})z^{n-1}H_{n-1}.  \tag{4.52}
\end{equation}%
Finally, we set z$^{-1}=\sqrt{t}$ which brings us back to (4.8). This time,
however, we can use results known in literature for $q-$Hermite polynomials
[41-43] in order to obtain at once 
\begin{equation}
\lim_{q\rightarrow 1^{-}}\frac{H_{n}(x\sqrt{\frac{1-q}{2}}\mid q)}{\sqrt{%
\frac{1-q}{2}}}=H_{n}(x),  \tag{4.53}
\end{equation}%
where $H_{n}(x)$ are the standard Hermitian polynomials. In view of (4.48),
(4.49), not surprisingly, the S-W polynomials are also reducible to $%
H_{n}(x) $. Details can be found in the same references.

\subsection{ASEP, q-deformed harmonic oscillator and spin chains}

In this subsection we would like to connect the results obtained thus far
with the XXX and XXZ spin chains. Although a connection \ with XXX \ spin
chain was established already at the beginning of this section, we would
like to arrive at the same conclusions using alternative (physically
inspired) arguments and methods. To understand the logic of our arguments we
encourage our readers to read Appendix\ A at this point. In it we provide a
self contained summary of results related to the asymmetric simple exclusion
process (ASEP), especially emphasizing its connection with static and
dynamic properties of XXX\ and XXZ spin chains.

ASEP was discussed in high energy physics literature, e.g. see [44], in
connection with random matrix ensembles. To avoid repeats, we would like to
use the results of Appendix A \ in order\ to consider the steady-state
regime only. To be in \ accord with literature on ASEP, we would like to
complicate matters by \ imposing some nontrivial boundary conditions.

In the steady -state regime equation (A.12) of Appendix A acquires the form
: $SC=\Lambda $. Explicitly, 
\begin{equation}
SC=p_{L\text{ }}ED-p_{R}DE.  \tag{4.54}
\end{equation}%
In the steady-state regime, the operator $S$ becomes an arbitrary c-number
[45]. In view of this, following Sasamoto [46] we rewrite (4.54) as 
\begin{equation}
p_{R}DE-p_{L}ED=\zeta \left( D+E\right) .  \tag{4.55}
\end{equation}%
Such operator equation should be supplemented by the boundary conditions
which are chosen to be as 
\begin{equation}
\alpha <W\mid E=\zeta <W\mid \text{ and }\beta D\mid V>=\zeta \mid V>. 
\tag{4.56}
\end{equation}%
The normalized steady-state probability for some configuration $\mathcal{C}$
can be written \ now as 
\begin{equation}
P(\mathcal{C})=\frac{<W\mid X_{1}X_{2}\cdot \cdot \cdot X_{N}\mid V>}{<W\mid
C^{N}\mid V>}  \tag{4.57}
\end{equation}%
with \ the operator $X_{i}$ being either $D$ or $E$ depending on wether the $%
i-$th site is occupied or empty. To calculate $P(\mathcal{C})$ we need to
determine $\zeta $ while assuming parameters $\alpha $ and $\beta $ to be
assigned. We demonstrate in Appendix A that it is possible to equate $\zeta $
to one \ so that, in agreement with [47], we obtain the following
representation of $D$ and $E$ operators:%
\begin{equation*}
D=\frac{1}{1-q}+\frac{1}{\sqrt{1-q}}a,\ \text{\ }E=\frac{1}{1-q}+\frac{1}{%
\sqrt{1-q}}a^{+}
\end{equation*}%
converting equation (4.54) into (4.28d). In view of this mapping into $q-$%
deformed oscillator algebra, we can expand both vectors $\mid V>$ and $%
<W\mid $ into a Fourier series, e.g. $\mid V>=\sum\nolimits_{m}\Omega
_{m}(V)\mid m>$ where, using (4.33), we put $\mid m>=\Psi _{m}.$ By
combining equations (4.33) and (4.56) and results of Appendix A we obtain
the following recurrence equation for $\Omega _{n}:$ 
\begin{equation}
\Omega _{n}(V)(\frac{1-q}{\beta }-1)=\Omega _{n+1}(V)\sqrt{1-q^{n+1}}. 
\tag{4.58}
\end{equation}%
Following [47] we assume that $<0\mid V>=1.$ Then, the above \ recurrence
produces 
\begin{equation*}
\Omega _{n}(V)=\frac{v^{n}}{\sqrt{(q,q)_{n}}},
\end{equation*}%
with \ parameter $v=\frac{1-q}{\beta }-1.$ Analogously, we obtain: 
\begin{equation*}
\Omega _{n}(W)=\frac{w^{n}}{\sqrt{(q,q)_{n}}},
\end{equation*}%
with $w=\frac{1-q}{\alpha }-1.$ Obtained results exhibit apparently singular
behavior for $q\rightarrow 1^{-}.$ These singularities are only apparent
since they cancel out when one computes quantities of\ physical interest
discussed in both [48] and [49]. As results of Appendix A indicate, such a
crossover is also nontrivial physically since it involves careful treatment
of the transition from XXZ to XXX antiferromagnetic spin chains. Hence, the
results obtained thus far enable us to connect the partition function (4.2)
(or (4.7)) with either XXX or XXZ spin chains but are not yet sufficient for
making an unambiguous choice between these two models. This task is
accomplished in the rest of this section.

\subsection{Crossover between the XXZ and XXX spin chains: connections with
the KPZ and EW equations and the lattice Liouville model}

\bigskip

Following Derrida and Malick [49], we notice that ASEP is the lattice
version of the famous Kradar-Parisi-Zhang \ (KPZ) equation [50]. The
transition $q\rightarrow 1^{-}$ corresponds to transition (in the sense of
renormalization group analysis) from the regime of ballistic deposition
whose growth is described by the KPZ \ equation to another regime described
by the Edwards-Wilkinson (EW) \ equation. In the context of ASEP (that is
microscopically) such a transition is discussed in detail in [51].
Alternative treatment is given in [49]. \ \ The task of obtaining the KPZ or
EW equations from those describing the ASEP is nontrivial and was
accomplished only very recently [oliveiraetal, Lazarides]. It is essential
for us that in doing so the rules of constructing the restricted solid-on
-solid (RSOS) models were invoked. From the work by Huse [54] it is known
that such models can be found in four thermodynamic regimes.The crossover
from the regime III to regime IV is described by the\ critical exponents of
Friedan, Qui and Shenker unitary CFT series [55]. The crossover from regime
III to regime IV happens to be relevant to crossover from the KPZ to EW
regime as we would like to explain now.

As results of \ Appendix\ A indicate, the truly asymmetric \ simple
exclusion process is associated with the XXZ model at the microscopic level
and with the KPZ equation/model at the macroscopic level. Accordingly, the
symmetric exclusion process is associated with the XXX model at the
microscopic level and with the EW equation/model \ at the macroscopic level.
At the level of Bethe ansatz for open XXZ chain with boundaries full details
of the crossover from the KPZ to EW regime were exhaustively worked out only
recently [56]. For the purposes of this work it is important to notice that
for certain values of parameters the Hamiltonian of \ open XXZ spin chain
model \footnote{%
That is equation (1.3) of [56].} \ with boundaries can be brought \ to the
following canonical form%
\begin{equation}
H_{XXZ}=\frac{1}{2}[\sum\limits_{j=1}^{N-1}(\sigma _{j}^{x}\sigma
_{j+1}^{x}+\sigma _{j}^{y}\sigma _{j+1}^{y}+\frac{1}{2}(q+q^{-1})\sigma
_{j}^{z}\sigma _{j+1}^{z})+\frac{1}{2}(q-q^{-1})(\sigma _{1}^{z}-\sigma
_{N}^{z})].  \tag{4.59}
\end{equation}%
In the case of ASEP we have $q=\sqrt{p_{R}/p_{L}}$ so that for physical
reasons parameter $q$ is not complex. However, mathematically, we can allow
for $q$ \ to be complex. In particular, following Pasquer and Saleur [57] we
can let $q=e^{i\gamma }$ with $\gamma =\dfrac{\pi }{\mu +1}.$ For such
values of $q$ use of finite scaling analysis applied to the spectrum of the
above defined Hamiltonian produces \ the central charge 
\begin{equation}
c=1-\frac{6}{\mu (\mu +1)}\text{ , }\mu =2,3,....  \tag{4.60}
\end{equation}%
of the unitary CFT \ series. Furthermore, if $e_{i}$ is the generator of the
Temperley-Lieb algebra\footnote{%
That is $e_{i}^{2}=e_{i},$ $e_{i}e_{i+1}e_{i}=q^{-1}e_{i}$ and $%
e_{i}e_{j}=e_{j}e_{i}$ for $\left\vert i-j\right\vert \geq 2.$}, then $%
H_{XXZ}$ can be rewritten as [58] 
\begin{equation}
H_{XXZ}=-\sum\limits_{j=1}^{N-1}[e_{i}-\frac{1}{4}(q+q^{-1})].  \tag{4.61}
\end{equation}%
This fact allows us to make immediate connections with quantum groups and
theory of knots and links. Below, in Section 5 we shall use different
arguments to arrive at similar conclusions. The results just described allow
us to connect the CFT and \ exactly integrable \ lattice models. If this is
the case, one can pose the following question: given the connection we just
described, can we write down explicitly the corresponding path integral
string-theoretic models reproducing results of exactly integrable lattice
models at and away from criticality? Before providing \ the answer in the
following subsection, we would like to conclude this subsection with a
partial answer. In particular, we would like to mention the work by Faddeev
and Tirkkonen [59] connecting the \textsl{lattice} Liouville model with the
spin 1/2 XXZ chain. Based on this result, it should be clear that in the
region \ $c\leq 1$ it is indeed possible by using combinatorial analysis
described above to make a link between the continuum and \ the discrete
Liouville theories\footnote{%
The matrix $c=1$ theories will be discussed separately below.}. It can be
made in such a way that, at least at crtiticality, the results of exactly
integrable 2 dimensional models are in agreement with those which are
obtainable field- theoretically. The domain $c>1$ is physically meaningless
because the models (\textsl{other} \textsl{than string-theoretic}) we
discussed in this section loose their physical meaning in this region. This
conclusion \ will be further reinforced in the next subsection.

\subsection{ASEP, vicious random walkers and string models}

\bigskip

We have discussed at length the role of vicious random walkers in derivation
of the Kontsevich-Witten (K-W) model in our previous work [60]. Forrester
[61] noticed that the random turns vicious walkers model is just a special
case of ASEP. Further details on connections between the ASEP, vicious
walkers, KPZ and random matrix theory can be found in the paper by Sasamoto
[62]. \ In the paper by Mukhi [63] it is emphasized that while the K-W model
is the matrix model representing $c<1$ bosonic string, \ the Penner matrix
model with imaginary coupling constant is representing $c=1$ \ Euclidean
string on the cylinder of \ (self-dual) radius $R=1\footnote{%
This was initially demonstrated by Distler and Vafa [64].}$ Furthermore,
Ghoshal and Vafa [65] have demonstrated that $c=1,R=1$ string is dual to the
topological string on a conifold singularity. We shall \ briefly discuss
this connection below. Before doing so, \ it is instructive to discuss the
crossover from $c=1$ to $c<1$ string models in terms of vicious walkers. To
do so we shall use some results from our work on K-W model and from the
paper by Forrester [61].

Thus, we would like to consider planar lattice where at the beginning we
place only one directed path P: from $(a,1)$ to $(b,N)$\footnote{%
Very much in the same way as discussed \ already in Section 2.}. The
information about this path can be encoded into multiset $Hor_{y}(P)$ of
y-coordinates of the horizontal steps of P. Let now%
\begin{equation}
w(P)=\prod\limits_{i=Hor_{y}(P)}x_{i}.  \tag{4.62}
\end{equation}%
Using these definitions, the extension of these results to an assembly of
directed random vicious walkers is given as a product: $W(\hat{P})\equiv
w(P_{1})\cdot \cdot \cdot w(P_{k}).$ Finally, the generating function for an
assembly of such walkers is given by%
\begin{equation}
h_{b-a}(x_{1},...,x_{N})=\sum\limits_{\hat{P}}W(\hat{P}),  \tag{4.63}
\end{equation}%
where $W(\hat{P})$ is made of monomials of the type $%
x_{1}^{m_{1}}x_{2}^{m_{2}}\cdot \cdot \cdot x_{N}^{m_{N}}$ provided that $%
m_{1}+\cdot \cdot \cdot +m_{N}=b-a.$ The following theorem [ , ] is of
central importance for calculation of such defined generating function.

Given integers $0<a_{1}<\cdot \cdot \cdot <a_{k\text{ \ }}$and $%
0<b_{1}<\cdot \cdot \cdot <b_{k}$ , let \textbf{M}$_{i,j}$ be the $k\times k$
matrix \textbf{M}$_{i,j}=h_{b_{j}-a_{i}}(x_{1},...,x_{N})$ then, 
\begin{equation}
\det \mathbf{M}=\sum\limits_{\hat{P}}W(\hat{P})  \tag{4.64}
\end{equation}%
where the sum is taken over all sequences ($P_{1},...,P_{k})\equiv \hat{P}$
of nonintersecting lattice paths $P_{i}:(a_{i},1)\rightarrow
(b_{i},N),i=1-k. $

Let now $a_{i}=i$ and $b_{j}=\lambda _{i}+j$ so that $1\leq i,j\leq k$ with $%
\lambda $ being a partition of $N$ with $k$ parts then, $\det \mathbf{M}%
=s_{\lambda }(x_{1},...,x_{N}),$ where $s_{\lambda }(\mathbf{x})$ is the
Schur polynomial. In our work [60 ] we demonstrated that in the limit $%
N\rightarrow \infty $ such defined Schur polynomial coincides with the
partition(generating) function for the Kontsevich model. \ Many additional
useful results related to Schur functions are discussed in our recent paper
[2].

To get results by Forrester requires\ us to apply some additional efforts.
These are worth discussing. Unlike the K-W case, this time, we need to
discuss \ the continuous random walks in the plane. Let $x$-coordinate
represent "space" while $y$-coordinate- "time". If initially ($t=0$) we had $%
k$-walkers in the positions $-L$ $<x_{1}<x_{2}<\cdot \cdot \cdot <x_{k\text{ 
}}<L,$ the same order should persist $\forall $ $t>0$. At each tick of the
clock each walker is moving either to the right or to the left with equal
probability $p$ (that is we are in the regime appropriate for the XXX spin
chain in the ASEP terminology). As before, let \textbf{x}$%
_{0}=(x_{1,0},...,x_{k,0})$ be the initial configuration of $k-$walkers and 
\textbf{x}$_{f}=(x_{1,f},...,x_{k,f})$ be \ the final configuration at time $%
t$. To calculate the total number of walks starting at $t=0$ at \textbf{x}$%
_{0}$ and ending at time $t$ at \textbf{x}$_{f}$ we need to know the
probability distribution $W_{k}(\mathbf{x}_{0}\rightarrow \mathbf{x}_{f};t)$
that the walkers proceed without bumping into each other. Should these
random walks be totally uncorrelated, we would obtain for the probability
distribution the standard Gaussian result:%
\begin{equation}
W_{k}^{0}(\mathbf{x}_{0}\rightarrow \mathbf{x}_{f};t)=\frac{\exp \{-(\mathbf{%
x}_{f}-\mathbf{x}_{0})^{2}/2Dt\}}{\left( 2\pi Dt\right) ^{k/2}}.  \tag{4.65}
\end{equation}%
In the present case the walks are restricted (correlated) so that the
probability should be modified. This modification can be found in the work
by Fisher and Huse [66]. These authors obtain%
\begin{equation}
W_{k}(\mathbf{x}_{0}\rightarrow \mathbf{x}_{f};t)=U_{k}(\mathbf{x}_{0},%
\mathbf{x}_{f};t)\frac{\exp \{-(\mathbf{x}_{f}^{2}+\mathbf{x}_{0}^{2})/2Dt\}%
}{\left( 2\pi Dt\right) ^{k/2}}  \tag{4.66}
\end{equation}%
with%
\begin{equation}
U_{k}(\mathbf{x}_{0},\mathbf{x}_{f};t)=\sum\limits_{g\in S_{k}}\varepsilon
(g)\exp [\frac{(\mathbf{x}_{f}\cdot g\mathbf{x}_{0})}{Dt}].  \tag{4.67}
\end{equation}%
In this expression $\varepsilon (g)=\pm 1,$ and the index $g$ runs over all
members of the symmetric group $S_{k}$. Mathematically, following Gaudin
[67], this problem can be looked upon as a problem of a random walk inside
the $k-$dimensional kaleidoscope (Weyl cone) usually complicated by
imposition of some boundary conditions at the walls of the cone. Connection
of such random walk problem with random matrices was discussed by Grabiner
[68] whose results were \ very recently improved and generalized by
Krattenthaller [69]. In the work by de Haro some applications of Grabiner's
results to high energy physics were considered [70]. Here we would like to
approach the same class of problems based on the results obtained in this
paper. In particular, some calculations made in [66] indicate that for $%
L\rightarrow \infty $ with accuracy up to $O$ ($L^{2}/Dt)$ it is possible to
rewrite $U_{k}(\mathbf{x}_{0},\mathbf{x}_{f};t)$ as follows:%
\begin{equation}
U_{k}(\mathbf{x}_{0},\mathbf{x}_{f};t)\simeq const\Delta (\mathbf{x}%
_{f})\Delta (\mathbf{x}_{0})/\left( Dt\right) ^{n_{k}}+O(L^{2}/Dt) 
\tag{4.68}
\end{equation}%
with $n_{k}=\left( 1/2\right) k(k-1)$ and $const=1/1!2!\cdot \cdot \cdot
(k-1)!$ and $\Delta (\mathbf{x})$ being the Vandermonde determinant, i.e.%
\begin{equation}
\Delta (\mathbf{x})=\prod\limits_{i<j}(x_{i}-x_{j}).  \tag{4.69}
\end{equation}%
Next, from standard texts in probability theory it is known that \textsl{%
non-normalized} expression, say, for $W_{k}^{0}(\mathbf{x}_{0}\rightarrow 
\mathbf{x}_{f};t)$ in the limit of long times provides the number of random
walks of $n$ steps (since $n\rightleftarrows t)$ from point $\mathbf{x}_{0}$
to point \textbf{x}$_{f}$. Hence, the same must be true for $W_{k}(\mathbf{x}%
_{0}\rightarrow \mathbf{x}_{f};t)$ and, therefore, $W_{k}(\mathbf{x}%
_{0}\rightarrow \mathbf{x}_{f};t)\approx \det \mathbf{M.}$ Consider such
walks for which $\mathbf{x}(t=0)\equiv \mathbf{x}_{0}=\mathbf{x}%
(t=t_{f})\equiv \mathbf{x}_{f}.$ Then, using (4.66) and (4.68) we obtain the
probability distribution for the Gaussian unitary ensemble [71], i.e.%
\begin{equation}
W_{k}(\mathbf{x}_{0}=\mathbf{x}_{f};t)=const\text{ }\Delta ^{2}(\mathbf{x}%
)\exp (-\mathbf{x}^{2}).  \tag{4.70}
\end{equation}%
Some additional manipulations (described in our work [60]) using this
ensemble lead directly to the K-W matrix model. Forrester [61] had
considered a related quantity: the probability that all $k$ vicious walkers
will survive at time $t_{f}.$ To obtain this probability requires
integration of $W_{k}(\mathbf{x}_{0}\rightarrow \mathbf{x}_{f};t)$ over the
simplex $\Delta $ defined by $-\ L<x_{1}<x_{2}<\cdot \cdot \cdot <x_{k\text{ 
}}<L\footnote{%
Such type of integration is described in detail in our papers, Parts I and
II, from which it follows that in the limit $L\rightarrow \infty $ such a
simplex integration can be replaced by the usual integration, i.e $%
\int\nolimits_{\Delta }\prod\limits_{i}^{k}dx_{i}\cdot \cdot \cdot
\rightleftarrows \frac{1}{k!}\int\limits_{-\infty }^{\infty
}\prod\limits_{i}^{k}dx_{i}\cdot \cdot \cdot $ in accord with Forrester.}.$
Without loss of generality, it is permissible to \ use $W_{k}(\mathbf{x}_{0}=%
\mathbf{x}_{f};t)$ instead of $W_{k}(\mathbf{x}_{0}\rightarrow \mathbf{x}%
_{f};t)$ in calculating such a probability. Then, the obtained result
coincides (up to a constant) with the partition function of topological
gravity $\mathcal{Z}(g),$ equation (3.1) of [72,73]\footnote{%
Since the hermitian matrix model given by (3.1) is just a partition function
for the Gaussian unitary ensemble [71].}). Furthermore, such defined
partition function can be employed to reproduce back the Hermite polynomial $%
H_{k}(x)$ defined by (4.53) which has an interpretation as the
wavefunction(amplitude) of the FZZT $D-$brane [72,73]. Specifically, we have%
\begin{eqnarray}
&<&\det (x-M)>=\left( \frac{g}{4}\right) ^{\frac{n}{2}}H_{k}(x\sqrt{\frac{1}{%
g}})  \notag \\
&=&\frac{1}{\mathcal{Z}(g)}\int dM\det (x-M)e^{-\frac{1}{g}trM^{2}}. 
\TCItag{4.71}
\end{eqnarray}%
This expression is a special case of Heine's formula representing monic
orthogonal polynomials through random matrices. In the above formula $k$ is
related to the size of Hermitian matrix $M$ and $g$ is the coupling constant.

Following Forrester [61], the result (4.66) can be treated more accurately
(albeit a bit speculatively) if, in addition to the parameter $D$ we
introduce \ another parameter $a$ - the spacing between random walkers at
time $t=0$. Furthermore, if the time direction is treated as space direction
(as it is commonly done for 1d quantum systems in connection with 2d
classical systems), then yet another parameter $\tau (k,t)$ should be
introduced which effectively renormalizes $D$. This eventually causes us to
replace $\mathcal{Z}(g)$ by the following integral (up to a constant)%
\begin{equation}
\mathcal{\hat{Z}}(g)=\prod\limits_{i=1}^{k}\int\limits_{-\infty }^{\infty
}dx_{i}\exp (-\frac{1}{2g}\ln
^{2}x_{i})\prod\limits_{i<j}(x_{i}-x_{j})^{2}\equiv \int dMe^{-\frac{1}{2g}%
tr(\ln M)^{2}}  \tag{4.72}
\end{equation}%
Tierz demonstrated [74] that $\mathcal{\hat{Z}}(g)$ (up to a constant) is
partition function of the Chern-Simons (C-S) field theory with gauge group $%
U(k)$ living on the 3-sphere $S^{3}.$ Okuyama [73] used (4.71) in order to
get analogous result for a D-brane amplitude in C-S model. Using Heine's
formula, he obtained the Stieltjes-Wiegert \ (S-W) polynomial, our equation
(4.47), which can be expressed via the Rogers-Szego polynomial according to
(4.49) and, hence, via the $q-$Hermite polynomial in view of the relation
(4.51). Since in the limit $q\rightarrow 1^{-}$the $q-$Hermite polynomial is
reducible to the usual Hermite polynomial according to (4.53), there should
be analogous procedure in going from the partition function $\mathcal{\hat{Z}%
}(g)$ to $\mathcal{Z}(g).$ Such a procedure can be developed, in principle,
by reversing arguments of Forrester. However, these arguments are much less
rigorous and physically transparent than those used in previous subsection
where we discussed the crossover from XXZ to XXX model. \ In view of the
results presented in the following section, we leave the problem of
crossover between the matrix ensembles outside the scope of this paper. To
avoid duplications, we refer our readers to the paper by Okuyama [73] where
details are provided relating our results to the topological A and B -branes.

\section{Gaudin model as linkage between the WZNW model and K-Z equations.
Recovery of \ the Veneziano-like amplitudes}

\ 

\subsection{General remarks}

\bigskip

We would like to remind to our readers that all results obtained thus far
can be traced back to our equation (4.7) defining the Rogers-Szego
polynomial which \ physically was interpreted as partition function for the
ferromagnetic P-F spin chain\footnote{%
The antiferromagnetic version of P-F spin chain is easily obtainable from
this ferromagnetic version as discussed in Section 3.}. In previous sections
numerous attempts\ were made to connect this partition function to various
models, even though already in Section 4.1 we came to the conclusion that in
the limit of infinitely long chains the antiferromagnetic version of P-F
spin chain can be \ replaced by the spin 1/2 antiferromagnetic XXX chain. If
this is so, then fom literature it is known that behaviour of such spin
chain is described by the $SU_{1}(2)$ WZNW model [30]. Hence, at the
physical level of rigor\ the problem of connecting Veneziano amplitudes to
physical model can be considered as solved. In this section we argue that at
the mathematical level of rigor this is not quite so yet. This conclusion
concerns not only problems dicussed in this paper but, in general, the
connection between the WZNW models, spin chains and K-Z equations. It is
true that K-Z equations and WZNW model \ are inseparable from each other
[30] but the extent\ to which spin chains can be directly linked to both \
the WZNW models and K-Z equations remains to be investigated. We would like
to do so in this section. For the sake of space, we shall discuss only the
most essential facts leaving (with few exceptions) many details and proofs
to literature.

Following Varchenko [8], we notice that the link between the K-Z equations
and WZNW models can be made\ only with help of\ the Gaudin model, while the
connection with spin chains can be made only by using the \textsl{quantum}
version of the K-Z equation. Such quantized version of \ the K-Z equation 
\textsl{is not} immediately connected with the standard WZNW model \ as
discussed in many places [8,75]. In this section, we would like to discuss
in some detail the Gaudin model and its relation to the P-F spin chain and,
hence to the Veneziano model formulated in Part II. We begin with summary of
facts \ related to this model.

\subsection{Gaudin magnets, K-Z equation and P-F spin chain}

\bigskip

Although theory of the Gaudin magnets plays an important role in topics such
as Langlands correspondence, Hitchin systems, etc.[76-78] in this work we do
not discuss these topics. Instead, we would like to focus only on issues of
\ immediate relevance to this paper.\ Gaudin came up with his magnetic chain
model in 1976 [67] being influenced by earlier works of Richardson [79, 80]
on exact solution of the BCS equations of superconductivity. This connection
\ with superconductivity will play an important role in what follows.

In physics literature\ all Gaudin-type models are based on the $SU(2)$
algebra of spin operators\footnote{%
In mathematics literature to be used below [8,75] the $SL(2,C)$ group is
used instead of \ its subgroup, $SU(2)$ [ 81].}. Instead of one Hamiltonian,
the set of commuting Hamiltonians of the type [82]%
\begin{equation}
H_{i}=\sum\limits_{j(\neq i)=1}^{N}\sum\limits_{\alpha =1}^{3}w_{ij}^{\alpha
}\sigma _{i}^{\alpha }\sigma _{j}^{\alpha }  \tag{5.1}
\end{equation}%
is used. In view of the fact that, by construction, $[H_{i},H_{j}]=0$, $%
3N(N-1),$ the coefficients $w_{ij}^{\alpha }$ should satisfy the following
equations%
\begin{equation}
w_{ij}^{\alpha }w_{jk}^{\gamma }+w_{ji}^{\beta }w_{ik}^{\gamma
}-w_{ik}^{\alpha }w_{jk}^{\beta }=0.  \tag{5.2}
\end{equation}%
These equations can be solved by imposing the antisymmetry requirement: $%
w_{ij}^{\alpha }=-w_{ji}^{\alpha }$ which can be satisfied by replacing $%
w_{ij}^{\alpha }$ by the unknown functions $w_{ij}^{\alpha }=f^{\alpha
}(z_{i}-z_{j})$ of \ difference between two new real parameters $z_{i}$ and $%
z_{j}$. It is only natural to make further restrictions based on requirement
that the $z-$component of the total spin $S^{3}$ =$\sum\nolimits_{i}\sigma
_{i}^{3}$ is conserved. This causes $w_{ij}^{1}=w_{ij}^{2}\equiv X_{ij}$ and 
$w_{ij}^{3}=Y_{ij}$ thus leading to equations 
\begin{equation}
Y_{ij}X_{jk}+Y_{ki}X_{jk}+X_{ki}X_{ij}=0.  \tag{5.3}
\end{equation}%
These constraint equations admit the following sets of solutions: 
\begin{equation}
X_{ij}=Y_{ij}=\frac{1}{z_{i}-z_{j}}\text{ \ \ (rational),}  \tag{5.4a}
\end{equation}%
\begin{equation}
X_{ij}=\frac{1}{\sin \left( z_{i}-z_{j}\right) }\text{ , }Y_{ij}=\cos \left(
z_{i}-z_{j}\right) \text{ \ (trigonometric),}  \tag{5.4b}
\end{equation}%
\begin{equation}
X_{ij}=\frac{1}{\sinh \left( z_{i}-z_{j}\right) },\text{ }Y_{ij}=\cosh
\left( z_{i}-z_{j}\right) \text{ (hyperbolic).}  \tag{5.4c}
\end{equation}%
While the first solution, (5.4a), to be used in this work, \ corresponds to
the long \ range analog of the standard $XXX$ spin chain, the remaining two
solutions correspond to the long range analog of the $XXZ$ spin chain.

Folloving Varchenko [8] we are now in the position to write down the K-Z
equation. For this purpose we combine equations (5.1) and (5.4a) \ and
reintroduce the coupling constant $g$ (so that $w_{ij}^{\alpha }\rightarrow
gw_{ij}^{\alpha })$ in such a way that the K-Z equation acquires the form%
\begin{equation}
(\kappa \frac{\partial }{\partial z_{i}}-H_{i}(z_{1},...,\text{ }z_{N}))\Phi
(z_{1},...,\text{ }z_{N})=0,\text{ \ \ \ \ }i=1,_{...},N,.  \tag{5.5}
\end{equation}%
where $\kappa =g^{-1}.$ This result requires several comments. First, \ from
the theory of WZNW models it is known that parameter $\kappa $ cannot take
arbitrary values. For instance, for $SU_{1}(2)$ WZNW model $\kappa =\frac{3}{%
2}$ [30]. Second, we can always rescale $z$-coordinates and to redefine the
Hamiltonian to make the constant arbitrary small. Apparently, this \ was
asssumed in the asymptotic analysis of the K-Z equation described in [7,8].
Third, if this is the case, then such analysis (to be used below) differs
essentially from other approaches connecting string models with spin chains,
e.g. see [83], because such a connection was made in these works for $SU(N)$%
-type magnets (or gauge theories) in the unphysical limit $N\rightarrow
\infty .$ Since for $SU(N)$ models $\kappa =\frac{1}{2}(k+N),$ in the limit $%
N\rightarrow \infty $ we have $\kappa \rightarrow \infty .$ The WKB-type
method of Reshetikhin and Varchenko (to be discussed below) fails exactly in
this limit.

With K-Z equation defined, we would like to make a connection between the
Gaudin and $P-F$ model. To a large extent this was already accomplished in
[84]. Following this reference, we define the spin Calogero (S-C) model as
follows 
\begin{equation}
H_{S-C}=\frac{1}{2}\sum\limits_{i}^{{}}(p_{i}^{2}+\omega
^{2}x_{i}^{2})+g\sum\limits_{\substack{ i<j  \\ }}^{{}}\frac{\vec{\sigma}%
_{i}\cdot \vec{\sigma}_{j}}{\left( z_{i}-z_{j}\right) ^{2}}  \tag{5.6}
\end{equation}%
to be compared with $\mathcal{H}$ in (3.2)\footnote{%
We added the oscillator-type potential absent in the original work [84] for
the sake of additional comparisons, e.g.with (3.4). In what follows such a
constraint is not essential and will be ignored.}. Using the rational form
of the Gaudin Hamiltonian, this result can be equivalently rewritten as 
\begin{equation}
H_{S-C}=\frac{1}{2}\sum\limits_{l}^{{}}(p_{l}^{2}+\omega ^{2}x_{l}^{2}+i%
\frac{g}{2}[p_{l},H_{l}]).  \tag{5.7}
\end{equation}%
That this is indeed the case can be seen by the following chain of arguments.

Consider the strong coupling limit ($g\rightarrow \infty )$ of $H_{S-C}$ so
that the kinetic term is a perturbation. Next, we consider the eigenvalue
problem for one of the Gaudin's Hamiltonians, i.e.%
\begin{equation}
H_{l}\Psi ^{(l)}=E^{(l)}\Psi ^{(l)},  \tag{5.8}
\end{equation}%
and apply the operator $ip_{l}$ to both sides of this equation. Furthermore,
consider in \textsl{this} limit the combination $H_{S-C}\Psi ^{(l)}.$
Provided that the eigenvalue problem (5.8) does have a solution, it is
always possible to Fourier expand ($ip_{l}\Psi ^{(l)})$ using as basis set $%
\Psi ^{(l)}.$ In such a case we end up with the eigenvalue problem for the
P-F spin chain in which the eigenfunctions are the same as for the Gaudin's
problem and the eigenvalues are $ip_{l}E^{(l)}.$ Physical significance of
this result \ will be discussed in detail below. Before doing so, we have to
make a connection between the K-Z equation (5.5) and Gaudin eigenvalue
equation (5.8) following [7,8].

We begin by replacing $SU(2)$ spin operators by $SL(2,C)\equiv sl_{2}$
operators $e$, $f$ and $h$ obeying following commutation relations 
\begin{equation}
\lbrack h,e]=2e;\text{ \ \ }[e,f]=h;\text{ \ }[h,f]=-2f.  \tag{5.9}
\end{equation}%
This \ Lie algebra was discussed in our previous work, Part II, in
connection with new models reproducing Veneziano amplitudes. \ In this work,
we shall extend these results following \ ideas of Richardson and Varchenko.

From [81] it is known \ that $SU(2)$ is just a subgroup of $sl_{2}.$
Introduce the Casimir element $\Omega $ $\in $ $sl_{2}\otimes sl_{2}$ via%
\begin{equation}
\Omega =e\otimes f+f\otimes e+\frac{1}{2}h\otimes h  \tag{5.10}
\end{equation}%
so that $\forall x\in sl_{2}$ it satisfies the commutation relation $%
[x\otimes 1+1\otimes x,$ $\Omega ]=0$ inside $\ theU(sl_{2})\otimes
U(sl_{2}) $ where $U(sl_{2})$ is the universal enveloping algebra of $%
sl_{2}. $ Consider the vector space $V=V_{1}\otimes V_{2}\otimes \cdot \cdot
\cdot \otimes V_{N}$. An element $x\in sl_{2}$ acts on $V$ as follows: $%
x\otimes 1\otimes 1\otimes \cdot \cdot \cdot \otimes 1+\cdot \cdot \cdot
+1\otimes 1\otimes \cdot \cdot \cdot \otimes x.$ For indices $1\leq i<j\leq
N $ \ let $\Omega ^{(i,j)}:V\rightarrow V$ be an operator which acts as $%
\Omega $ on i-th and j-th position and as identity on all others, then the
K-Z equation can be written as 
\begin{equation}
\kappa \frac{\partial }{\partial z_{i}}\Phi =\sum\limits_{j\neq i}\frac{%
\Omega ^{(i,j)}}{z_{i}-z_{j}}\Phi \text{, }\ i=1,...,N.  \tag{5.11}
\end{equation}%
In the simplest case, the K-Z equation is defined in the domain $%
U=\{(z_{1},...,z_{N})\in \mathbf{C}^{N}\mid z_{i}\neq z_{j}\}.$

From \ now on we shall use equation (5.11) instead of (5.5). To connect K-Z
equation with the XXX Gaudin magnet we shall use a kind of WKB method
developed by Reshetikhin and Varchenko [7] and summarized in lecture notes
by Varchenko [8]. Following these authors, we \ shall look for a solution of
(5.11) in the form ($\kappa \rightarrow 0):$%
\begin{equation}
\Phi (\mathbf{z},\kappa )=e^{\frac{1}{\kappa }S(\mathbf{z})}\{f_{0}(\mathbf{z%
})+\kappa f_{1}(\mathbf{z})+\cdot \cdot \cdot \},  \tag{5.12}
\end{equation}%
where $\mathbf{z}=\{z_{1},...,z_{N}\}$, $S(\mathbf{z})$ is some scalar
function (to be described below) and $f_{j}(\mathbf{z})$, $j=0,1,2,...$, are 
$V-$valued functions. Provided that the function $S$ is known, $V-$valued
functions can be recursively determined (as it is done in WKB analysis).
Specifically, given that $H_{i}=\sum\limits_{j\neq i}\dfrac{\Omega ^{(i,j)}}{%
z_{i}-z_{j}},$ we obtain,%
\begin{equation}
H_{i}f_{0}(\mathbf{z})=\frac{\partial S}{\partial z_{i}}f_{0}(\mathbf{z}) 
\tag{5.13}
\end{equation}%
to be compared with (5.8). Next we get 
\begin{equation}
H_{i}f_{1}(\mathbf{z})=\frac{\partial S}{\partial z_{i}}f_{1}(\mathbf{z})+%
\frac{\partial f_{0}}{\partial z_{i}}  \tag{5.14}
\end{equation}%
and so on. \ Since the function $S(\mathbf{z})$ (the Shapovalov form) plays
\ an important role in these calculations, we would like to discuss it in
some detail now.

\subsection{The Shapovalov form}

\bigskip

Let us consider the following auxiliary problem. Let $A(x)$ and $B(x)$ be
some pre assigned polynomials of degree $n$ and $n-1$ respectively. Find a
polynomial $C(x)$ of degree $n-2$ such that the differential equation%
\begin{equation}
A(x)y^{^{\prime \prime }}(x)-B(x)y^{\prime }(x)+C(x)y(x)=0  \tag{5.15}
\end{equation}%
has solution which is polynomial of preassigned degree $k$. Such polynomial
solution is called the Lame$^{\prime }$ function. Stieltjes [7,8] proved the
following

Theorem. Let $A$ and $B$ be given polynomials of degree $n$ and $n-1$,
respectively so that $B(x)/A(x)=\sum\nolimits_{j=1}^{n}\dfrac{m_{j}}{x-x_{j}}
$. Then there is a polynomial $C$ of degree $n-2$ and a polynomial solution $%
y(x)=\prod\nolimits_{i=1}^{k}(x-x_{i})$ of (5.15) if and only if $\mathbf{%
\check{x}=}(x_{1},...,x_{k})$ is the critical point of the function%
\begin{equation}
\Phi
_{k,n}(x_{1},...,x_{k};z_{1},...,z_{n})=\prod\limits_{j=1}^{k}\prod%
\limits_{i=1}^{n}(x_{j}-z_{i})^{-m_{i}}\prod\limits_{1\leq i<j\leq
k}(x_{i}-x_{j})^{2}.  \tag{5.16}
\end{equation}%
A point $\mathbf{\check{x}}$\textbf{\ }is critical for\textbf{\ }$\Phi (%
\mathbf{x})$ if all its first derivatives vanish at it.

We would like now to make a connection between the Shapovalov form $S$ and
the results just obtained. $S$ is symmetric bilinear form on previously
introduced space $V$ such that $S(v,v)=1$, $S(hx,y)=S(x,hy),$ $%
S(ex,y)=S(x,fy)$ where $h,e,f$ are defined in (5.9). Furthermore, $S(\Omega
(x_{1}\otimes x_{2}),y_{1}\otimes y_{2})=S(x_{1}\otimes x_{2},\Omega
(y_{1}\otimes y_{2}))$ $\forall x_{1},y_{1}\in V_{1}$ and $\forall
x_{2},y_{2}\in V_{2}.$ As result, we obtain,%
\begin{equation}
S(H_{i}x,y)=S(x,H_{i}y)\text{ }\forall x,y\in V.  \tag{5.17}
\end{equation}%
Next, let $m$ be some nonnegative integer and $V_{m}$ be the irreducible\
Verma module with the highest weight $m$ and the highest weight singular
vector $v_{m},$ i.e.%
\begin{equation}
hv_{m}=mv_{m},\text{ }ev_{m}=0.  \tag{5.18}
\end{equation}%
Consider a tensor product $V\equiv V^{\otimes M}=V_{m_{1}}\otimes \cdot
\cdot \cdot \otimes V_{m_{n}}$ so that $M=(m_{1},...,m_{n}).$ $\forall
V_{m_{i}}$ vectors $v_{m_{i}}$, $fv_{m_{i}},f^{2}v_{m_{i}},\cdot \cdot \cdot
,f^{m_{i}}v_{m_{i}}$ form a basis of $V_{m_{i}}$ \footnote{%
According to [8] in all subsequent calculations it is sufficient to use%
\textsl{\ the finite} Verma module, i.e. $L_{m}=V_{m}/<f^{m+1}v_{m}>.$ This
restriction is in accord \ with our previous calculations, e.g. see Part II,
Section 8, where such a restriction originates from the Lefschetz
isomorphism theorem used in conjunction with supersymmetric model
reproducing Veneziano amplitudes.}so that the Shapovalov form is orthogonal
with respect to such a basis and is decomposable as $S=S_{m_{1}}\otimes
\cdot \cdot \cdot \otimes S_{m_{n}}$ \ Let, furthermore, $%
J=(j_{1},...,j_{n}) $ be a set of nonnegative integers such that $%
j_{1}+\cdot \cdot \cdot +j_{n}=k$ where $k$ is the same as in (5.16) and $%
0\leq j_{i}\leq m_{l}$. This allows us to define the vectors $%
f^{J}v_{M}=f^{j_{1}}v_{m_{1}}\otimes \cdot \cdot \cdot \otimes
f^{j_{n}}v_{m_{n}}$ . \ These vectors $\{f^{J}v_{M}\}$ are by construction
orthogonal with respect to the Shapovalov form and provide a basis for the
space $V^{\otimes M}.$ Introduce the \textsl{weight} of a partition $A$ as $%
\left\vert A\right\vert =a_{1}+a_{2}+...$ then, in view of (5.18), we define
the singular vector $f^{J}v_{M}$ via 
\begin{equation}
h(f^{J}v_{M})=(\left\vert M\right\vert -2\left\vert J\right\vert )f^{J}v_{M}%
\text{ , \ \ }e\text{(}f^{J}v_{M})=0  \tag{5.19}
\end{equation}%
of weight $\left\vert M\right\vert -2\left\vert J\right\vert .\footnote{%
This fact can be easily undestood from the properties of $sl_{2}$ Lie
algebra representations since it is known, [8] and Part II, that for the
module of highest weight $m$ we have $h(f^{k}v_{m})=(m-2k)(f^{k}v_{m}).$}$
The Bethe ansatz vectors $\mathcal{V}$ for the Gaudin model can be defined
now as 
\begin{equation}
\mathcal{V(}\mathbf{\check{x}},\mathbf{z})=\sum\limits_{J}A_{J}(\mathbf{%
\check{x}},\mathbf{z})f^{J}v_{M}  \tag{5.20}
\end{equation}%
where \textbf{\v{x}} \ is a critical point of $\Phi (\mathbf{x,z})$ which
was defined by (5.16). A function $A_{J}(\mathbf{\check{x}},\mathbf{t})$ is
defined as follows%
\begin{equation}
A_{J}(\mathbf{\check{x}},\mathbf{z})=\sum\limits_{\sigma \in \mathcal{P(}%
k;J)}\prod\limits_{i=1}^{k}\frac{1}{x_{i}-z_{\sigma (i)}}  \tag{5.21}
\end{equation}%
with $\mathcal{P(}k;J)$ being the set of maps $\sigma $ from the $%
\{1,...,k\} $ to $\{1,...,n\}$. Finally, using these definitions it is
possible to prove that%
\begin{equation}
S(\mathcal{V(}\mathbf{\check{x}},\mathbf{z}),\mathcal{V(}\mathbf{\check{x}},%
\mathbf{z}))=\det_{1\leq i,j\leq k}(\frac{\partial ^{2}}{\partial
x_{i}\partial x_{j}}\ln \Phi _{k,n}(\check{x}_{1},...,\check{x}%
_{k};z_{1},...,z_{n})).  \tag{5.22}
\end{equation}%
The equations determining critical points 
\begin{equation}
\frac{1}{\Phi _{k,n}(\mathbf{x}^{0},\text{ }\mathbf{z}^{0})}\frac{\partial }{%
\partial x_{i}}\Phi _{k,n}(\mathbf{x}(\mathbf{z}),\text{ }\mathbf{z})\mid _{%
\mathbf{z}=\mathbf{z}^{0}}=0  \tag{5.23}
\end{equation}%
are the Bethe ansatz equations for the Gaudin model. Using these equations
the eigenvalue equation (5.13) for the Gaudin model now acquires the
following form%
\begin{equation}
H_{i}(\mathbf{z}^{0})\mathcal{V(}\mathbf{x}^{0},\mathbf{z}^{0})=\frac{%
\partial }{\partial z_{i}}\ln \Phi _{k,n}(\mathbf{x}(\mathbf{z}),\text{ }%
\mathbf{z})\mid _{\mathbf{z}=\mathbf{z}^{0}}\mathcal{V(}\mathbf{x}^{0},%
\mathbf{z}^{0}).  \tag{5.24}
\end{equation}%
In the next subsection we shall sudy in some detail the Bethe ansatz
equation (5.23).This will allow us to define eigenvalues in (5.24)
explicitly.

\subsection{ Mathematics \ and physics of the Bethe ansatz equations for XXX
Gaudin model according to\ works by Richardson. Connections with the
Veneziano model}

\bigskip Using (5.16) in (5.23) produces the following set of\ the Bethe
ansatz equations:%
\begin{equation}
\sum\limits_{i=1}^{n}\frac{m_{i}}{x_{j}-z_{i}}=\sum\limits_{\substack{ i=1 
\\ i\neq j}}^{k}\frac{2}{x_{j}-x_{i}},\text{ \ \ }j=1,...,k.  \tag{5.25}
\end{equation}%
To understand the physical meaning of these equations \ we \ shall use
extensively results of\ two key papers by Richardson [79,80]. To avoid
duplications, and for the sake of space, our readers are encoraged to read
thoroughly these papers. \ Although \ originally they were written having
applications to nuclear physics in mind, they are no less significant for
condensed matter [82] and atomic physcs [85]. \ Because of this, only
nuclear physics terminology will be occasionally used. At the time of
writing of these papers, QCD was still in its infancy. Accordingly, no
attempts were made to apply Richardson's results to QCD. Recently,
Ovchinnikov [86] have conjectured that the Richardson-gaudin equations can
be useful for development of color superconductivity in QCD [87].
Incidentally, in the same paper [87] it is emphasized that such type of
superconductivity can exist only if the number of colors is not too large,
e.g. N$_{c}$ =3. This fact is in accord with remarks made in Section 5.2
regarding the validity of the WKB-type methods in the limit N$\rightarrow
\infty $ for the K-Z equation.

Thus, following Richardson [80], \ we consider the system of interacting
bosons described by the (pairing) Hamiltonian\footnote{%
In the paper with Sherman [79] Richardson explains in detail how one can map
the fermionic (pairing) system into bosonic.}%
\begin{equation}
H=\sum\nolimits_{l}\varepsilon _{l}\hat{n}_{l}-\frac{g}{2}%
\sum\nolimits_{ll^{^{\prime }}}A_{l}^{+}A_{l}.  \tag{5.26}
\end{equation}%
Here we have $\hat{n}_{l}=\sum\limits_{\mathbf{k}(\varepsilon _{\mathbf{k}%
}=\varepsilon _{l})}a_{\mathbf{k}}^{+}a_{\mathbf{k}}$ , $A_{l}^{+}=\sum%
\limits_{\mathbf{k}(\varepsilon _{\mathbf{k}}=\varepsilon _{l})}a_{\mathbf{k}%
}^{+}a_{-\mathbf{k}}^{+}$ and $A_{l}=\sum\limits_{\mathbf{k}(\varepsilon _{%
\mathbf{k}}=\varepsilon _{l})}a_{-\mathbf{k}}a_{\mathbf{k}}.$ It is assumed
that the single-particle spectrum $\{\varepsilon _{l}\}$ is such that $%
\varepsilon _{l}<\varepsilon _{l+1}$ $\forall l$ and that the degeneracy of $%
l$-th level is $\Omega _{l}$ so that the sums (over \textbf{k} ) each
contain $\Omega _{l}$ terms. It is assumed furthermore that the system
possesses the time-reversal symmetry implying $\varepsilon _{\mathbf{k}%
}=\varepsilon _{-\mathbf{k}}$. The operators $a_{\mathbf{k}}^{+}$ and $a_{%
\mathbf{k}}$ obey usual commutation rules for bosons, i.e. $[a_{\mathbf{k}%
},a_{\mathbf{k}^{\prime }}^{+}]=\delta _{\mathbf{kk}^{\prime }}$. The sign
of the coupling constant in principle can be both positive and negative. We
shall work, however, with more physically interesting case of negative
coupling (so that $g$ in (5.26) is actually $\left\vert g\right\vert $).

An easy computation using commutation rule for bosons produces the following
results%
\begin{equation}
\lbrack \hat{n}_{l},A_{l^{^{\prime }}}^{+}]=2\delta _{ll^{\prime
}}^{{}}A_{l}^{+},  \tag{5.27a}
\end{equation}%
\begin{equation}
\lbrack A_{l},A_{l^{\prime }}^{+}]=2\delta _{ll^{\prime }}^{{}}(\Omega _{l}+2%
\hat{n}_{l}),  \tag{5.27b}
\end{equation}%
\begin{equation}
\lbrack \hat{n}_{l},A_{l^{^{\prime }}}^{{}}]=-2\delta _{ll^{\prime
}}^{{}}A_{l}^{{}}.  \tag{5.27c}
\end{equation}%
If we make a replacement \ of $\hat{n}_{l}$ in (5.27a) and (5.27.c) by $%
\frac{\Omega _{l}}{2}+\hat{n}_{l}\equiv $ $\dfrac{\mathbf{\hat{n}}_{l}}{4}$
and keep the same notation in the r.h.s. of (5.27b) we shall arrive at the $%
sl_{2}$ Lie algebra isomorphic to that given in (5.9). The same Lie algebra
was uncovered and used in our Part II for description of new models
describing Veneziano amplitudes.Because of this, we would like now to
demonstrate that the rest of arguments of Part II can be implemented now in
the present context thus making the P-F model (which is derivative of the
Richardson-Gaudin XXX model) correct model related to Veneziano amplitudes.

Following Richardson [80], we notice that the model described by Hamiltonian
(5.26) and algebra (5.27) admit two types of excitations: those which are
associated with the unpaired particles and those with coupled pairs. The
unpaired $\nu -$particle state is defined by the following two equations%
\begin{equation}
\hat{n}\mid \varphi _{\nu }>=\nu \mid \varphi _{\nu }>,  \tag{5.28}
\end{equation}%
\begin{equation}
A_{l}\mid \varphi _{\nu }>=0\text{ }\forall l.  \tag{5.29}
\end{equation}%
Here, $\hat{n}=\sum\nolimits_{l}\hat{n}_{l}$ so that, in fact, 
\begin{equation}
\hat{n}_{l}\mid \varphi _{\nu }>=\nu _{l}\mid \varphi _{\nu }>  \tag{5.30}
\end{equation}%
and, therefore, $\nu =\sum\nolimits_{l}\nu _{l}$. Furthermore%
\begin{equation}
H\mid \varphi _{\nu }>=\sum\nolimits_{l}\varepsilon _{l}\nu _{l}\mid \varphi
_{\nu }>.  \tag{5.31}
\end{equation}%
Following Richardson, we want to demonstrate that parameters $\varepsilon
_{l}$ in (5.31) can be identified with parameters $z_{l}$ in the Bethe
equations (5.25). Because of this, the eigenvalues for the P-F chain are
obtained as described \ in Section 5.2., that is 
\begin{equation}
E_{i}^{(\mathcal{P}-\mathcal{F})}=\frac{\partial }{\partial \varepsilon _{i}}%
\sum\nolimits_{l}\varepsilon _{l}\nu _{l}=\nu _{i}.  \tag{5.32}
\end{equation}%
These are eigenvalues of $\hat{n}_{l}$ defined in (5.30). Furthermore, this
eigenvalue equation is \textsl{exactly the same} as was used in Part II,
Section 8, with purpose of reproducing Veneziano amplitudes. Moreover,
equations (5.28) and (5.29) have the same mathematical meaning as equations
(5.19) defining the Verma module. Because of this, we follow Richardson's
paper to describe this module in physical terms. By doing so additional
comparisons will be made between the results of Part II and works by
Richardson. Since the Hamiltonian (5.26) describes two kinds of particles:
a) pairs of particles (whose total linear and angular momentum is zero) and,
b) unpaired particles (that is single particles which do not interact with
just described pairs), the total number of (quasi) particles is $n=N+\nu 
\footnote{%
In Richardson's paper we find instead: $n=2N+\nu .$ This is, most likely, a
misprint as explained in the text.}$. Since we redefined the number operator
as $\frac{\Omega _{l}}{2}+\hat{n}_{l}\equiv $ $\dfrac{\mathbf{\hat{n}}_{l}}{4%
}\equiv \mathbf{\hat{N}}_{l}$ \ we expect that , once the correct state
vector describing excitations is found, equation (5.30) should be replaced
by the analogous equation for $\mathbf{\hat{N}}_{l}$ whose eigenvalues will
be $\frac{\Omega _{l}}{2}+\nu _{l}$.\footnote{%
These amendments are not present in Richardson's paper but they are in
accord with its content.}

A simple minded way of creating such a state is by constructing the
following state vector $A_{l_{1}}^{+}\cdot \cdot \cdot A_{l_{N}}^{+}\mid
\varphi _{\nu }>$ . This vector does not possess the needed symmetry of the
problem. To create the state vector (actually, the Bethe vector of the type
given by (5.20))\ of correct symmetry one should introduce a linear
combination \ of $A_{l}^{+}$ operators according to the following
prescription:%
\begin{equation}
B_{\alpha }^{+}=\sum\limits_{l}u_{\alpha }(l)A_{l}^{+},\text{ }\alpha
=1,...,N  \tag{5.33}
\end{equation}%
with constants $\ u_{\alpha }(l)$ to be determined below. The (unnormalized)
Bethe-type vectors are given then \ as $\mid \psi >=B_{1}^{+}\cdot \cdot
\cdot B_{N}^{+}\mid \varphi _{\nu }>$ and, accordingly, instead of (5.31),
we obtain 
\begin{equation}
H\mid \psi >=(\sum\nolimits_{l}\varepsilon _{l}\nu _{l})\mid \psi
>+[H,B_{1}^{+}\cdot \cdot \cdot B_{N}^{+}]\mid \varphi _{\nu }>.  \tag{5.34}
\end{equation}%
The task now lies in calculating the commutator and to determine the
constants $u_{\alpha }(l).$ Details can be found in Richardson's paper [80].
The final result looks \ as follows%
\begin{eqnarray}
H &\mid &\psi >-E\mid \psi >  \TCItag{5.35} \\
&=&\sum\limits_{\alpha =1}^{N}(\prod\limits_{\gamma \neq \alpha }B_{\gamma
}^{+})\sum\limits_{l}A_{l}^{+}[(2\varepsilon _{l}-E_{\alpha })u_{\alpha
}(l)+\sum\limits_{l^{\prime }}(\Omega _{l^{\prime }}+2\hat{n}_{l^{\prime
}})u_{\alpha }(l^{\prime })+4g\sum\limits_{\beta (\beta \neq \alpha
)}^{{}}M_{\beta \alpha }]\mid \varphi _{\nu }>.  \notag
\end{eqnarray}%
By requiring the r.h.s. of this equation to be zero we arrive at the
eigenvalue equation 
\begin{equation}
H\mid \psi >=E\mid \psi >,\text{ where }E=\sum\limits_{l}\varepsilon _{l}\nu
_{l}+\sum\limits_{\alpha =1}^{N}E_{\alpha }.  \tag{5.36}
\end{equation}%
Furthermore, this requirement after several manipulations leads us to the
Bethe ansatz \ equations\footnote{%
It should be noted that in the original paper [80] the sign in front of the
3rd term in the l.h.s. is positive. This is because Richardson treats both
positive and negative couplings simultaneously. Equation (5.37a) is in
agreement with (3.24) of Richardson-Sherman paper [79] where the case of
negative coupling (pairing) is treated.} 
\begin{equation}
\frac{1}{2g}+\sum\limits_{\beta (\beta \neq \alpha )}^{N}\frac{2}{E_{\beta
}-E_{\alpha }}-\sum\limits_{l=1}^{L}\frac{\Omega _{l}/2+\nu _{l}}{%
2\varepsilon _{l}-E_{\alpha }}=0,\text{ }\alpha =1,...,N,  \tag{5.37a}
\end{equation}%
as well to the explicit form of coefficients $u_{\alpha }(l):u_{\alpha
}(l)=1/(2\varepsilon _{l}-E_{\alpha })$ and the matrix elements $M_{\alpha
,\beta }$ (since, by construction, $u_{\alpha }(l)u_{\beta }(l)=M_{\alpha
,\beta }u_{\alpha }(l)+M_{\beta ,\alpha }u_{\beta }(l)).$ In the limit $%
g\rightarrow 0$ we expect $E_{\alpha }\rightarrow 2\varepsilon _{l}$ $\ and$ 
$\Omega _{l}\rightarrow 0$ in accord with (5.28)-(5.30). Therefore, we
conclude that $\frac{\Omega _{l}}{2}+\nu _{l}$ is an eigenvalue of the
operator \textbf{\^{N}}$_{l}$ acting on $\mid \psi >$ in accord with remarks
made before. In the opposite limit: $g\rightarrow \infty $ \ the system of
equations (5.37a) \ will coincide with (5.25) upon obvious identifications: $%
x_{\alpha }\rightleftarrows E_{\alpha },2\varepsilon _{l}\rightleftarrows
z_{l},N\rightleftarrows k,L\rightleftarrows n$ and $\Omega _{l}/2+\nu
_{l}\rightleftarrows m_{l}.$ Next, in view of (5.32) and (5.36) we obtain
the following result for the occupation numbers:%
\begin{eqnarray}
\tilde{\Omega}_{i} &\equiv &E_{i}^{(\mathcal{P-F})}=\frac{\partial }{%
\partial \varepsilon _{i}}[\sum\limits_{l}\varepsilon _{l}\nu
_{l}+\sum\limits_{\alpha =1}^{N}E_{\alpha }]  \notag \\
&=&\nu _{i}+\sum\limits_{\alpha =1}^{N}\frac{\partial E_{\alpha }}{\partial
\varepsilon _{i}}.  \TCItag{5.38}
\end{eqnarray}%
Based on the results just obtained, it should be clear that, actually, $%
E_{i}^{(\mathcal{P-F})}$ =$\nu _{i}+\frac{\Omega _{i}}{2}$ so that $\frac{%
\Omega _{i}}{2}=\sum\limits_{\alpha =1}^{N}\frac{\partial E_{\alpha }}{%
\partial \varepsilon _{i}}.$ Richardson [jmp] \ cleverly demonstrated that
the combination $\sum\limits_{\alpha =1}^{N}\frac{\partial E_{\alpha }}{%
\partial \varepsilon _{i}}$ must be an integer.

Consider now a special case: $N=1$. Evidently, for this case, the derivative 
$\frac{\partial E_{\alpha }}{\partial \varepsilon _{i}}$ should \ also be an
integer. For different $\varepsilon _{i}^{\prime }s$ these may, in general,
be different integers. This fact has some physical significance to be
explained below.

To simplify matters, by analogy with theory of superconducting grains [82],
we assume that $\ $the$\ $energy $\varepsilon _{i}$ can be written as $%
\varepsilon _{i}=d(2i-L-1),$ $i=1,2,...,L.$ The adjustable parameter $d$ \
measures the level spacing for the unpaired \ particles in the limit $%
g\rightarrow 0$. With such simplification, we obtain the following BCS-type
equation using (5.37) (for $N=1$):%
\begin{equation}
\sum\limits_{l=1}^{L}\frac{\tilde{\Omega}_{l}}{2\varepsilon _{l}-E}=\frac{1}{%
G},  \tag{5.39}
\end{equation}%
where $G$ is the rescaled coupling constant. Such \ an equation was
discussed in the seminal paper by Cooper [88] which paved a way to the BCS
theory of superconductivity. To solve this equation, let now $%
F(E)=\sum\limits_{l=1}^{L}\tilde{\Omega}_{l}(2\varepsilon _{l}-E)^{-1}$ so
that (5.39) is reduced to%
\begin{equation}
F(E)=G^{-1}.  \tag{5.40}
\end{equation}%
This equation can be solved graphically as depicted below


\begin{figure}[tbp]
\begin{center}
\includegraphics[width=2.34 in]{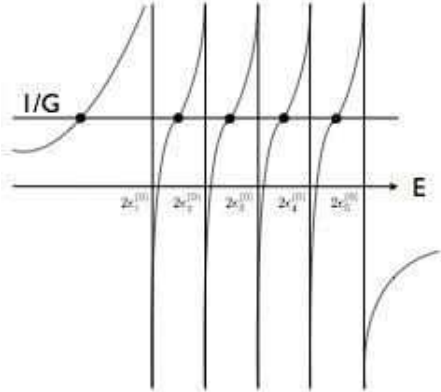}
\end{center}
\caption{Graphical solution of the equation (5.40)}
\end{figure}

As can be seen from Fig.1, solutions to this equation for $G=\infty $ \ can
be read off from the $x$ axis. In addition, if needed, \ for any $N\geq 1$
the system of equations (5.37a) can be rewritten in a similar BCS-like form
if we introduce the renormalized coupling constant $G_{\alpha }$ via 
\begin{equation}
G_{\alpha }=G[1+2G\sum\limits_{\beta (\beta \neq \alpha )}^{N}\frac{1}{%
E_{\beta }-E_{\alpha }}]^{-1}\text{ }  \tag{5.41}
\end{equation}%
so that \ now we obtain:%
\begin{equation}
F(E_{\alpha })=G_{\alpha }^{-1},\alpha =1,...,N.  \tag{5.37b}
\end{equation}%
This sustem of equations can be solved iteratively, beginning with equation
(5.40). There is, however, better way of obtaing these solutions. In view of
equations (5.15), (5.16) and (5.23) solutions $\{E_{\alpha }\}$ of (5.37.b)
are the roots of the Lame$^{\prime }-$type function which is \ obtained as
solution of (5.15). Surprisingly, this fact known to mathematicians for a
long time has been recognized in nuclear physics literature only very
recently [89].

\subsection{Emergence of the Veneziano-like amplitudes \ as consistency
condition for $N=1$ solutions of the K-Z equations. Recovery of the
pion-pion scattering amplitude}

\bigskip

Since results for the Richardson-Gaudin (R-G) model are obtainable from the
corresponding solutions of the K-Z equations in this subsection we would
like to explain why $N=1$ solution of the Bethe-Richardon equations can be
linked with the Veneziano-like amplitudes describing the pion-pion
scattering. In doing so, we shall by pass the P-F model since, anyway, it is
obtainable from the R-G model.

Thus, we begin again with equations (5.10)-(5.11). We would like to look at
the special class of solutions of (5.11) for which the parameter $\left\vert
J\right\vert $ in Verma module (5.19) is equal to one. This corresponds
exactly to the case $N=1$. Folloving Varchenko [8], by analogy with (5.16)\
we introduce the function $\Phi (\mathbf{z},t)$ via 
\begin{equation}
\Phi (\mathbf{z},t)=\prod\limits_{1\leq i<j\leq L}(z_{i}-z_{j})^{\dfrac{%
m_{i}m_{j}}{\kappa }}\prod\limits_{l=1}^{L}(t-z_{l})^{-\dfrac{m_{l}}{\kappa }%
}.  \tag{5.42}
\end{equation}%
It is a multivalued function at \ points of its singularities at $%
z_{1},...,z_{L}.$ Using this function, \ we define the set of 1-forms via%
\begin{equation}
\omega _{j}=\Phi (\mathbf{z},t)\frac{dt}{t-z_{j}},\text{ \ }j=1,...,L, 
\tag{5.43}
\end{equation}%
and the vector $\mathbf{I}^{(\gamma )}$ of integrals $\mathbf{I}^{(\gamma
)}=($I$_{1},...,$I$_{L})\equiv (\int\nolimits_{\gamma }\omega
_{1},...,\int\nolimits_{\gamma }\omega _{L})$ with $\gamma $ being a
particular Pochhammer countour: \ a double loop winding around \ any two
points $z_{\alpha }$, $z_{\beta }$ taken from the set $z_{1},...,z_{L}.$
Deatails can be found in [8,75].

We want now to design the singular Verma module for the K-Z equations using
\ equation (5.19) and results just presented. Taking into account the
following known relations:%
\begin{equation*}
a)\text{ }ef^{k}v_{m}=k(m-k+1)f^{k-1}v_{m},\text{ and }b)\text{ }%
hf^{k}v_{m}=(m-2k)f^{k}v_{m}
\end{equation*}%
for the Lie algebra $sl_{2}$ also used in Part II, Section 8, and taking
into account that in the present ($N=1$) case the basis vectors $%
f^{J}v_{M}=f^{j_{1}}v_{m_{1}}\otimes \cdot \cdot \cdot \otimes
f^{j_{n}}v_{m_{n}}$ acquire the following form: $f^{\mathbf{1}%
}v_{M}=v_{m_{1}}\otimes \cdot \cdot \cdot \otimes fv_{m_{s}}\otimes \cdot
\cdot \cdot \otimes v_{m_{n}},s=1,...,L,$ provided that $m_{i}^{\prime }s$
are the same as in (5.25) (or (5.42)), \ the singular vector for such a
Verma module is given by 
\begin{equation}
w(\gamma )=\sum\limits_{s=1}^{L}I_{s}v_{m_{1}}\otimes \cdot \cdot \cdot
\otimes fv_{m_{s}}\otimes \cdot \cdot \cdot \otimes v_{m_{n}}.  \tag{5.44}
\end{equation}%
In view of the Lie algebra relations just introduced, we obtain $e\cdot w=0$
or, explicitly, 
\begin{equation}
\sum\limits_{s=1}^{L}m_{s}I_{s}=0.  \tag{5.45}
\end{equation}%
Hence, for a fixed Pochhammer contour $\gamma $ there are $L-1$ independent
basis vectors $\{w^{i}\}$. They represent $L-1$ independent solutions of the
K-Z equation of the type $k=1$ (or $N=1$). Let now $z_{i}^{\prime }s$ be
ordered in such a way that $z_{1}<\cdot \cdot \cdot <z_{L}.$ Furthermore, in
view their physical interpretation described in previous section, these $%
z_{i}^{\prime }s$ can be chosen to be equidistant. Consider then a special
set of Pochhamer contours $\{\gamma _{i}\}$ \ \ around points $z_{i}$ and $%
z_{i+1},$ $i=1,2,...,L-1$ \ and consider the matrix $\mathbf{M}$ made of
integrals of the type \ $M_{j}^{i}=-\dfrac{m_{j}}{\kappa }%
\int\nolimits_{\gamma _{i}}\omega _{j}$ then, any ($k=1)-$ type solution $%
\phi ^{i}(i=1,2,...,L-1)$ of the K-Z equation can be represented as 
\begin{equation}
\phi ^{i}=\sum\limits_{j}M_{j}^{i}w^{j},\text{ }i=1,2,...,L-1.  \tag{5.46}
\end{equation}%
From linear algebra it is known that in order for these K-Z solutions to be
independent \ we have to require that $\det \mathbf{M}$ $\neq 0.$ The proof
of this fact is given in Appendix B. Calculation of the determinant of $%
\mathbf{M}$ is described in detail in [8] with the result:%
\begin{equation}
\det \mathbf{M=\pm }\text{A}\frac{\Gamma (1-\frac{m_{1}}{\kappa })\cdot
\cdot \cdot \Gamma (1-\frac{m_{L}}{\kappa })}{\Gamma (1-\frac{\left\vert
M\right\vert }{\kappa })}  \tag{5.47}
\end{equation}%
with $\pm $A being some known constant\footnote{$\pm $A=$\prod\limits 
_{\substack{ 1\leq i,j\leq L  \\ (i\neq j)}}(z_{i}-z_{j})^{\dfrac{-m_{j}}{%
\kappa }}$} and $\Gamma (x)$ being Euler's gamma function. For $L=2$ without
loss of generality one can choose $z_{1}=0$ and $z_{2}=1,$ then \ in thus
obtained determinant one easily can recognize the Veneziano-type $\pi
^{+}\pi ^{-}$ scattering amplitude used in the work by Lovelace [90]. We
have discussed this amplitude previously in connection with mirror symmetry
issues [91]. This time, however, we would like to discuss other topics.

In particular, we notice first that all mesons are made of two quarks.
Specifically, we have $u\bar{d}$ for $\pi ^{+},d\bar{u}$ for $\pi ^{-}$ and $%
d\bar{d}$ for $\pi ^{0}.$ These are very much like the Cooper pairs with $q%
\bar{q}$ \ quark pairs contributing to the Bose condensate \ which was
created as result of spontaneous chiral symmetry breaking. As in the case of
more familiar Bose condensate, in addition to the ground state we expect to
have a tower of the excited states made of such quark pairs. Experimentally,
these are interpreted as more massive mesons. Such excitations are ordered
by their energies, angular momentum and, perhaps, by other quantum numbers
which can be taken into account if needed. Color confinement postulate makes
such a tower infinite. Evidently, the Richardson-Gaudin (R-G) model fits
ideally this qualitative picture. Equation (5.40) describes excitations \ of
such Cooper-like pairs (even in the limit: $G\rightarrow \infty )$ as can be
seen from Fig.1$.$ In the P-F model the factor $\tilde{\Omega}_{i}$ plays
effectively the role of energy as discussed already in this work and Part
II. Therefore, in view of (5.38), it is appropriate to write: $\tilde{\Omega}%
_{i}=f(E_{i}),$ with $E_{i}$ being the R-G energies. Although the explicit
form of such $f-$dependence may be difficult to obtain, for our purposes it
is sufficient only to know that such a dependence does exist. This then
allows us to make an identification: $\tilde{\Omega}_{i}\rightleftarrows 
\dfrac{m_{i}}{\kappa }$ consistent with Varchenko's results, e.g. compare
his Theorem 3.3.5 (page 35) with Theorem 6.3.2. (page 90) [8]. But, we had
established that $\tilde{\Omega}_{i}$ is an integer, therefore, $\dfrac{m_{i}%
}{\kappa }$ should be also an integer. This creates some apparent problems.
For instance, when $\left\vert M\right\vert =\kappa $, the determinant, $%
\det \mathbf{M,}$ becomes zero implying that solutions of K-Z equation
become interdependent. This fact has physical significance to be discussed
below and in Section 6. To do so we use some results from our Part I. In
particular, a comparison between 
\begin{equation}
\sin \pi z=\pi z\prod\limits_{k=1}^{\infty }(1-\left( \frac{k}{z}\right)
)(1+\left( \frac{k}{z}\right) )  \tag{5.48}
\end{equation}%
and 
\begin{equation}
\frac{1}{\Gamma (z)}=ze^{-Cz}\prod\limits_{k=1}^{\infty }(1+\left( \frac{k}{z%
}\right) )e^{-\dfrac{z}{k}}  \tag{5.49}
\end{equation}%
where $C$ is some known constant, \ tells us immediately that not only $%
\left\vert M\right\vert =\kappa $ will cause $\det \mathbf{M=}0$ but also $%
\left\vert M\right\vert =\kappa (k+1),k=0,1,2,...$ Accordingly, the
numerator of (5.47) will create poles whenever $\dfrac{m_{i}}{\kappa }=1.$
Existence of \ independent K-Z solutions is \textsl{not} destroyed if,
indeed, such poles do occur. These facts allow us to relabel $\dfrac{m_{i}}{%
\kappa }$ as $\alpha (s)$ (or $\alpha (t)$ or $\alpha (u)$, etc.) as it is
done in high energy physics with continuous parameters s, t, u,... replacing
discrete i's, different for different $\Gamma $ functions in the numerator
of \ (5.47). In the simplest case, this allows us to reduce the determinant
in (5.47) to the form used by Lovelace, i.e.%
\begin{equation}
\det \mathbf{M=-}\lambda \frac{\Gamma (1-\alpha (s))\Gamma (1-\alpha (t))}{%
\Gamma (1-\alpha (s)-\alpha (t))}.  \tag{5.50}
\end{equation}%
If, as usual, we parametrize $\alpha (s)=\alpha (0)+\alpha ^{\prime }s$,
then equation $1=\alpha (s)+\alpha (t)$ causes the $\det \mathbf{M}$ to
vanish. This also fixes parameter $\alpha (0)$: $\alpha (0)=1/2.$ This
result was obtained \ by Adler long before sting theory emerged and is known
as Adler's \ selfconsistency condition [92]. With such "gauge fixing", one
can fix the slope $\alpha ^{\prime }$\ as well \ if one notices that\ the
experimental data allow us to make a choice: $1=\alpha (m_{\rho }^{2}).$This
leads to: $\alpha ^{\prime }=\frac{1}{2m_{\rho }^{2}}\sim 0.885(Gev^{-2})$
in accord with observations.$.$

The obtained \ results are not limited to study of excitations of just one
"supeconducting" pair of quarks. In princile, any finite amount of such
pairs can be studied. In such a case the result for $\det \mathbf{M}$
becomes considerably more complicated but the connections with one
dimensional magnets become even more explicit. We plan to discuss these
issues in future publications.

\section{Discussion. \ Unimaginable ubiquity of Veneziano-like amplitudes in
Nature}

\subsection{General remarks}

\bigskip In the Introduction, following Heisenberg, we posed the question:
Is combinatorics of observational data sufficient for recovery of the
underlying unique microscopic model? That is, can we have the complete
understanding of such a model based on information provided by
combinatorics? As we demonstrated, especially in Section 4, this task is
impossible to accomplish without imposing additional constraints \ which,
normally, are not dictated by the combinatorics only. In Section 5 we
demonstrated that, even accounting for such constraints, the obtained
results could be in conflict with rigorous mathematics. Last but not the
least, since Veneziano amplitudes gave birth to string theory one can pose a
question: Is these Veneziano (or Veneziano-like) amplitudes, perhaps
corrected to account for particles with spin, contain enough information
(analytical, number-theoretic, combinatorial, etc.) that allows restoration
of the underlying microscopic model uniquely? \ The answer is: "No"! \ In
the rest of this section we explain why.

\subsection{ Random fragmentation and coagulation processes and the
Dirichlet distribution}

We begin by recalling some known facts from the probability theory. For
instance, we recall that the stationary Maxwell distribution for velocities
of particles in the gas is just of Gaussian-type. \ It can be obtained \ as
a stationary solution of the Boltzmann's dynamical equation maximizing
Boltzmann's entropy\footnote{%
As discussed recently in our work [93] on the Poincar$e^{\prime }$ and
geometrization conjectures.}.The question arises: Is it possible to find
(discrete or continuous) dynamical equations which will provide known
probability laws as stable stationary solutions? This task will involve
finding of \ dynamical equations along with the corresponding Boltzmann-like
entropies which \ will reach their maxima at respective equilibria for these
dynamical equations. We are certainly not in the position in this closing
section of our paper to discuss this problem in full generality. Instead, we
focus our attention only on processes wihich are described by the so called
Dirichlet distributions.These originate from the integral (equation (2.8) of
Part I) attributed to Dirichlet, that is 
\begin{equation}
\mathcal{D}(x_{1},...,x_{n+1})=\idotsint\limits_{\substack{ u_{1}\geq 0,...,%
\text{ }u_{n}\geq 0  \\ u_{1}+\cdot \cdot \cdot +u_{n}\leq 1}}%
u_{1}^{x_{1}-1}\cdot \cdot \cdot u_{n}^{x_{n}-1}(1-u_{1}-\cdot \cdot \cdot
-u_{n})^{x_{n+1}-1}du_{1}\cdot \cdot \cdot du_{n}.  \tag{6.1}
\end{equation}%
A random vector $(\mathbf{X}_{1},...,\mathbf{X}_{n})\in \mathbf{R}^{n}$ \
such that $\mathbf{X}_{i}\geq 0$ $\forall i$ and $\sum\limits_{i=1}^{n}$%
\textbf{\ \ }$\mathbf{X}_{i}$\textbf{\ }$\mathbf{=}1$\textbf{\ }is said to
be Dirichlet distributed with parameters ($x_{1},...,x_{n};x_{n+1})$
[nevzorov] if the probability density function for $(\mathbf{X}_{1},...,%
\mathbf{X}_{n})$ is given by 
\begin{eqnarray}
P_{\mathbf{X}_{1},...,\mathbf{X}_{n}}(u_{1},...,u_{n}) &=&\frac{\Gamma
(x_{1}+\cdot \cdot \cdot +x_{n+1})}{\Gamma (x_{1})\cdot \cdot \cdot \Gamma
(x_{n+1})}u_{1}^{x_{1}-1}\cdot \cdot \cdot
u_{n}^{x_{n}-1}(1-\sum\limits_{i=1}^{n}u_{i})^{x_{n+1}-1}  \notag \\
&\equiv &\frac{\Gamma (x_{1}+\cdot \cdot \cdot +x_{n+1})}{\Gamma
(x_{1})\cdot \cdot \cdot \Gamma (x_{n+1})}u_{1}^{x_{1}-1}\cdot \cdot \cdot
u_{n}^{x_{n}-1}u_{n+1}^{x_{n+1}-1},\text{ provided that \ }u_{n+1}  \notag \\
&=&1-u_{1}-\cdot \cdot \cdot -u_{n}  \TCItag{6.2}
\end{eqnarray}%
To get some physical feeling of just defined distribution, we notice the
following peculiar aspects of this distribution. First, for any discrete
distribution, we know that the probability $p_{i}$ must be normalized, that
is $\sum\nolimits_{i}p_{i}=1.$ Thus, the Dirichlet distribution is dealing
with averaging of the probabilities! Or, better, is dealing with the problem
of effectively selecting the most optimal discrete probability. The most
primitive of these probabilities is the binomial probability given by 
\begin{equation}
p_{m}=\left( 
\begin{array}{c}
n \\ 
m%
\end{array}%
\right) p^{m}(1-p)^{n-m},\text{ \ }m=0,1,2,....,n\text{.}  \tag{6.3}
\end{equation}%
If $X$ is random variable obeying this law of probability then, the
expectation \ $E(X)$ is calculated as 
\begin{equation}
E(X)=\sum\limits_{m=1}^{n}mp_{m}=np\equiv \mu .  \tag{6.4}
\end{equation}%
Consider such a distribution in the limit: $n\rightarrow \infty .$ In this
limit, if we write $p=\mu /n$ , then the Poisson distribution is obtained as 
\begin{equation}
p_{m}=\frac{\mu ^{m}}{m!}e^{-\mu }.  \tag{6.5}
\end{equation}%
Next, we notice that $m!=\Gamma (m+1),$ furthermore, we replace $m$ by real
valued variable $\alpha $ and $\mu $ by $x$. This allows us to introduce the
gamma distribution \ with exponent $\alpha $ whose probability density is 
\begin{equation}
p_{X}(x)=\frac{1}{\Gamma (\alpha )}x^{\alpha -1}e^{-x}  \tag{6.6}
\end{equation}%
for some gamma distributed random variable $X$. Finally, we would like to
demonstrate now how the Dirichlet distribution can be represented through
gamma distributions. Since the gamma distribution originates from the
Poisson distribution, sometimes in literature the Dirichlet distribution is
called the Poisson-Dirichlet (P-D) distribution [14]. To demonstrate
connection between the Dirichlet and gamma distributions is relatively easy.
Following Kingman [14], consider a set of positive independent gamma
distributed random variables: $Y_{1},...,Y_{n+1}$ with exponents $\alpha
_{1},...,\alpha _{n+1}.$ Furthermore, consider \ $Y=Y_{1}+\cdot \cdot \cdot
+Y_{n+1}$ and construct \ a vector \textbf{u} with components: $u_{i}=\frac{%
Y_{i}}{Y}$. Then, since $\sum\nolimits_{i=1}^{n+1}u_{i}$ =$1,$ the
components of this vector are Dirichlet distributed and, in fact,
independent of $Y$. \ Details are given in Appendix C.

Such described Dirichlet distribution is an equilibrium measure in various
fields ranging from spin glasses to computer science, from linguistics to
genetics, from forensic science to economics, etc. Many useful references\
involving these and other applications can be found in [9-11]. Furtheremore,
\ most of fragmentation and coagulation processes involve the P-D
distribution as their equilibrium measure. Some applications of general
theory of these processes to to nuclear and particle physics were initiated
in a series of papers by Mekjian, e.g. see [12] and references therein. To
avoid duplications, we would like to rederive some of Mekjian results
differently in order to exibit their connections with previous sections.

\subsection{The Ewens sampling formula and Veneziano amplitudes}

This formula was discussed by Mekjian in [94] without any reference to P-D
distribution. It is discussed in many other places, including Ewens own
monograph [95]. Our exposition follows work by Watterson [96] where he
considers a simple \ P-D average of \ monomials of the type generated by the
individual terms in the expansion\footnote{%
Very recently Watterson's results were successfully applied to some problems
in economics [97].} \ 
\begin{equation}
\mathbf{u}^{n}=(u_{1}+\cdot \cdot \cdot
+u_{k})^{n}=\sum\limits_{n=(n_{1},...,n_{k})}\frac{n!}{n_{1}!n_{2}!\cdot
\cdot \cdot n_{k}!}u_{1}^{n_{1}}\cdot \cdot \cdot u_{k}^{n_{k}}.  \tag{6.7}
\end{equation}%
Such type of expansion was used in Part I (equations (2.9),(2.11)) for
calculation of multiparticle Veneziano amplitudes. Not surprisingly,
Watterson's calculation also results in the multiparticle Veneziano
amplitude which upon multiplication by some combinatorial factor in a well
defined limit produces the Ewens sampling formula playing \ a major role in
genetics. \ Although in Appendix D we reproduce the Ewens sampling formula
without use of the P-D distribution, Kingman\ [98] demonstrated that "A
sequence of populations has the Evens sampling property if and only if it
has the P-D limit\footnote{%
That is to say, that the Ewens sampling formula implies the P-D distribution
and vice versa. In the context of high energy physics it is appropriate to
mention that the law of conservation of energy-momentum reflected in (2.4)
leads to the P-D distribution or, equivalently, to the Veneziano formula for
multiparticle amplitudes.}". Hence, we expect that our readers will consult
the Appendix D prior to reading of what follows. Furthermore, since the
vector \textbf{u} is P-D distributed, it is appropriate to mention at this
point that equation (6.7) genetically represents the Hardy-Weinberg law [95]
for mating species\footnote{%
E.g. see Wikipedia where it is known as Hardy-Weinberg principle.}. Hence,
the Ewens sampling formula provides a refinement of this law accounting for
mutations.

Considear a special case of (6.2) for which $x_{1}=x_{2}=\cdot \cdot \cdot
x_{K+1}=\varepsilon $ and let $\varepsilon =\theta /K$ with parameter $%
\theta $ to be defined later. Then, (6.2) is converted to 
\begin{eqnarray}
P_{\mathbf{X}_{1},...,\mathbf{X}_{K}}(u_{1},...,u_{K}) &\equiv &\phi _{k}(%
\mathbf{u})=\frac{\Gamma ((K+1)\varepsilon )}{\left[ \Gamma (\varepsilon )%
\right] ^{K+1}}\prod\limits_{i=1}^{K+1}u_{i}^{\varepsilon -1}\text{ provided
that }1  \notag \\
\text{ } &=&\sum\nolimits_{i=1}^{K+1}u_{i}  \TCItag{6.8}
\end{eqnarray}%
In view of (6.7), consider an average $P(n_{1},...,n_{K})$ over the simplex $%
\Delta $ (defined by $\sum\nolimits_{i=1}^{K+1}u_{i}=1)$ given by%
\begin{equation}
P(n_{1},...,n_{K})=\frac{n!}{n_{1}!n_{2}!\cdot \cdot \cdot n_{K}!}%
\idotsint\limits_{\Delta }u_{1}^{n_{1}}\cdot \cdot \cdot u_{K}^{n_{K}}\phi
_{k}(\mathbf{u})du_{1}\cdot \cdot \cdot du_{K}.  \tag{6.9}
\end{equation}%
A straightforward calculation produces:%
\begin{equation}
P(n_{1},...,n_{K})=\frac{n!}{n_{1}!n_{2}!\cdot \cdot \cdot n_{K}!}\frac{%
\Gamma ((K+1)\varepsilon )}{\left[ \Gamma (\varepsilon )\right] ^{K+1}}%
\prod\limits_{i=1}^{K}\frac{\Gamma (\varepsilon +n_{i})}{\Gamma
((K+1)\varepsilon +n)}  \tag{6.10}
\end{equation}%
to be compared with (5.47). Evidently, the parameter $\varepsilon $ can be
identified with $\kappa $ in $(5.47)$ and, if we select $\theta $ to be a
positive integer, then by replacing $n_{i}^{\prime }s$ with $-n_{i}^{\prime
}s$ we reobtain \ back (5.47) (up to a constant). To obtain the Ewens
sampling formula (equation (D.6)) from (6.10) few additional steps are
required. These are: a) we have to let $K\rightarrow \infty $ while allowing
many of $n_{i}^{\prime }s$ in (6.7) to become zero (this explains meaning of
the word "sampling"), b) we have to order remaining $n_{i}^{\prime }s$ \ in
such a way that $n_{(1)}\geq n_{(2)}\geq \cdot \cdot \cdot \geq
n_{(k)}>0,0,...,0,$ c) we have to cyclically order the remaining $%
n_{i}^{\prime }s$ in a way explained in the Appendix D by introducing $%
c_{i}^{\prime }s$ as numbers of remaining $n_{(i)}^{\prime }s$ \ which are
equal to $i$. That is we have to make a choice between representing $%
r=\sum\nolimits_{i=1}^{k}n_{(i)}$ or $r=\sum\nolimits_{i=1}^{r}ic_{i}$ under
condition that $\ k=\sum\nolimits_{i=1}^{r}c_{i},$ d) finally, just like in
the case of Bose (Fermi) statistics, we have to multipy the r.h.s. of (6.10)
by the obviously looking combinatorial factor $M=K!/[(c_{1}!\cdot \cdot
\cdot c_{r}!)((K-k)!]$. Under such conditions we obtain: $\Gamma
((K+1)\varepsilon )\simeq \Gamma (\theta ),\Gamma ((K+1)\varepsilon
+r)=\Gamma (\theta +r),\frac{\Gamma (\varepsilon +n_{(i)})}{n_{(i)!}}=\frac{1%
}{n_{(i)}}.$ Less trivial is the result: $K!/[(K-k)!\left[ \Gamma
(\varepsilon )\right] ^{k}]\rightarrow \theta ^{k}.$ Evidently, the factor $%
\dfrac{n!}{n_{1}!n_{2}!\cdot \cdot \cdot n_{K}!}$ \ in (6.10) now should be
replaced by $\dfrac{r!}{n_{(1)}\cdot \cdot \cdot n_{(k)}}.$ Finally, a
moment of thought causes us to replace $n_{(i)}^{\prime }s$ by $\ i^{c_{i}}$%
\footnote{%
This is so because the $c_{i}$ numbers count how many of $n_{(i)}^{\prime }s$
\ are equal to $i$.} in order to arrive at the Ewens sampling formula:%
\begin{equation}
P(k;n_{(1)},...,n_{(k)})=\frac{r!}{[\theta ]^{r}}\prod\limits_{i=1}^{r}\frac{%
\theta ^{c_{i}}}{i^{c_{i}}c_{i}!}  \tag{6.11}
\end{equation}%
in agreement with (D.6). This derivation was made without any reference to
genetics and is completely model-independent. To demonstrate connections
with high energy physics in general and with Veneziano amplitudes in
particular, \ we would like to explain the rationale behind this formula
using absolute minimum facts from genetics.

Genetic information is stored in \textsl{genes}. These are some segments (%
\textsl{locuses}) of the double stranded DNA molecule. This fact allows us
to think about the DNA molecule as a world line for mesons made of a pair of
quarks. \ Phenomenologically, the DNA is essentially the \textsl{chromosome}%
. Humans and many other species are \textsl{diploids}. This means that they
need for their reproduction (meiosis) \textsl{two} sets of chromosomes-one
from each parent. \ Hence, we can think of meiosis as\ process analogous to
the meson-meson scattering. \ We would like to depict this process
graphically to emphasize the analogy. Before doing so we need to make few
remarks. First, the life cycle for diploids is rather bizarre. Each cell of
a grown up organism contains 2 sets of chromosomes. The maiting, however,
requires this rule to be changed. The \textsl{gametes} (sex cells) from each
parent carry only one set of chromosomes (that is such cells are \textsl{%
haploid }!). The existence of 2 sets of chromosomes makes individual
organism unique because of the following. Consider, for instance, a specific
trait, e.g. "tall" vs "short". Genetically this property in encoded in some
gene\footnote{%
Or in many genes, but we talk about a given gene for the sake of argument.}.
A particular realization of the gene (causing the organism to be, say, tall)
is called "\textsl{allele}". Typically, there are 2 alleles -one for each of
the chromosomes in the two chromosome set. For instance, T and t (for "tall"
and "short"), or T and T or t and t or, finally, t and T (sometimes order
matters). Then, if father donates 50\% of T cells and 50\% of t cells and
mother is doing to do the same, the offspring is likely going to have either
TT composition with probability 1/4, or tt (with probability 1/4) or tT
(with probability 1/4) and, finally, tt with probability 1/4. But, one of
the alleles is usually dominant (say, T) so that we will see 3/4 of tall
people in the offspring and 1/4 short. \ What we just described is the
essence of the Hardy-Weinberg law based, of course, on the original works by
Mendel. Details can be found in genetics literature [95].

Let us concentrate our attention on a particular locus so that the genetic
character(trait) of a particular individual is described by specifying its
two genes at that locus. For $N$ individuals in the population there are $2N$
chromosomes containing such a locus. For each allele, one is interested in
knowing the proportion of $2N$ chromosiomes at which the gene is realized as
this allele. This gives a probability distribution over the set of possible
alleles which describes a genetic make-up of the population (as far as we
are \ only looking at some specific locus). The problem now is to model the 
\textsl{dynamical process} by which this distribution changes in time from
generation to generation accounting for mutations and selection (caused by
the environment). Mutation can be caused just by chane of one nucleotide
along the DNA strand\footnote{%
The so called "Single Nucleotide Polymorphism" (SNP) which is detectable
either electrophoretically or by DNA melting experiments, etc.}. Normally,
the mutant allele is independent of its parent since once the mutation took
place it is very unlikely that the corrupt message means anything at all.
Hence, the mutant can be either "good" (fit) or "bad" (unfit) for life and
its contribution can be ignored. If $u$ is the probability of mutation per
gene per generation then, the parameter $\theta =4Nu$ in (6.11). With this
information , we are ready to restore the rest of the genetic content of
Watterson's paper [96]. In particular, random P-D variables $\mathbf{X}_{1},%
\mathbf{X}_{2},...,\mathbf{X}_{K}$ denote the allele relative frequences in
a population consisting of $K$ alleles. Evidently, by construction, they are
Dirichlet-distributed. Let $K\rightarrow \infty $ and let $k$ be an
experimental sample of representative frequencies $k\ll K.$ The composition
of such a sample will be random, both because of the nature of the sampling
process and because the population itself is subject to random fluctuations.
For this reason we averaged the Hardy-Weinberg distribution (6.7) over the
P-D distribution in order to arrive at the final result (6.11). This result
is an equilibrium result. Its experimental verification can be found in
[ewens, watterson2]. It is of interest to arrive at it dynamically \ along
the lines discussed in Section 6.2. This is accomplished in the next
subsection but in a different context. \ Based on the facts just discussed
and comparing them with those of Section 2 and Part II, it should be clear
that both genetics and physics of meson scattering have the same
combinatorial origin$.$ All random processes involving decompositions $%
r=\sum\nolimits_{i=1}^{k}n_{(i)}$ (or $r=\sum\nolimits_{i=1}^{r}ic_{i}$) are
the P-D processes [9]. \ 

To conclude this subsection, we would like to illustrate graphically why
genetics and physics of hadrons have many things in common. This is done
with help of the figures 2 through 4.


\begin{figure}[tbp]
\begin{center}
\includegraphics[width=2.4 in]{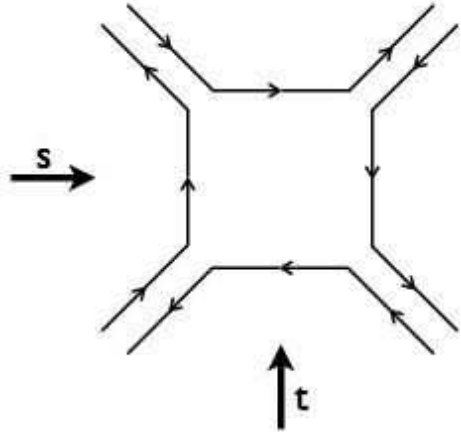}
\end{center}
\caption{The simplest duality diagram describing meson-meson scattering
[99]. The same picture \ describes "collision" of two parental DNA's during
meiosis and can be seen directly under the electron microscope. E.g.see
Fig.2.3 in [100], page 18.}
\end{figure}

\bigskip


\begin{figure}[tbp]
\begin{center}
\includegraphics[width=2.65 in]{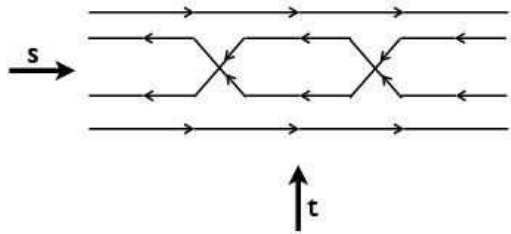}
\end{center}
\caption{Non -planar loop Pomeron diagram for meson-meson scattering [101].
The same diagram describe homologous DNA recombination, e.g. see fig.2.2 in
[100], page 17.}
\end{figure}


\begin{figure}[tbp]
\begin{center}
\includegraphics[width=2.65 in]{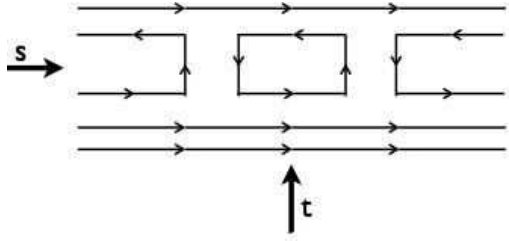}
\end{center}
\caption{Planar loop meson-baryon scattering duality diagram.The same
diagram describes the interaction $scattering$ between the triple and double
stranded DNA helices [102]}
\end{figure}

\subsection{Stochastic models for \ second order chemical reaction kinetics
involving Veneziano-like amplitudes}

\bigskip

In Section 4 and Appendix A we demonstrated the impotant role of \ the ASEP
in elucidating the correct physics. Historically, however, long before the
ASEP was formulated, \ the role of stochastic processes in chemical kinetics
\ was already recognized. A nice summary is contained in the paper by
McQuarrie [103]. The purpose of this subsection is to connect the results in
chemical kinetics with those in genetics in order to reproduce Veneziano (or
Veneziano-like) amplitudes as an equilibrium measures for the underlying
chemical/biological processes.

Following Darvey et al [104] we consider a chemical reaction $A+B%
\begin{array}{c}
k_{1} \\ 
\rightleftarrows \\ 
k_{-1}%
\end{array}%
C+D$ analogous to the meson-meson scattering processes which triggered the
discovery of \ the Veneziano amplitudes. Let the respective concentrations
of the reagents be $a,b,c$ and $d$. Then, according to rules of chemical
kinetics, we obtain the following "equation of motion"%
\begin{equation}
\frac{da}{dt}=-k_{1}ab+k_{-1}cd.  \tag{6.12}
\end{equation}%
This equation has to be supplemented with the initial condition. It is
obtained by accounting for the mass conservation. Specifically, let the
initial concentrations of reagents be respectively as $\alpha =A(0),\beta
=B(0),\gamma =C(0)$ and $\delta =D(0).$ Then, evidently, $\alpha +\beta
+\gamma +\delta =a+b+c+d,$ \ provided that \ for all times $a\geq 0,b\geq
0,c\geq 0$ and $d\geq 0$ (to be compared with equations (2.1), (2.3)).
Accounting for these facts, \ equation (6.12) can be rewritten as 
\begin{equation}
\frac{da}{dt}=(k_{-1}-k_{1})a^{2}-[k_{1}(\beta -\alpha )+k_{-1}(2\alpha
+\gamma +\delta )]a+k_{-1}(\alpha +\gamma )(\alpha +\beta ).  \tag{6.13}
\end{equation}%
Theis is thus far is standard result of chemical kinetics. The new element
emerges when one claims that the the variables $a,b,c$ and $d$ are random
but are still subject to the mass conservation. Then, as we know already
from previous subsections, we are dealing with the P--D-type process. New
element now lies in the fact that this process is dynamical. Following
Kingman [105] we would like to formulate it in precise mathematical terms.
For this purpose, we introduce the vector \textbf{p}(t)=(p$_{1}$(t),..., p$%
_{k}$(t)) such that it moves randomly on the simplex $\Delta $ defined by%
\begin{equation}
\Delta =\{\mathbf{p}(t);p_{j}\geq 0,\sum\nolimits_{i=1}^{k}p_{i}=1\} 
\tag{6.14}
\end{equation}%
In our case the possible states of the system at time $t$ which could lead
to a new state specified by $a,b,c,d$ at time $t+\Delta t$ involving not
more than one transformation in the time interval $\Delta t$ are [104]%
\begin{equation}
\left( 
\begin{array}{cccc}
a+1 & b+1 & c-1 & d-1 \\ 
a-1 & b-1 & c+1 & d+1 \\ 
a & b & c & d%
\end{array}%
\right) .  \tag{6.15}
\end{equation}%
In writing this matrix, following [104], we assume that random variables $%
a,b,c$ and $d$ are integers, just like in (2.3),(2.4). By analogy with
equations of motion of Appendix A, \ using (6.15) we obtain,%
\begin{eqnarray}
P(a,b,c,d;t+\Delta t)-P(a,b,c,d;t) &=&[k_{1}(a+1)(b+1)P(a+1,b+1,c-1,d-1;t) 
\notag \\
&&+k_{-1}(c+1)(d+1)P(a-1,b-1,c+1,d+1;t)  \notag \\
&&-(k_{1}ab+k_{-1}cd)P(a,b,c,d;t)]\Delta t+O(\Delta t)  \TCItag{6.16}
\end{eqnarray}%
In view of the fact that the motion is taking place on the simplex $\Delta $
it is sufficient to look at the stochastic dynamics of just one variable,
say, $a$ (very much like in the deterministic equation (6.13). This replaces
(6.16) by the following result:%
\begin{eqnarray}
\frac{d}{dt}P_{a}(t) &=&k_{1}[(a+1)(a+1+\beta -\alpha
)P_{a+1}(t)+k_{-1}[(\gamma +\alpha -a+1)(\delta +\alpha -a+1)P_{a-1}(t) 
\notag \\
&&-[k_{1}a(\beta -\alpha +a)+k_{-1}(\gamma +\alpha -a)(\delta +\alpha
-a)]P_{a}(t);\text{ \ provided that}  \notag \\
\text{\ \ }P_{\alpha }(0) &=&1,\text{ \ \ }\alpha =a\text{ and }P_{\alpha
}(0)=0\text{ if }a\neq \alpha .  \TCItag{6.17}
\end{eqnarray}%
To solve this equation we introduce the generating function $G(x,t)$ via%
\begin{equation*}
G(x,t)=\sum\limits_{a=0}P_{a}(t)x^{a}
\end{equation*}%
and use this function in (6.17) to obtain the following Fokker--Plank-type
equation%
\begin{eqnarray}
\frac{\partial }{\partial t}G(x,t) &=&x(1-x)(k_{1}-xk_{-1})\frac{\partial
^{2}}{\partial x^{2}}G+(1-x)[k_{1}(\beta -\alpha +1)  \notag \\
&&+k_{-1}(2\alpha +\gamma +\delta -1)x]\frac{\partial }{\partial x}G  \notag
\\
&&-k_{-1}(\alpha +\gamma )(\alpha +\delta )(1-x)G(x,t)  \TCItag{6.18}
\end{eqnarray}%
This equation admits separation of variables: $G(x,t)=S(x)T(t)$ with
solution for $T(t)$ in the expected form: $T(t)=exp(-\lambda _{n}k_{1}t)$
leading to the equation for $S(x)$%
\begin{equation}
x(1-x)(1-Kx)\frac{d^{2}}{dx^{2}}S(x)+[\beta -\alpha +1+K(2\alpha +\gamma
+\delta -1)x](1-x)\frac{d}{dx}S-[K(\alpha +\gamma )(\alpha +\beta
)(1-s)-\lambda _{n}]S(x)=0  \tag{6.19}
\end{equation}%
This equation is of Lame-type discussed in Section 5 (e.g.see (5.15)) and,
therefore, its solution should be a polynomial in $x$ of degree at most $%
\varpi $ where $\varpi $ should be equal to the minimum of ($\alpha +\gamma
,\alpha +\delta ,\beta +\gamma ,\delta +\delta ).$ As in quantum \
mechanics, this implies that the spectrum of eigenvalues $\lambda _{n}$ is \
discrete, finite and nondegenerate. Among these eigenvalues there must be $%
\lambda _{0}=0$ since such an eigenvalue corresponds to the time-independent
solution of (6.19) corresponding to true equilibrium. Hence, for this case
we obtain instead of (6.19) the following final result:%
\begin{equation}
x(1-Kx)\frac{d^{2}}{dx^{2}}S(x)+[\beta -\alpha +1+K(2\alpha +\gamma +\delta
-1)x]\frac{d}{dx}S-[K(\alpha +\gamma )(\alpha +\beta )]S=0  \tag{6.20}
\end{equation}%
where $K=k_{-1}/k_{1}.$ This constant can be eliminated from (6.20) if we
rescale $x:x\rightarrow Kx.$ After this, equation acquires the standard
hypergeometric form%
\begin{equation}
x(1-x)\frac{d^{2}}{dx^{2}}S(x)+[\beta -\alpha +1+(2\alpha +\gamma +\delta
-1)x]\frac{d}{dx}S(x)-(\alpha +\gamma )(\alpha +\beta )S(x)=0.  \tag{6.21}
\end{equation}%
In [105] Kingman obtained the Fokker-Planck type equation analogous to our
(6.18) describing the dynamical peocess whose stable equilibrium is
described by (6.21) ( naturally, with different coefficients) and leads to
the P-D distribution (6.2) essential for obtaining Ewens sampling formula.
Instead of reproducing his results in this work, we would like to connect
them with results of our Section 5. \ For this purpose,we begin with the
following observation.

\subsubsection{ \ Quantum mechanics, hypergeometric\ functions and P-D
distribution}

In our works [2,3] we \ provided detailed explanation of the fact \ that all
\ exactly solvable 2-body quantum mechanical problems involve different
kinds of special functions obtainable from the Gauss hypergeometric
funcftion whose integral representation is given by 
\begin{equation}
F(a,b,c;z)=\frac{\Gamma (c)}{\Gamma (b)\Gamma (c-b)}\int%
\limits_{0}^{1}t^{b-1}(1-t)^{c-b-1}(1-zt)^{-a}dt.  \tag{6.22}
\end{equation}%
As it is well known from quantum mechanics, in the case of disctete spectrum
all quantum mechanical problems involve orthogonal polynomials.The question
then arises: under what conditions on coefficients ($a,b$ and $c$) infinite
hypergeometric series whose integral representation is given by (6.22) can
be reduced to a finite polynomial? This happens, for instance, if we impose
the \textsl{quantization condition}: $-a=0,1,2,....$ In such a case we can
write $(1-zt)^{-a}=\sum\nolimits_{i=1}^{-a}(_{i}^{-a})(-1)^{i}(zt)^{i}$ \
and use this finite expansion in (6.22). In view of (6.2) we obtain the
convergent generating function for the Dirichlet distribution (6.2). Hence, 
\textsl{all \ known quantum mechanical} \textsl{problems involving discrete
spectrum} \textsl{are effectively examples of the P-D stochasic processes}.%
\footnote{%
For hypergeometric functions of multiple arguments this was recently shown
in [106].} Next, we are interested in the following. Given this fact, can we
include the determinantal formula (5.47) into this quantization scheme? \
Very fortunately, this can be done. as explained in the next subsection..

\subsubsection{Hypergeometric functions, Kummer series expansions and
Veneziano amplitudes}

In view of just introduced quantization condition, the question arises: is
this the only condition reducing the hypergeometric function to a polynomial
? More broadly: what conditions on coefficients $a,b$ and $c$ \ should be
imposed so that the function $F(a,b,c;z)$ becomes a polynomial? The answer
to this question was provided by Kummer in the first half of 19th century
[107]. We would like to summarize his results and to connect them with
determinantal formula (5.43). By doing so we shall reobtain Veneziano
amplitudes for \ chemical process described by (6.21).

According to general theory of hypergeometric equations [107], the infinite
series for hypergeometric function degenerates to a polynomial if one of the
numbers%
\begin{equation}
a,b,c-a\text{ or \ }c-b  \tag{6.23}
\end{equation}%
is an integer. This condition is equivalent to a condition that, at least
one of eight numbers $\pm (c-1)\pm (a-b)\pm (a+b-c)$ is an odd number.
According to general theory of hypergeometric functions of multiple
arguments summarized in Section 5, the $k=1$ -type solutions can be obtained
using 1-forms (5.43) accounting for \ singular module constraint (5.45). in
the form given by equation (5.42). In the case of Gauss-type hypergeometric
functions, relations of the type given by (5.45) were obtained by Kummer who
found 24 interdependent solutions. Evidently, this number is determined by
the number of independent Pochhamer contours [107]. Therefore, among these
he singled out 6 (generating these 24) and among these 6 he established that
every 3 of them are related to each other via equation of the type (5.45).

Let us denote these 6 functions as $u_{1},...,u_{6}$ then, we can represent,
say, $u_{2}$ and $u_{6}$ using $u_{1}$ and $u_{5}$ as basis set. We can do
the same with $u_{1}$ and $u_{5}$ by representing them through $u_{2}$ and $%
u_{6}$ and, finally, we can connect $u_{3}$ and $u_{4}$ with $u_{1}$ and $%
u_{5}.$ Hence, it is sufficient to consider, say, $u_{2}$ and $u_{6}.$ We
obtain, 
\begin{equation}
\left( 
\begin{array}{c}
u_{2} \\ 
u_{6}%
\end{array}%
\right) =\left( 
\begin{array}{cc}
M_{1}^{1} & M_{2}^{1} \\ 
M_{1}^{2} & M_{2}^{2}%
\end{array}%
\right) \left( 
\begin{array}{c}
u_{1} \\ 
u_{5}%
\end{array}%
\right) ,  \tag{6.24}
\end{equation}%
with $M_{1}^{1}=\dfrac{\Gamma (a+b-c+1)\Gamma (1-c)}{\Gamma (a+1-c)\Gamma
(b-c+1)};$ $M_{2}^{1}=\dfrac{\Gamma (a+b+1-c)\Gamma (c-1)}{\Gamma (a)\Gamma
(b)};M_{1}^{2}=\dfrac{\Gamma (c+1-a-b)\Gamma (1-c)}{\Gamma (1-a)\Gamma (1-b)}%
;M_{2}^{2}=\dfrac{\Gamma (c+1-a-b)\Gamma (c-1)}{\Gamma (c-a)\Gamma (c-b)}.$
The determinant of this matrix becomes zero if either two rows or two
columns become the same. For instance, we obtain:%
\begin{equation}
\dfrac{\Gamma (a)\Gamma (b)}{\Gamma (c-1)}=\dfrac{\Gamma (a-c+1)\Gamma
(b-c+1)}{\Gamma (1-c)}\text{ and }\frac{\Gamma (c-a)\Gamma (c-b)}{\Gamma
(c-1)}=\frac{\Gamma (1-a)\Gamma (1-b)}{\Gamma (1-c)}.  \tag{6.25}
\end{equation}%
For $c=1$ we obtain an identity. From [darvey] we find that (6.21) admits 2
independent solutions:%
\begin{equation}
S(x)=\left\{ {}\right. \frac{\text{either }F(-\alpha -\gamma ,-\alpha
-\delta ,\beta -\alpha +1;Kx),\text{ for }\beta \geq \alpha \text{ }}{\text{%
or }\left( Kx\right) ^{\alpha -\beta }F(-\beta -\gamma ,-\beta -\delta
,\alpha -\beta +1;Kx),\text{ for }\beta \leq \alpha \text{ .}}  \tag{6.26}
\end{equation}%
Hence, the condition $c=1$ \ in (6.25) causes two solutions for $S(x)$ to
degenerate into one polynomial solution, provided that we make an
identification: $\beta =\alpha $ in (6.26). Notice that \ to obtain this
result there is no need to impose an extra condition: $a=b\footnote{%
Here $a$ and $b$ have the same meaning as in (6.22) and should not be
confused with concentrations.}$ (or, in our case, which is the same as $%
\gamma =\delta ).$

This makes sence physically both in chemistry and in high energy physics. In
the case of high energy physics, if the Veneziano amplitudes are used for\
description of, say, $\pi \pi $ scattering, in Part I (page 54)\ it is
demonstrated that processes for which "concentrations "$a=b$ cause this
amplitude to vanish. The Veneziano condition: $a+b+c=-1((1.5)$ of Part I)
has its analog in chemistry where it plays the same role, e.g. of mass
conservation. In the present case we have $\alpha +\beta +\gamma +\delta
=const$ and the Veneziano-like amplitude obtainable from (6.25),(6.26) is
given now by 
\begin{equation}
V_{c}(a,b)=\frac{\Gamma (-\alpha -\gamma )\Gamma (-\alpha -\delta )}{%
-c\Gamma (-c)}\mid _{c=1}  \tag{6.27}
\end{equation}%
In view of known symmetry of the hypergeometric function: $%
F(a,b,c;x)=F(b,a,c;x)$, we also have: $V_{c}(b,a)=V_{c}(a,b).$ This is
compatible with the symmetry for Veneziano amplitude. Combining (5.47) with
(6.27) we have the following options: a) $\alpha =0,\gamma =1,\delta
=1,2,...;b)$ $\alpha =1,\gamma =0,\delta =0,1,2,...$ These conditions are
compatible with those in (1.19) of Part I for Veneziano amplitudes. Finally,
in view of (6.22), these are quantization conditions for resonances as
required..

\ 

\textbf{A}$\boldsymbol{.}$ \textbf{Basics of ASEP}$\ \ $

\ 

\textbf{A.1.} \textbf{Equations of motion and spin chains}

\ 

The one dimensional asymmetric simple exclusion process (ASEP) had been
studied for some time [108]. The purpose of this Appendix is to summarize
the key features of this process which are of immediate relevance to the
content of this paper. To this purpose, following Sch\"{u}tz [109], we shall
briefly describe the ASEP with sequential updating. Let $B_{N}:=%
\{x_{1},...,x_{N}\}$ be a set of sites of one dimensional lattice arranged
at time $t$ in such a way that $x_{1}<x_{2}$\ $<\cdot \cdot \cdot <x_{N}$.
It is expected that each time update will not destroy this order.

\ Consider first the simplest case of $N=1$. Let $p_{R}$ $(p_{L})$ be the
probability of a particle located at the site $x$ to move to the right
(left) then, after transition to continuous time, the master equation for
the probability $P(x;t)$ can be written as follows%
\begin{equation}
\frac{\partial }{\partial t}P(x;t)=p_{R}P(x-1;t)+p_{L}P(x+1;t)-P(x;t). 
\tag{A.1}
\end{equation}%
Assuming that $P(x;t)=\exp (-\varepsilon t)P(x)$ so that that

$P(x;t)=\int\limits_{0}^{2\pi }dp\exp (-\varepsilon t)f(p)\exp (ipx),$ $p\in
\lbrack 0,2\pi ),$ we obtain the dispersion relation for the energy $%
\varepsilon (p):$%
\begin{equation}
\varepsilon (p)=p_{R}(1-e^{-ip})+p_{L}(1-e^{ip}).  \tag{A.2}
\end{equation}%
The initial condition $P(x;0)=\delta _{x,y}$ determines $f(p)=e^{-ipy}/2\pi $
and yields finally%
\begin{equation}
P(x;t\shortmid y;0)=\frac{1}{2\pi }\int\limits_{0}^{2\pi }dpe^{-\varepsilon
(p)t}e^{-ipy}e^{ipx}=e^{-(q+q^{-1})Dt}q^{x-y}I_{x-y}(2Dt),  \tag{A.3}
\end{equation}%
where $q=\sqrt{p_{R}/p_{L}}$, $D=\sqrt{p_{R}p_{L}}$ and $I_{n}(2Dt)$ is the
modified Bessel function. These results can be easily extended to the case $%
N=2$. Indeed, for this case \ we obtain the following equation of motion%
\begin{eqnarray}
\varepsilon P(x_{1},x_{2})
&=&-p_{R}(P(x_{1}-1,x_{2})+P(x_{1},x_{2}-1)-2P(x_{1},x_{2}))  \notag \\
&&-p_{L}(P(x_{1}+1,x_{2})+P(x_{1},x_{2}+1)-2P(x_{1},x_{2}))  \TCItag{A.4}
\end{eqnarray}%
which should be suppllemented by the boundary condition%
\begin{equation}
P(x,x+1)=p_{R}P(x,x)+p_{L}P(x+1,x+1)\text{ }\forall x.  \tag{A.5}
\end{equation}%
Imposition of this boundary condition allows us to look for a solution of
(A.4) in the\ (Bethe ansatz) form%
\begin{equation}
P(x_{1},x_{2})=A_{12}e^{ip_{1}x_{1}}e^{ip_{2}x_{2}}+A_{21}e^{ip_{2}x_{1}}e^{ip_{1}x_{2}}
\tag{A.6}
\end{equation}%
yielding $\varepsilon (p_{1},p_{2})=\varepsilon (p_{1})+\varepsilon (p_{2}).$
Use of the boundary condition (A.5) fixes the ratio (the S-matrix) $%
S_{12}=A_{12}/A_{21}$ as follows:%
\begin{equation}
S(p_{1},p_{2})=-\frac{p_{R}+p_{L}e^{ip_{1}+ip_{2}}-e^{ip_{1}}}{%
p_{R}+p_{L}e^{ip_{1}+ip_{2}}-e^{ip_{2}}}.  \tag{A.7}
\end{equation}%
To connect this result with the quantum spin chains,\ we consider the case
of symmetric hopping first. In this case we have $p_{R}=p_{L}=1/2$ so that
(A.7) is reduced to 
\begin{equation}
S_{XXX}(p_{1},p_{2})=-\frac{1+e^{ip_{1}+ip_{2}}-2e^{ip_{1}}}{%
1+e^{ip_{1}+ip_{2}}-2e^{ip_{2}}}  \tag{A.8}
\end{equation}%
from which we can recognize the $S$ matrix for XXX spin 1/2 Heisenberg
ferromagnet [67]. If $p_{R}\neq p_{L}$, to bring (A.7) in correspondence
with the spin chain $S-$ matrix requires additional efforts. Following Gwa
and Spohn [110] we replace the complex numbers $e^{ip_{1}}$ and $e^{ip_{2}}$
in (A.7) respectively by z$_{1}$ and z$_{2}.$ In such a form \ (A.7) exactly
\ coincides with the S-matrix \ obtained by Gwa and Spohn\footnote{%
E.g. see their equation (3.5).}. After this we can rescale z$_{i}$ $(i=1,2)$
as follows: $z_{i}=\sqrt{\frac{q}{p}}\tilde{z}_{i}.$ Substitution of such an
asatz into (A.7) leads to the result 
\begin{equation}
S_{XXZ}(\tilde{z}_{1},\tilde{z}_{2})=-\frac{1+\tilde{z}_{1}\tilde{z}%
_{2}-2\Delta \tilde{z}_{1}}{1+\tilde{z}_{1}\tilde{z}_{2}-2\Delta \tilde{z}%
_{2}},  \tag{A.9}
\end{equation}%
provided that $2\Delta =1/\sqrt{p_{L}p_{R}}$. For $p_{R}=p_{L}=1/2$ we
obtain $\Delta =1$ as required for the XXX chain.\footnote{%
It should be noted though that such a parametrization is not unique. For
instance, following [56] it is possible to choose a slightly different
parametrization, e.g. $\Delta =-\frac{1}{2}(q+q^{-1}),$ where $q=\sqrt{%
p_{R}/p_{L}}.$} If, however, $p_{R}\neq p_{L},$ then, the obtained $S$%
-matrix \ coincides with that known for the XXZ model [goden] if we again
relabel \~{z}$_{i}$ by $e^{ip_{i}}$ which is always permissible since the
parameter $p$ is determined by the Bethe equations (to be discussed below)
anyway.

In the case of XXZ spin chain it is customary to think about the massless $%
-1\leq \Delta \leq 1$ and massive $\left\vert \Delta \right\vert >1$ regime.
The massless regime describes various CFT discussed in the text while the
massive regime describes massive excitations away from criticality. As
Gaudin had demonstrated [67], for XXZ chain it is sufficient to consider
only $\Delta >0$ domain \ which makes XXZ model perfect for uses in ASEP.
The cases $\Delta =0$ and $\Delta \rightarrow \infty $ also physically
interesting: the first corresponds to the XY model and the second to the
Ising model.

Once the S-matrix is found, \ the $N-$ particle solution can be easily
constructed [109]. For instance, for N=3 we write%
\begin{eqnarray*}
\Psi (x_{1},x_{2},x_{3}) &=&\exp
(ip_{1}x_{1}+ip_{2}x_{2}+ip_{3}x_{3})+S_{21}\exp
(ip_{2}x_{1}+ip_{1}x_{2}+ip_{3}x_{3}) \\
&&+S_{32}S_{31}\exp
(ip_{2}x_{1}+ip_{3}x_{2}+ip_{1}x_{3})+S_{21}S_{31}S_{32}\exp
(ip_{3}x_{1}+ip_{2}x_{2}+ip_{3}x_{3}) \\
&&+S_{31}S_{32}\exp (ip_{3}x_{1}+ip_{1}x_{2}+ip_{2}x_{3})+S_{32}\exp
(ip_{1}x_{1}+ip_{3}x_{2}+ip_{2}x_{3}),
\end{eqnarray*}%
etc. This result is used instead of $f(p)$ in (A.3) so that the full
solution is given by 
\begin{equation}
P(x_{1},...,x_{N};t\shortmid y_{1},...,y_{N};0)=\prod\limits_{l=1}^{N}\frac{1%
}{2\pi }\int\limits_{0}^{2\pi }dp_{l}e^{-\varepsilon
(p_{l})t}e^{-ip_{l}y_{l}}\Psi (x_{1},...,x_{N}).  \tag{A.11}
\end{equation}%
The abobe picture should be refined as follows. First, the particle sitting
at $x_{i}$ will move to the right(left) only if the nearby site is not
occupied. Hence, the probabilities $p_{R}$ and $p_{L}$ can have values
ranging from 0 to 1. For instance, for the totally asymmetric exclusion
process (TASEP) particle can move to the right with probability 1 if the
neighboring site to its right is empty. Othervice the move is rejected. \
Since under such circumstances particle can never move to the left, there
must be a particle source located next to the leftmost particle position and
the particle sink located immediately after the rightmost position. After
imposition of emission and absorption rates for these sources and sinks, we
end up with the Bethe ansatz complicated by the imposed boundary conditions.
Although in the case of solid state physics \ these conditions are normally
assumed to be periodic, in the present case, they should be chosen among the
solutions of the Sklyanin \ boundary equation [45, 58]. At more intuitive
level of presentation compatible with results just discussed, the Bethe
ansatz for XXZ chain accounting for the boundary effects is given in the
pedagogically written paper by Alcaraz et al [111].

\bigskip\ 

\textbf{A.2.} \textbf{Dynamics of ASEP and operator algebra }

\ 

To make these results useful for the main text, few additional steps are
needed. For this purpose we shall follow works by Sasamoto and Wadati [112]
and Stinchcombe and Sch\"{u}tz [45]. In doing so we rederive many of their
results differently.

We begin with observation that the state of one dimensional lattice
containing $N$ sites can be described in terms of a string of operators $D$
and $E,$ \textbf{\ }where $D$ stands for the occupied and $E$ for empty $k-$%
th position along the $1d$ lattice. The non normalized probability (of the
type given in (A.11)) can then be presented as a sum of terms like this $\
<EEDEDDD\cdot \cdot \cdot E>$ \ to be discussed in more details below.

Let $C=D+E$ be the time- independent\textbf{\ }operator. Then, for the
operator $D$ to be time-dependent the following commutation relations should
hold%
\begin{equation}
SC+\dot{D}C=\Lambda ,  \tag{A.12a}
\end{equation}%
\begin{equation}
CS-C\dot{D}=\Lambda ,  \tag{A.12b}
\end{equation}%
\begin{equation}
\dot{D}D+D\dot{D}=[D,S].  \tag{A.12c}
\end{equation}%
If $\Lambda =p_{L}CD$\textbf{\ -}$p_{R}DC$\textbf{\ +}$\mathbf{(}%
p_{R}-p_{L})D^{2}$ \ (or $\Lambda =p_{L\text{ }}ED-p_{R}DE$ ),\ it is
possible to determine $S$ using equations (A.12) so that we obtain,%
\begin{equation}
\dot{D}=\frac{1}{2}[\Lambda ,C^{-1}],  \tag{A.13a}
\end{equation}%
\textbf{\ }%
\begin{equation}
S=\frac{1}{2}\{\Lambda ,C^{-1}\},  \tag{A.13b}
\end{equation}%
provided that 
\begin{equation}
\Lambda C^{-1}D=DC^{-1}\Lambda  \tag{A.13c}
\end{equation}%
with \{ , \} being an anticommutator.

As before, let us consider the case $p_{R}=p_{L}$\textbf{\ }$\mathbf{=}\frac{%
1}{2}$ first$.$ This condition leads to $\Lambda =\frac{1}{2}[C,D]$. It is
convenient at this stage to introduce an operator $D_{n}$\textbf{\ =}$%
C^{n-1} $\textbf{\ }$DC^{-n}$\textbf{\ \ }and its Fourier\textbf{\ }%
transform $\mathcal{D}_{p}$\textbf{\ =}$\sum\nolimits_{n}D_{n}\exp (ipn).$
Using (A.13a) with $\Lambda $ just defined leads to the following equation
of motion for $\mathit{D}_{n}:$%
\begin{equation}
\dot{D}_{n}=\frac{1}{2}[D_{n+1}+D_{n-1}-2D_{n}]  \tag{A.14}
\end{equation}%
to be compared with (A.1). Such a comparison produces at once $\mathcal{D}%
_{p}(t)=\exp (-\varepsilon (p))\mathcal{D}_{p}(0)$\textbf{\ }so that \textbf{%
\ }$\varepsilon (p)=1-\cos p$ as before\footnote{%
E.g. see (A.2) with $p_{L}=p_{R}=1/2$}. Consider now equation (A.13c). Under
conditions $p_{R}=p_{L}$\textbf{\ }$\mathbf{=}\frac{1}{2}$ it can be written
as

$CDC^{-1}D\mathbf{\ +}DC^{-1}DC\mathbf{\ =}2D^{2}$\textbf{\ }or,\textbf{\ }%
equivalently, as\footnote{%
Since using definition of $D_{n}$ we have: $D_{n+1}=CD_{n}C^{-1}$ and $%
D_{n-1}=C^{-1}D_{n}C.$}\textbf{\ }%
\begin{equation}
D_{n+1}D_{n}+D_{n}D_{n-1}=2D_{n}D_{n}.  \tag{A.15}
\end{equation}%
Following Gaudin [67], we consider a formal expansion $D_{n}D_{m}$\ \ =$%
\alpha \exp (ip_{1}n+ip_{2}m)+\beta $\ $\exp (ip_{2}n+ip_{1}m)$\ \ and use
it in the previous equation in order to obtain:%
\begin{eqnarray}
&&\alpha \exp (ip_{1}(n+1)+ip_{2}n)+\beta \ \exp (ip_{2}(n+1)+ip_{1}n) 
\notag \\
&&+\alpha \exp (ip_{1}n+ip_{2}(n-1))+\beta \ \exp (ip_{2}n+ip_{1}(n-1)) 
\notag \\
&=&2\alpha \exp (ip_{1}n+ip_{2}n)+2\beta \ \exp (ip_{2}n+ip_{1}n). 
\TCItag{A.16}
\end{eqnarray}%
From here we \ also obtain:\ \ \ 
\begin{equation*}
(\alpha \exp (ip_{1}n+ip_{2}n)(\exp (ip_{1})+\exp (-ip_{2})-2)+\beta \exp
(ip_{1}n+ip_{2}n)(\exp (ip_{2})+\exp (-ip_{1})-2)=0
\end{equation*}%
and, therefore, 
\begin{equation}
S(p_{1},p_{2})\equiv \frac{\alpha }{\beta }=-\frac{1+\exp
(i(p_{1}+p_{2}))-2\exp (ip_{1})}{1+\exp (i(p_{1}+p_{2}))-2\exp (ip_{2})}\exp
(i(p_{2}-p_{1}))  \tag{A.17}
\end{equation}%
to be compared with (A.8).\ An extra factor $\exp (i(p_{2}-p_{1}))$ can be
actually dropped from the $S-$matrix in view of the following chain of
arguments.

\ Introduce the correlation function as follows%
\begin{eqnarray}
P(x_{1},...,x_{N};t &\shortmid &y_{1},...,y_{N};0)\equiv Z_{N}^{-1}Tr[%
\mathit{D}_{1}(t)\cdot \cdot \cdot \mathit{D}_{N}(t)C^{N}]  \notag \\
&=&\prod\limits_{l=1}^{N}\frac{1}{2\pi }\int\limits_{0}^{2\pi
}dp_{l}e^{-\varepsilon (p_{l})t}e^{-ip_{l}y_{l}}\Psi (p_{1},...,p_{N}), 
\TCItag{A.18}
\end{eqnarray}%
\ where $\Psi (p_{1},...,p_{N})=$\ $Z_{N}^{-1}Tr[\mathcal{D}_{p_{1}}(0)\cdot
\cdot \cdot \mathcal{D}_{p_{N}}(t)C^{N}]$ and $Z_{N}=tr[C^{N}].$ \ In
arriving at this result the definition of $\mathcal{D}_{p}(t),$ was used
along with\ the fact that $\mathcal{CD}_{p}\mathcal{C}^{-1}=\ e^{-ip}$\ $%
\mathcal{D}_{p}$. Also, the invariance of the trace under cyclic
permutations and the translational invariance of the correlation function
implying that $\Psi (p_{1},...,$ $p_{N})\neq 0$ only if $\sum\nolimits_{i}$ $%
p_{i}$\ \ $=0$ \ was taken into account. These conditions are sufficient for
obtaining the \ Bethe ansatz equations \ 
\begin{equation}
\exp (ip_{i}N)=\prod\limits_{j=1}^{N}\tilde{S}(p_{i},p_{j})\text{ }\forall
i\neq j,  \tag{A.19}
\end{equation}%
\ where $\tilde{S}(p_{i},p_{j})$ is the same $S-$matrix as in (A.17), except
of the missing factor $\exp (i(p_{i}-p_{j}))$\ which is dropped in view of
translational invariance\footnote{%
Surely, in case when the effects of boundaries should be accounted, this
factor should be treated depending on the kind of boundary conditions
imposed.}.\ \ 

Extension of these results to the case $p_{R}\neq p_{L}$ is nontrivial.
Because of this, we would like to provide some details not shown in the
cited references. In particular, contary to claims made in [112], we would
like to demonstrate that the system of equations (A.12) obtained in [45] is
equivalent to the system of equations 
\begin{equation}
\lbrack C,S]=0,  \tag{A.20a}
\end{equation}%
\begin{equation}
C\dot{D}+CT-SD=-p_{L}CD+p_{R}DC,  \tag{A.20b}
\end{equation}%
\begin{equation}
\dot{D}C+DS-TC=p_{L}CD-p_{R}DC,  \tag{A.20c}
\end{equation}%
\begin{equation}
\dot{D}D+D\dot{D}=[T,D]  \tag{A.20d}
\end{equation}%
obtained in [112] with the purpose of describing asymmetric processes.

To make a comparison between (A.12) and (A.20) we notice that (A.20) has
operators $S$ and $T$ \ which cannot be trivially identified with those
present in (A.12). Hence, the task lies in making such an identification.
For this purpose if we assume that $S$ in (A.20) is the same as in (A.12)
then, \ in view of (A.20a), by subtracting (A.12b) from (A.12a) we obtain: 
\begin{equation}
\dot{D}C+C\dot{D}=0.  \tag{A.21}
\end{equation}%
This leads to either $\dot{D}=C^{-1}\dot{D}C$ or $\dot{D}=-C\dot{D}C^{-1}.$
Therefore, taking into account that, by construction, $C$ is
time-independent, we obtain: $D=$ $-CDC^{-1}+\Theta ,$ where $\Theta $ is
some diagonal time-independent matrix operator.

Next, using these results we multiply (A.20b) from the right by $C^{-1}$ and
(A.20c) by $C^{-1}$ from the left, and add them together in order to arrvie
at equation (19) of [sasam], i.e.

\bigskip 
\begin{equation}
2\dot{D}=p_{R}C^{-1}DC+p_{L}CDC^{-1}-(p_{R}+p_{L})D.  \tag{A.22}
\end{equation}%
Also, by multiplying this result from the right by $D$ we obtain equation
(20) of [112], that is 
\begin{equation}
0=p_{R}DC^{-1}DC+p_{L}CDC^{-1}D-(p_{R}+p_{L})D^{2},  \tag{A.23}
\end{equation}%
provided that $[T,D]=0.$ \ That this is indeed the case can be seen from the
same reference where the following result for $T$ is obtained: 
\begin{equation}
2T=(2+p_{R}-p_{L})D+p_{R}C^{-1}DC-p_{L}CDC^{-1}.  \tag{A.24}
\end{equation}%
Using it, we obtain: $[T,D]=0,$ in view of the fact that $[C^{-1}DC,D]=0$
and $[CDC^{-1},D]=0$ since $D=-CDC^{-1}+\Theta $ as we have\ already
demonstrated. Furthermore, (A.24) can be straightforwardly obtained by
subtracting (A.20c) (multiplyed by $C^{-1}$ from the rihgt) from (A.20a)
(multiplyed by C$^{-1}$ from the left). Thus, \ contary to the claims made
in [112], equations (A.12) and (A.20) are, in fact, equivalent.
Nevertheless, as claimed in [112], the system of equations (A.20) is easier
to connect with the Bethe ansatz formalism.

Indeed, using already known fact that $D_{n}$\textbf{\ =}$C^{n-1}DC^{-n}$
equation (A.22) can be brought into the form: 
\begin{equation}
\dot{D}_{n}=\frac{1}{2}[p_{R}D_{n+1}+p_{L}D_{n-1}-(p_{R}+p_{L})D_{n}]. 
\tag{A.25}
\end{equation}%
This result is formally in agreement with previously obtained (A.14) for the
fully symmetric case. The authors of [112] have chosen such a normalization
for probabilities $p_{R}$ and $p_{L}$ that for symmetric case $p_{R}=p_{L}=1$
(instead of $p_{R}=p_{L}=1/2).$ To restore the normally accepted condition $%
p_{R}=p_{L}=1/2$ requires only to rescale time appropriately. \ This
observation is consitent with the fact that the analog of equation (A.15)
(which plays the central role in the Bethe ansatz-type calculations)
obtained with help of (A.23) is given by 
\begin{equation}
p_{L}D_{n+1}D_{n}+p_{R}D_{n}D_{n-1}=(p_{R}+p_{L})D_{n}D_{n}  \tag{A.26}
\end{equation}%
which holds true wether we choose $p_{R}=p_{L}=1$ or $p_{R}=p_{L}=1/2%
\footnote{%
It should be noted though that the authors of [112] have erroneously
obtained (e.g. see their equation (23)) \ $%
p_{R}D_{n}^{2}+p_{L}D_{n+1}^{2}=(p_{R}+p_{L})D_{n}D_{n+1}$ instead of our
(A.26).}.$ Obtained results allow us to reobtain the $S-$matrix for the XXZ
model in a way \ already described.

\ 

\textbf{A.3.} \textbf{Steady- state and q- algebra for the deformed harmonic
oscillator}

\ 

Using (4.55) we have%
\begin{equation}
p_{R}DE-p_{L}ED=\zeta \left( D+E\right)  \tag{A.27}
\end{equation}%
Let now $D=A_{1}+B_{1}a$ and $E=A_{2}+B_{2}a^{+}$ where $A_{i}$ and $B_{i}$, 
$i=1,2,$ are some c-numbers. Substituting these expressions back to (A.27)
we obtain the following set of equations%
\begin{equation}
\zeta (A_{1}+A_{2})-\varepsilon A_{1}A_{2}=C,  \tag{A.28a}
\end{equation}%
where $C$ is some constant to be determined below, and%
\begin{equation}
\zeta B_{1}=\varepsilon B_{1}A_{2},  \tag{A.28b}
\end{equation}%
\begin{equation}
\zeta B_{2}=\varepsilon B_{2}A_{1}.  \tag{A.28c}
\end{equation}%
From here we obtain: $A_{1}=A_{2}=A=\zeta /\varepsilon ,$ with $B_{1},$ $%
B_{2}$ being yet arbitrary c-numbers and $\varepsilon =p_{R}-p_{L}$. \ We
can determine these numbers by comparing our results with those in [47].
This allows us to select $B_{1}=B_{2}=\frac{\xi }{\sqrt{1-q}},$ $\dfrac{%
\zeta ^{2}}{\varepsilon }=\frac{\xi ^{2}}{1-q}=C,$ $q=\dfrac{p_{L}}{p_{R}}$
so that we obtain:%
\begin{equation}
D=\frac{1}{1-q}+\frac{1}{\sqrt{1-q}}a,  \tag{A.29a}
\end{equation}%
\begin{equation}
E=\frac{1}{1-q}+\frac{1}{\sqrt{1-q}}a^{+}  \tag{A.29b}
\end{equation}%
and, finally,%
\begin{equation}
aa^{+}-qa^{+}a=1  \tag{A.29c}
\end{equation}%
in accord with (4.28d).

\ \ 

\bigskip\ \textbf{B.Linear independence of solutions of K-Z equation}

Linear independence of solutions of K-Z equation is based on the following
arguments. Consider change of the basis%
\begin{equation}
\mathbf{\tilde{e}}^{j}=A_{i}^{j}\mathbf{e}^{i}\text{ , }i,j=1,2,...,n 
\tag{B.1}
\end{equation}%
in $\mathbf{R}^{n}$. Using this result, consider the exterior product%
\begin{equation}
\mathbf{\tilde{e}}^{1}\wedge \cdot \cdot \cdot \wedge \mathbf{\tilde{e}}%
^{n}=[\det \mathbf{A]e}^{1}\wedge \cdot \cdot \cdot \wedge \mathbf{e}^{n}. 
\tag{B.2}
\end{equation}%
Next, suppose, that \ the vectors $\mathbf{\tilde{e}}^{j}$ are lineraly
-dependent. In particular,this means that 
\begin{equation}
\mathbf{\tilde{e}}^{n}=\alpha _{1}\mathbf{\tilde{e}}^{1}+\cdot \cdot \cdot
+\alpha _{n-1}\mathbf{\tilde{e}}^{n-1}  \tag{B.3}
\end{equation}%
for some nonzero $\alpha _{i}^{\prime }s.$\ Using this expansion in (B.2) we
obtain%
\begin{equation}
\mathbf{\tilde{e}}^{1}\wedge \cdot \cdot \cdot \wedge \mathbf{\tilde{e}}%
^{n-1}\wedge (\alpha _{1}\mathbf{\tilde{e}}^{1}+\cdot \cdot \cdot +\alpha
_{n-1}\mathbf{\tilde{e}}^{n-1})\equiv 0  \tag{B.4}
\end{equation}%
implying $[\det \mathbf{A]=}0.$ Convesely, if $[\det \mathbf{A]\neq }0$%
\textbf{\ }then, vectors $\mathbf{\tilde{e}}^{j}$ \ are linearly independent.

\ 

\bigskip \textbf{C}. \textbf{Connections between the gamma and Dirichlet
distributions}

\bigskip

Using results of our Part I, especially, equation (3.27), such a connection
can be easily established. Indeed, consider $n+1$ independently \
distributed \ random gamma variables with exponents $\alpha _{1},...,\alpha
_{n+1}$. The joint probability density for such variables is given by 
\begin{equation}
p_{Y_{1}},..._{Y_{n+1}}(s_{1},...,s_{n+1})=\frac{1}{\Gamma (\alpha _{1})}%
\cdot \cdot \cdot \frac{1}{\Gamma (\alpha _{n+1})}s_{1}^{\alpha _{1}-1}\cdot
\cdot \cdot s_{n+1}^{\alpha _{n+1}-1}.  \tag{C.1}
\end{equation}%
Let now $s_{i}=t_{i}t,$ \ where $t_{i}$ are chosen in such a way that $%
\sum\nolimits_{n=1}^{n+1}t_{i}=1.$ Then, using such a substitution into
(C.1) we obtain at once: 
\begin{eqnarray}
p_{u_{1}},..._{u_{n+1}}(t_{1},...,t_{n+1}) &=&[\int\limits_{0}^{\infty
}t^{\alpha -1}e^{-t}]\frac{1}{\Gamma (\alpha _{1})}\cdot \cdot \cdot \frac{1%
}{\Gamma (\alpha _{n+1})}t_{1}^{\alpha _{1}-1}\cdot \cdot \cdot
t_{n+1}^{\alpha _{n+1}-1}\text{ provided that }1  \notag \\
&=&\sum\nolimits_{n=1}^{n+1}t_{i}.  \TCItag{C.2}
\end{eqnarray}%
Since $\alpha =\alpha _{1}+\cdot \cdot \cdot +\alpha _{n+1}$, we obtain: $%
\int\limits_{0}^{\infty }t^{\alpha -1}e^{-t}=\Gamma (\alpha _{1}+\cdot \cdot
\cdot +\alpha _{n+1})$ so that the density of probability (C.2) is indeed of
Dirichlet-type\ given by (6.2).

\textbf{D. Some facts from combinatorics of the symmetric group }$S_{n}$

\textbf{\bigskip }Suppose we have a finite set X. $\forall x\in \mathbf{X}$
consider a bijection $\mathbf{X}\longrightarrow $X made of some permutation
sequence: $x,$ $\pi (x),\pi ^{2}(x),...$ \ Because the set is finite, we
must have $\pi ^{m}(x)=x$ for some $m\geq 1.$ A sequence $(x,$ $\pi (x),\pi
^{2}(x),..,\pi ^{m-1}(x))=C_{m}$ is called a \textsl{cycle of length m}. The
set $\mathbf{X}$ can be subdivided into \textsl{disjoint} product of cycles
so that any permutation $\pi $ is just a product of these cycles. Normally
such a product is not uniquely defined. To make it uniquely defined, we have
to assume that the set \textbf{X} is ordered according to a certain rule.
The, standard cycle representation can be constructed by requiring that a)
each cycle is written with its largest element first, and b) the cycles are
written in increasing order of their respective largest elements. Let $N$ be
some integer and consider a decomposition of $N$ as $N=\sum%
\nolimits_{i=1}^{K}n_{i}$ . We say that $\ \mathit{n\equiv }%
(n_{0},...,n_{K}) $ is \textsl{partition} of $N$ (or $n\vdash N)$ \ The same
result can be achieved if, instead we would consider the following
decomposition of $N$: $N=\sum\nolimits_{i=1}^{N}ic_{i\text{ }}$ where,
according to our conventions, we have $c_{i}\equiv c_{i}(\pi )$ is the
number of cycles of length $i$. The total number of cycles then is given by $%
K$=$\sum\nolimits_{i=1}^{N}c_{i\text{ }}.$ Define a number $\tilde{S}(N,K)$
as the number of permutations of \textbf{X} with exactly $K$ cycles. Then,
the \textsl{Stirling} number of the \textsl{first }kind can be defined as $%
S(N,K):=(-1)^{N-K}\tilde{S}(N,K).$ The numbers $\tilde{S}(N,K)$ can be
obtained recursively using the following recurrence relation%
\begin{equation}
\tilde{S}(N,K)=(N-1)\tilde{S}(N-1,K)+\tilde{S}(N-1,K-1),\text{ }N,K\geq 1;%
\text{ }\tilde{S}(0,0)=1.  \tag{D.1}
\end{equation}%
Use of this recurrence allows us to obtain the following important result%
\begin{equation}
\sum\limits_{K=0}^{N}\tilde{S}(N,K)x^{K}=x(x+1)(x+2)\cdot \cdot \cdot
(x+N-1).  \tag{D.2}
\end{equation}%
Let now $x=1$ in $(D.2),$ then we can define the probability $p(K;N)=\tilde{S%
}(N,K)/N!$ Furthermore, one can define yet another probability by
introducing a notation $[x]^{N}=x(x+1)(x+2)\cdot \cdot \cdot (x+N-1).$ Then,
we obtain: 
\begin{equation}
\sum\limits_{K=0}^{N}\tilde{S}(N,K)\frac{x^{K}}{[x]^{N}}=\sum%
\limits_{K=0}^{N}P_{K}(N;x)=1.  \tag{D.3}
\end{equation}%
Such defined probability $P_{K}(N;x)$ can be further rewritten in view of
the famous result by Cauchy. To obtain his result, we introduce the
generating function%
\begin{equation}
\mathcal{F}_{K}^{N}(\mathbf{x)=}\sum\limits_{_{\substack{ %
K=\sum\nolimits_{i=1}^{N}c_{i\text{ }}  \\ N=\sum\nolimits_{i=1}^{N}ic_{i%
\text{ }}}}}\frac{N!}{1^{c_{1}}c_{1}!2^{c_{2}}c_{2}!\cdot \cdot \cdot
N^{c_{N}}c_{N}!}x_{1}^{c_{1}}\cdot \cdot \cdot x_{N}^{c_{N}}  \tag{D.4a}
\end{equation}%
and require that $\tilde{S}(N,K)\mathbf{x}^{K}=\mathcal{F}_{K}^{N}(\mathbf{x)%
}$. This can happen only if 
\begin{equation}
\sum\limits_{K=0}^{N}\sum\limits_{_{\substack{ K=\sum\nolimits_{i=1}^{N}c_{i%
\text{ }}  \\ N=\sum\nolimits_{i=1}^{N}ic_{i\text{ }}}}}\frac{N!}{%
1^{c_{1}}c_{1}!2^{c_{2}}c_{2}!\cdot \cdot \cdot N^{c_{N}}c_{N}!}=1 
\tag{D.4b}
\end{equation}%
Thus, we obtain 
\begin{equation}
\tilde{S}(N,K)=\prod\limits_{i=1}^{K}\frac{N!}{i^{c_{i}}c_{i}!}\text{ , \
provided that }K=\sum\nolimits_{i=1}^{N}c_{i\text{ }}\text{ and }%
N=\sum\nolimits_{i=1}^{N}ic.  \tag{D.5}
\end{equation}%
In these notations the Ewens sampling formula acquires the following
canonical form%
\begin{equation}
P_{K}(N;x)\equiv \frac{x^{K}}{[x]^{N}}\prod\limits_{i=1}^{K}\frac{N!}{%
i^{c_{i}}c_{i}!}\text{ \ provided that }K=\sum\nolimits_{i=1}^{N}c_{i\text{ }%
}\text{ and }N=\sum\nolimits_{i=1}^{N}ic.  \tag{D.6}
\end{equation}

\bigskip

\bigskip

\textbf{References}

\bigskip

[1] \ \ P.Dirac, \textit{Lectures on quantum field theory},

\ \ \ \ \ \ Yeshiva University Press (1996).

[2] \ \ A.Kholodenko, \textit{Heisenberg honeycombs solve Veneziano puzzle},

\ \ \ \ \ \ \ hep-th/0608117.

[3] \ \ A.Kholodenko, \textit{Quantum signatures of Solar System dynamics},

\ \ \ \ \ \ arXiv.0707.3992.

[4] \ \ A.Kholodenko, \textit{New strings for old Veneziano amplitudes I.}

\ \ \ \ \ \ \textit{Analytical treatment}, J.Geom.Phys.\textbf{55} (2005) 50.

[5] \ \ A.Kholodenko, \textit{New strings for old Veneziano amplitudes II.}

\ \ \ \ \ \ \textit{Group-theoretic treatment, }J.Geom.Phys.\textbf{56}
(2006)1387.

[6] \ \ A.Kholodenko, \textit{New strings for old Veneziano amplitudes III.}

\ \ \ \ \ \ \ \textit{Symplectic treatment,} J.Geom.Phys.\textbf{56} 92006)
1433.

[7] \ \ \ N.Reshetikhin and A.Varchenko, \textit{Quasiclassical asymptotics
of }

\ \ \ \ \ \ \ \textit{solutions of KZ equations}, in Geometry, topology and
physics for

\ \ \ \ \ \ \ Raoul Bott, p. 293, International Press (1995).

[8] \ \ \ A.Varchenko, \textit{Special functions, KZ type equations, and }

\ \ \ \ \ \ \ \textit{representation theory}, AMS Publishers (2003).

[9] \ \ J.Bertoin, \textit{Random fragmentation and coagulation processes},

\ \ \ \ \ \ \ Cambridge University Press, Cambridge U.K. (2006).

[10] \ J.Pitman, Combinatorial stochasic processes,

\ \ \ \ \ \ \ Springer-Verlag, Berlin (2006).

[11] \ R. Arratia, A.Barbour and S.Tavare, \textit{Logarithmic combinatorial 
}

\textit{\ \ \ \ \ \ structures: a probabilistic approach, }European
Mathematical Society,

\ \ \ \ \ \ Z\"{u}rich (2003).

[12] \ A.Mekjian, \textit{Model for studing brancing processes, multiplicity}

\ \ \ \ \ \ \ \textit{distribution and non-Poissonian fluctuations in heavy
-ion collisions,}

\ \ \ \ \ \ \ PRL \textbf{86} (2001) 220.

[13] \ R.Stanley, \textit{Combinatorics and commutative algebra}, Birkh\"{a}%
user,

\ \ \ \ \ \ \ Boston (1996).

[14] \ J. Kingman, \textit{Poisson processes}, Clarendon Press, Oxford
(1993).

[15] \ S.Ghorpade and G.Lachaud, \textit{Hyperplane sections of Grassmannians%
}

\ \ \ \ \ \ \textit{and the number of MDS linear codes}, Finite Fields \&

\ \ \ \ \ \ Their Applications \textbf{7 (}2001) 468\textbf{.}

[16] \ R.Stanley, \textit{Enumerative combinatorics}, Vol.1,

\ \ \ \ \ \ \ Cambridge University Press, Cambridge, U.K. (1999).

[17] \ S.Mohanty, \textit{Lattice path counting and applications},

\ \ \ \ \ \ \ Academic Press, New York (1979).

[18] \ R.Bott and L.Tu, \textit{Differential forms in algebraic topology},

\ \ \ \ \ \ \ Springer-Verlag, Berlin (1982).

[19] \ J.Schwartz, \textit{Differential geometry and topology}, Gordon

\ \ \ \ \ \ \ and Breach, Inc., New York (1968).

[20] \ \ M.Stone, \textit{Supersymmetry and quantum mechanics of spin},

\ \ \ \ \ \ \ Nucl.Phys. \textbf{B 314} (1989) 557.

[21] \ O.Alvarez, I Singer and P.Windey, \textit{Quantum mechanics and}

\ \ \ \ \ \ \ \ \textit{the\ geometry of the Weyl character formula,}

\ \ \ \ \ \ \ \ Nucl.Phys.\textbf{B337} (1990) 467.

[22] \ \ A.Polyakov, \textit{Gauge fields and strings},

\ \ \ \ \ \ \ \ Harwood Academic Publ., New York (1987).

[23]. \ A.Polychronakos, \textit{Exact spectrum of SU(n) spin chain with}

\ \ \ \ \ \ \ \textit{\ inverse square exchange, }Nucl.Phys. \textbf{B419}
(1994) 553.

[24] \ \ H. Frahm, \textit{Spectrum of a spin chain with invese square
exchange},

\ \ \ \ \ \ \ \ J.Phys.\textbf{A 26} (1993) L473.

[25] \ \ \ A.Polychronakos, \textit{Generalized statistics in one dimension},

\ \ \ \ \ \ \ \ \ hep-th/9902157.

[26] \ \ \ A.Polychronakos, \textit{Physics and mathematics of Calogero
particles},

\ \ \ \ \ \ \ \ \ hep-th/0607033.

[27] \ \ \ K.Hikami, \textit{Yangian symmetry and Virasoro character in a
lattice}

\ \ \ \ \ \ \textit{\ \ spin }\ \textit{system with long -range interactions}%
,

\ \ \ \ \ \ \ \ Nucl.Phys.B 441 (1995) 530.

[28] \ \ E.Melzer, \textit{The many faces of a character,} hep-th/9312043.

[29] \ \ R.Kedem, B.McCoy and E.Melzer, \textit{The sums of Rogers, Schur
and }

\ \ \ \ \ \ \ \ \textit{Ramanujian and the Bose-Fermi correspondemnce in 1+1 
}

\ \ \ \ \ \ \ \ \textit{dimensional quantum field theory, hep-th/9304056.}

[30] \ \ A.Tsvelik, Quantum field theory in condensed matter physics,

\ \ \ \ \ \ \ \ Cambridge University Press, Cambridge U.K. (2003).

[31] \ \ J.Goldman and J.C.Rota, \textit{The number of subspaces of a vector
space,}

\ \ \ \ \ \ \ \ \ in Recent Progress in Conbinatorics,

\ \ \ \ \ \ \ \ p.75, Academic Press, New York (1969).

[32] \ \ \ V.Kac and P.Cheung, \textit{Quantum calculus},
Springer-Verlag,Berlin (2002).

[33] \ \ G.Andrews, \textit{The theory of partitions},

\ \ \ \ \ \ \ \ Addison-Wesley Publ.Co., London (1976).

[34] \ \ D.Galetti, \textit{Realization of the q-deformed harmonic oscillator%
}:

\ \ \ \ \ \ \ \ \textit{Rogers-Szego and Stiltjes-Wiegert polynomials},

\ \ \ \ \ \ \ \ Brazilian Journal of Physics \textbf{33} (2003)148.

[35] \ \ M.Chaiichian, H.Grosse and P.Presnajer, \textit{Unitary
representations}

\ \ \ \ \ \ \ \textit{\ of the q-oscillator algebra}, J.Phys.A\textbf{\ 27}
(1994) 2045.

[36] \ \ R. Floreani and L.Vinet, \textit{Q-orthogonal polynomials and the
oscillator}

\ \ \ \ \ \ \ \textit{\ quantum group}, Lett.Math.Phys. \textbf{22} (1991)
45.

[37] \ \ H.Karabulut, \textit{Distributed \ Gaussian polynomials as
q-oscillator }

\ \ \ \ \ \ \ \textit{eigenfunctions},

\ \ \ \ \ \ \ \ J.Math.Phys.\textbf{47} (2006) 013508.

[38] \ \ A.Macfarlane, \textit{On q-analogues of the quantum harmonic
oscillator}

\ \ \ \ \ \ \ \textit{\ and }\ \textit{the quantum group SU(2)}$_{q},$
J.Phys.A\textbf{\ 22} (1989) 4581.

[39] \ H.Karabulut and E.Siebert, \textit{Distributed \ Gaussian polynomials}

\ \ \ \ \ \ \ \textit{\ and associated \ Gaussian quadratures,} J.Math.Phys.%
\textbf{38} (1997) 4815.

[40] \ \ A.Atakishiev and Sh.Nagiyev, \textit{On the Rogers-Szego polynomials%
},

\ \ \ \ \ \ \ \ J.Phys.\textbf{A 27} (1994) L611.

[41] \ \ G.Gasper and M.Rahman, \textit{Basic hypergeometric series},

\ \ \ \ \ \ \ \ Cambridge University Press, Cambridge, U.K. (1990).

[42] \ \ R.Koekoek and R.Swarttouw, \textit{The Askey- scheme of
hypergeometric}

\ \ \ \ \ \ \ \textit{orthogonal polynomials and its q-analogs, }%
arXiv:math/9602214.

[43] \ \ M.Ismail, Classical and quantum orthogonal polynomials of one

\ \ \ \ \ \ \ variable, Cambridge University Press, Cambridge, U.K. (2005).

[44] \ \ T.Nagao and T.Sasamoto,\textit{\ Asymmetric simple exclusion
process and}

\ \ \ \ \ \ \ \ \textit{\ modified random matrix ensembles}, Nucl.Phys.%
\textbf{B 699} (2004) 487.

[45] \ \ R.Stinchcombe and G.Schutz, \textit{Application of operator
algebras to }

\ \ \ \ \ \ \ \ \textit{stochastic dynamics and Heisenberg chain}, PRL 
\textbf{75} (1995) 140.

[46] \ \ \ T.Sasamoto, \textit{One-dimensional partially asymmetric simple
exclusion}

\ \ \ \ \ \ \ \ \ \textit{process with open boundaries: orthogonal
polynomials approach},

\ \ \ \ \ \ \ \ J.Phys.\textbf{A 32} (1999) 7109.

[47] \ \ \ R.Blythe, M.Ewans, F.Colaiori and F.Essler, \textit{Exact
solution of}

\ \ \ \ \ \ \ \textit{\ a partially }\ \textit{asymmetric exclusion model
using a deformed }

\ \ \ \ \ \ \ \ \textit{oscillator algebra, }J.Phys.\textbf{A 33} (2000)
2313.

[48] \ \ \ B.Derrida, M.Ewans, V.Hakim and V.Pasquer, \textit{Exact solution 
}

\ \ \ \ \ \ \ \ \textit{of a 1d asymmetric exclusion model using a matrix
formulation},

\ \ \ \ \ \ \ \ \ J.Phys \textbf{A 26} (1993) 1493.

[49] \ \ \ B.Derrida and K.Malick, \textit{Exact diffusion constant for the
one-}

\ \ \ \ \ \ \ \ \ \textit{dimensional \ partially asymmetric exclusion model}%
,

\ \ \ \ \ \ \ \ J.Phys.\textbf{A 30} (1997) 1031.

[50] \ \ \ M.Kardar, G.Parisi and Yi-Ch.Zhang, \textit{Dynamic scaling of
groving}

\ \ \ \ \ \ \ \textit{\ interfaces}, PRL \textbf{56} (1986) 889.

[51] \ \ \ T.Sasamoto, S.Mori and M.Wadati, \textit{One-dimensional
asymmetric }

\ \ \ \ \ \ \ \ \ \ \textit{exclusion model with open boundaries},

\ \ \ \ \ \ \ \ \ J.Phys.Soc.Japan \textbf{65} (1996) 2000.

[52] \ \ \ T. Oliviera, K.Dechoum, J.Redinz and F. Aarao Reis, \textit{%
Universal}

\ \ \ \ \ \ \ \ \textit{\ and nonuniversal }\ \textit{features of the
crossover from linear to}

\ \ \ \ \ \ \ \ \textit{\ nonlinear interface growth,}

\ \ \ \ \ \ \ \ \ Phys.Rev. \textbf{E 74} (2006) 011604.

[53] \ \ \ A. Lazarides, \textit{Coarse-graining a restricted solid-on
-solid model},

\ \ \ \ \ \ \ \ \ Phys.Rev.\textbf{E 73} (2006) 041605.

[54] \ \ \ D.Huse, \textit{Exact exponents for infinitely many new
multicritical points},

\ \ \ \ \ \ \ \ \ Phys.Rev.\textbf{B 30} (1984) 3908.

[55] \ \ \ D.Friedan, Z.Qui and S.Shenker, \textit{Conformal invariance,
unitarity}

\ \ \ \ \ \ \ \ \ \textit{and critical exponents in two dimensions}, PRL 
\textbf{52} (1984) 1575.

[56] \ \ \ J.de\ Gier and F.Essler, \textit{Exact spectral gaps of the
asymmetric}

\ \ \ \ \ \ \ \ \textit{\ exclusion process with open boundaries, }%
J.Sat.Mech.(2006) P12011\textit{.}

[57] \ \ \ V.Pasquer and H.Saleur, \textit{Common structures beetween finite 
}

\ \ \ \ \ \ \ \ \ \textit{systems \ and conformal field theories through
quantum groups},

\ \ \ \ \ \ \ \ \ Nucl.Phys. \textbf{B 330} (1990) 523.

[58] \ \ \ C.Gomez, m.Ruiz-Altaba and G.Sierra,

\ \ \ \ \ \ \ \ \ \textit{Quantum groups in two-dimensional physics},

\ \ \ \ \ \ \ \ \ Cambridge University Press, Cambridge, U.K. (1996).

[59] \ \ \ L.Faddeev and O.Tirkkonen, \textit{Connections of the Liouville
model}

\ \ \ \ \ \ \ \ \textit{\ and XXZ spin chain}, Nucl.Phys.\textbf{B453}
(1995) 647.

[60] \ \ \ A.Kholodenko, \ \textit{Kontsevich-Witten model from \ 2+1
gravity: }

\ \ \ \ \ \ \ \ \textit{New exact combinatorial solution}, J.Geom.Phys.%
\textbf{43} (2002) 45.

[61]\ \ \ P.Forrester, \textit{Visious random walkers in the limit of a large%
}

\ \ \ \ \ \ \ \ \textit{number of walkers, }J.Stat.Phys. \textbf{56} (1989)
767.

[62] \ \ T.Sasamoto, \textit{Fluctuations of the one-dimensional asymmetric}

\ \ \ \ \ \ \ \ \textit{exclusion process using random matrix techniques,}

\ \ \ \ \ \ \ \ J.Stat.Mech. (2007) P07007.

[63] \ \ A.Mukherjee and S.Mukhi, \textit{c=1 matrix models: equivalences }

\ \ \ \ \ \ \ \ \textit{and open-closed string duality}, JHEP \textbf{0510}
(2005) 099.

[64] \ \ J.Distler and C.Vafa, \textit{A critical matrix model at c=1},

\ \ \ \ \ \ \ \ \ Mod.Phys.Lett.\textbf{A 6} (1991) 259.

[65] \ \ D.Ghoshal and C.Vafa, \textit{c=1 strting as the topological theory}

\ \ \ \ \ \ \ \textit{\ on a conifold, }Nucl.Phys. \textbf{B 453} (1995) 121.

[66] \ \ D.Huse and M.Fisher, \textit{Commensurate melting, domain walls,}

\ \ \ \ \ \ \ \textit{\ \ and dislocations}, Phys.Rev.\textbf{B29} (1984)
239.

[67] \ \ M.Gaudin, \textit{La function d'onde de Bethe,}

\ \ \ \ \ \ \ \ Masson, Paris (1983).

[68] \ \ D.Grabiner, \textit{Brownian motion in a Weyl chamber,}

\ \ \ \ \ \ \ \ \textit{non -colliding particles, and random matrices},

\ \ \ \ \ \ \ \ arXiv: math.RT/9708207.

[69] \ \ C. Krattenhaler, \textit{Asymptotics for random walks in alcoves }

\ \ \ \ \ \ \ \ \textit{of affine Weyl groups}, arXiv: math/0301203.

[70] \ \ S.de Haro, \textit{Chern-Simons theory, 2d Yang -Mills, }

\ \ \ \ \ \ \ \ \textit{and Lie algebra wanderers}, Nucl.Phys. \textbf{B730}
(2005) 313.

[71] \ \ M.Mehta, \textit{Random matrices}, Elsevier, Amsterdam (2004).

[72] \ \ J.Maldacena, G.Moore, N.Seiberg and D.Shih,

\ \ \ \ \ \ \ \ \textit{Exact vs. semiclassical target space of the minimal
string,}

\ \ \ \ \ \ \ \ JHEP \textbf{0410} (2004) 020.

[73] \ \ K.Okuyama, \textit{D-brane amplitudes in topological string on
conifold,}

\ \ \ \ \ \ \ \ Phys.Lett.\textbf{B 645} (2007) 273.

[74] \ \ M.Tierz, \textit{Soft matrix models and Chern-Simons partition
functions},

\ \ \ \ \ \ \ \ Mod.Phys. \textbf{A 19} (2004) 1365.

[75] \ \ P.Etingof, I.Frenkel and A.Kirillov Jr., \textit{Lectures on
representation }

\ \ \ \ \ \ \ \textit{theory and Knizhnik-Zamolodchikov equations},

\ \ \ \ \ \ \ AMS Publishers, Providence, R.I. (1998).

[76] \ \ A.Chervov and D.Talalaev, \textit{Quantum spectral curves, quantum}

\ \ \ \ \ \ \ \textit{integrable systems and the geometric Langlands
correspondence},

\ \ \ \ \ \ \ arXiv: hep-th/0604128.

[77] \ E.Frenkel, \textit{Langlands correspondence for loop groups},

\ \ \ \ \ \ \ Cambridge University Press, Cambridge, U.K. (2007).

[78] \ \ E.Frenkel and E.Witten, \textit{Geometric endoscopy and mirror
symmetry},

\ \ \ \ \ \ \ arXiv: 0710.5939.

[79] \ R.Richardson and N.Sherman, \textit{Exact eigenvalues of the
pairing-force }

\ \ \ \ \ \ \ \textit{Hamiltonian}, Nucl.Phys.\textbf{52} (1964) 221.

[80] \ R.Richardson, \textit{Exactly solvable many-boson model}, JMP \textbf{%
9} (1968) 1327.

[81] \ N.Vilenkin, \textit{Special functions and theory of group
representations},

\ \ \ \ \ \ \ Nauka, Moscow (1991).

[82] \ J.Dukelsky, S.Pittel and G.Sierra, \textit{Exactly solvable }

\ \ \ \ \ \ \textit{Richardson-Gaudin \ models for many-body }

\ \ \ \ \ \ \textit{quantum systems}, Rev.Mod.Phys. \textbf{76} (2004) 643.

[83] \ A.Gorsky, \textit{Gauge theories as string teories: the first results}%
,

\ \ \ \ \ \ \ arXiv: hep-th/0602184.

[84] \ A.Balantekin, T.Dereli and Y.Pehlivan, \textit{Exactly solvable
pairing model}

\ \ \ \ \ \ \ \textit{using an extension of Richardson-Gaudin approach},

\ \ \ \ \ \ \ Int.J.Mod.Phys.\textbf{E 14} (2005) 47.

[85] \ A.Ushveridze, \textit{Quasi-exactly solvable models in quantum
mechanics},

\ \ \ \ \ \ \ IOP Publishing Ltd., Philadelphia (1994).

[86] \ A.Ovchinnikov, \textit{Exactly solvable discrete BCS-type Hamiltonians%
}

\ \ \ \ \ \ \ \textit{and the six-vertex model}, Nucl.Phys. \textbf{B707}
(2002) 362.

[87] \ M.Alford, A.Schitt, K.Rajagopal and Th.Schafer,

\ \ \ \ \ \ \ \textit{Color superconductivity in quark matter},

\ \ \ \ \ \ \ arXiv: 0709.4635.

[88] \ L.Cooper, \textit{Boubd electron pairs in a degenerate electron gas},

\ \ \ \ \ \ \ Phys.Rev.\textbf{104} (1956) 1189.

[89] \ A.Balantekin, J.de Jesus and Y.Pehlivan, \textit{Spectra and symmetry
in}

\ \ \ \ \ \ \textit{\ nuclear pairing}, Phys.Rev. C \textbf{75} (2007)
064304.

[90] \ C.Lovelace, \textit{A novel application of Regge trajectories},

\ \ \ \ \ \ \ Phys.Lett.B \textbf{28} (1968) 264.

[91] \ A.Kholodenko, \textit{Traces of mirror symmetry in Nature},

\ \ \ \ \ \ \ International Math.Forum \textbf{3} (2008) 151.

[92] \ \ S.Adler, Consistency conditions on a strong interactions

\ \ \ \ \ \ \ \ implied by a partially conserved axial-vector current,

\ \ \ \ \ \ \ Phys.Rev. \textbf{137} (1965) B1022.

[93] \ \ A.Kholodenko, \textit{Towards physically motivated proofs}

\ \ \ \ \ \ \ \textit{of the Poincare and geometrization conjectures,}

\ \ \ \ \ \ \ J.Geom.Phys.\textbf{58} (2008) 259.

[94] \ \ A.Mekjian, \textit{Cluster distribution in physics and genetic}

\ \ \ \ \ \ \ \ \textit{diversity}, Phys.Rev.A \textbf{44}(1991) 8361.

[95] \ \ W.Ewens, \textit{Mathematical population genetics},

\ \ \ \ \ \ \ \ Springer-Verlag, Berlin (2004).

[96] \ \ G.Watterson, \textit{The stationary distribution of the}

\ \ \ \ \ \ \ \ \textit{infinitely-many neutral alleles diffusion model},

\ \ \ \ \ \ \ \ J.Appl.Probability \textbf{13} (1976) 639.

[97] \ \ M.Aoki, \textit{Open models of share markets with two }

\ \ \ \ \ \ \ \ \textit{dominant types of participants}, J.of Economic

\ \ \ \ \ \ \ \ Behaviour \& Organization \textbf{49} (2002) 199.

[98] \ \ J.Kingman, \textit{The population structure associated }

\ \ \ \ \ \ \ \ \textit{with the Ewens sampling formula},

\ \ \ \ \ \ \ \ Theoretical Pop.Biology \textbf{11} (1977) 274.

[99] \ \ H.Harari, \textit{Duality diagrams}, PRL \textbf{22} (1969) 562.

[100] \ D.Leach, \textit{Genetic recombination},

\ \ \ \ \ \ \ \ Blackwell Science Ltd., Oxford, U.K. (1996).

[101] \ P.Freund, \textit{Two component duality and strings},

\ \ \ \ \ \ \ \ arXiv: 0708.1983.

[102] \ S.Mirkin, \textit{Structure and biology of H DNA},

\ \ \ \ \ \ \ \ in Triple Helix Forming Oligonucleotides, p195,

\ \ \ \ \ \ \ \ Kluwer Academic, Boston (1999).

[103] \ D.McQuarrie, \textit{Stochastic approach to chemical kinetics},

\ \ \ \ \ \ \ \ J.Appl.Prob. \textbf{4} (1967) 413.

[104] \ I.Darvey, B.Ninham and P.Staff, \textit{Stochastic models for}

\ \ \ \ \ \ \ \ \textit{second-order chemical reaction kinetics. }

\ \ \ \ \ \ \ \ \textit{The equilibrium state}, J.Chem.Phys.\textbf{45}
(1966) 2145.

[105] \ J.Kingman, \textit{The dynamics of neutral mutation},

\ \ \ \ \ \ \ \ \ Proc.Roy.Soc. London A\textbf{\ 363} (1978) 135.

[106] \ \ A. Lijoi and E.Regazzini, \textit{Means of a Dirichlet process}

\ \ \ \ \ \ \ \ \ \textit{and multiple hypergeometric functions},

\ \ \ \ \ \ \ \ \ Ann.Probability \textbf{32} (2004) 1469.

[107] \ \ H.Bateman and A.Erdelyi, \textit{Higher transcendental functions},

\ \ \ \ \ \ \ \ \ \ Vol.1. McGraw Hill, New York (1953).

[108] \ \ R. Blythe and M.Evans, \textit{Nonequilibrium steady states of }

\ \ \ \ \ \ \ \ \ \textit{matrix-product form:} a solver guide, J.Phys.A 
\textbf{40} (2007) R333.

[109] \ \ G.Sch\"{u}tz, \textit{Exact solution of the master equation for
the }

\ \ \ \ \ \ \ \ \ \textit{asymmetric exclusion process}, J.Stat.Phys. 
\textbf{88} (1997) 427.

[110] \ \ L.Gwa and H.Spohn, \textit{Bethe solution for the dynamical-scaling%
}

\ \ \ \ \ \ \ \ \textit{\ exponent of the noisy Burgers equation,}
Phys.Rev.A \textbf{46} (1992) 844.

[111] \ \ F.Alcaraz, M.Barber, M.Batchelor, R.Baxter and G.Quispei,

\ \ \ \ \ \ \ \ \textit{\ Surface exponents of the quantum XXZ,
Ashkin-Teller and}

\ \ \ \ \ \ \ \ \ \textit{Potts models}, J.Phys.A \textbf{20} (1987) 6397.

[112] \ \ T.Sasamoto and M.Wadati, \textit{Dynamic matrix product ansatz and}

\ \ \ \ \ \ \ \ \textit{\ Bethe ansatz equation for asymmetric exclusion
process }

\ \ \ \ \ \ \ \ \ \textit{with periodic boundary}, J.Phys.Soc.Jpn. \textbf{66%
} (1997) 279.

\textit{\ \ \ \ \ \ }

\bigskip

\bigskip

\bigskip

\bigskip

\bigskip

\bigskip

\end{document}